\documentclass[prx,twocolumn,superscriptaddress,showpacs,nofootinbib,longbibliography]{revtex4-2}
\usepackage{graphicx}
\usepackage[normalem]{ulem}
\usepackage{braket}
\usepackage{epsfig}
\usepackage{amssymb,amsmath,amsfonts,hyperref}
\usepackage{wasysym}
\usepackage{latexsym}
\usepackage{xcolor}
\definecolor{je}{rgb}{0.858, 0.188, 0.478}
\definecolor{so}{rgb}{0.19,0.32,1.52}
\usepackage{verbatim}
\usepackage{chemformula}
\usepackage{tikz}
\usepackage{lipsum}
\newcommand{\uu}{\mathcal{U}}
\newcommand{\vv}{\mathcal{V}}
\def\Tr{\mathrm{Tr}}
\newcommand{\AB}{{\texorpdfstring{\text{AB}}{ab}}}
\newcommand{\ABB}{{\texorpdfstring{\text{AB$^*$}}{abb}}}
\begin{document}

\title{Statistical mechanics of dimers on quasiperiodic Ammann-Beenker tilings}

\author{Jerome Lloyd}\thanks{These authors contributed equally.} 
\affiliation{Rudolf  Peierls  Centre  for  Theoretical  Physics, Parks Road, Oxford  OX1  3PU,  United  Kingdom}
\affiliation{School of Physics and Astronomy, University of Birmingham, Edgbaston Park Road, Birmingham, B15 2TT, United Kingdom}
\affiliation{Department of Theoretical Physics, University of Geneva, 24 quai Ernest-Ansermet, 1211 Genève 4, Switzerland}
\author{Sounak Biswas}\thanks{These authors contributed equally.} 
\affiliation{Rudolf Peierls Centre for Theoretical  Physics, Parks Road, Oxford OX1 3PU, United Kingdom}
\author{Steven H. Simon}
\affiliation{Rudolf Peierls Centre for Theoretical Physics, Parks Road, Oxford OX1 3PU, United Kingdom}
\author{S. A. Parameswaran}
\affiliation{Rudolf Peierls Centre for Theoretical Physics, Parks Road, Oxford OX1 3PU, United Kingdom}
\author{Felix Flicker}
\affiliation{Rudolf Peierls Centre for Theoretical Physics, Parks Road, Oxford OX1 3PU, United Kingdom}
\affiliation{School of Physics and Astronomy, Cardiff University, The Parade, Cardiff CF24 3AA, United Kingdom}

\begin{abstract}
We study classical dimers on two-dimensional quasiperiodic Ammann-Beenker (\AB) tilings. Using the discrete scale-symmetry of quasiperiodic tilings,
we prove that each infinite tiling admits `perfect matchings', where every vertex is touched by one dimer. We show the appearance of so-called \emph{ monomer pseudomembranes}. These are sets of edges which collectively host exactly one dimer, which bound certain eightfold-symmetric regions of the
tiling. Regions bounded by pseudomembranes are matched together in a way that resembles perfect matchings of the tiling itself. These structures
emerge at all scales, suggesting the preservation of collective dimer fluctuations over long distances. We provide numerical evidence, via Monte Carlo
simulations, of dimer correlations consistent with power laws over a hierarchy of different lengthscales. We also find evidence of rich monomer
correlations, with monomers displaying a pattern of attraction and repulsion to different regions within pseudomembranes, along with signatures of
deconfinement within certain annular regions of the tiling.
\end{abstract}

\vskip2pc
\maketitle

%
\section{Introduction}
\label{sec:introduction}
%

Dimer models have long attracted interest as elegant routes to capture the interplay of local constraints and lattice geometry. The classical dimer problem is a combinatorial one, counting the number of configurations of hard rods on a lattice. It was solved in the early sixties for planar graphs by Pfaffian techniques~\cite{Kasteleyn61,Fisher61,FisherTemperley61}, a solution which exposed a link to the free fermion models. Much of the early interest was indeed spurred by connections to the planar Ising model~\cite{fisher1966dimer, hurst1960new}, which famously falls into this class. Classical dimers have continued to play an important role within physics, mathematics, and computer science communities over the last century owing to their ubiquity in problems of constraint satisfaction, optimization, and  combinatorics~\cite{Kenyon_ICTP,Moessner_Raman,Lovasz_Plummer,Chalker_review}. Many interesting questions remain concerning the role of dimensionality, geometry and topology of the underlying graph on the emergent physics.

Perhaps the principal reason to study classical dimer systems, however, is to better understand their quantum counterparts. Quantum dimer models were introduced by Rokhsar and Kivelson~\cite{RokhsarKivelson88,KivelsonEA87,MoessnerSondhi01,Moessner_Raman} as effective descriptions of short-range resonating valence bond physics in high-temperature superconductors~\cite{FazekasAnderson74,Anderson87}. They have since outgrown this original motivation and now rank among the paradigmatic models of quantum statistical mechanics. They are known to host a rich variety of phases and phase transitions~\cite{MoessnerSondhiChandra,LeungChiuRunge,Syljuasen,RalkoPoiblancMoessner,NikolicSenthil,FradkinEA04,VishwanathEA04}, including both gapped and gapless quantum spin liquids~\cite{MoessnerSondhi01, RalkoEA05,MisguichSerbanPasquier,Ivanov,MoessnerSondhiRVB_3D,HuseEA03,HermeleEA04} and deconfined quantum critical points~\cite{MoessnerSondhiFradkin,SenthilEA1,*SenthilEA2,SandvikDCP}, whose emergent gauge structure and fractionalized excitations have particularly intuitive descriptions in terms of dimers~\cite{Moessner_Raman}. More recently, their local constraints have been proposed as a route to glassy quantum dynamics and slow thermalization~\cite{Oakes_Garrahan_Powell,Lan_Powell,Theveniaut_etal,Feldmeier_Pollman_Knap}. The classical dimer model provides basic intuition toward the quantum problem. Furthermore, at exactly one point of the zero-temperature phase diagram --- the critical Rokhsar-Kivelson point --- the wavefunction is given by the equal amplitude superposition of all classical dimer states\footnote{This statement is strictly only true if the ground state is non-degenerate, but is simply generalised to the case with degeneracies.}. The ground state dimer correlations at the Rokhsar-Kivelson point match the infinite temperature correlations of the classical model. The classical dimer model thus serves as an important starting point to the quantum model. 

Here we consider classical dimer models on bipartite graphs~\cite{Kasteleyn61,Kasteleyn63,Fisher61,FisherTemperley61,HeilmannLieb,Fendley_Moessner_Sondhi,RaghavanHenley,KrauthMoessner03,HuseEA03,Alet_etal, Wang_Wu_Kagome,Wu, Alet_etal,Wildeboer20,DijkgraafEA07,Kenyon02,Kenyon99,KenyonEA00,KenyonOkounkov07,KenyonOkounkov06}. A graph is a set of vertices connected by edges. It is bipartite if its vertices can be partitioned into two mutually exclusive sets such that there are no edges between vertices belonging to the same set (same bipartite `charge'). Dimers are placed on the edges such that each vertex connects to zero or one dimers (a hard-core constraint). This defines a dimer covering, or \emph{matching}. An unmatched vertex not connected to a dimer is termed a monomer. A monomer-free configuration, if one exists, is called a perfect matching.

The perfect matchings of graphs admitting planar embeddings can be counted exactly using the previously mentioned Pfaffian techniques. In practice such methods amount to diagonalisation of the so-called Kasteleyn matrix, which in many cases can be done analytically via Fourier analysis: if the graph lacks translational invariance, however, no simple analytical form is guaranteed to exist. Numerical solution becomes computationally demanding for large systems (generally we are interested in the thermodynamic limit), and does not offer the same insight. A more intuitive perspective is afforded by the height representation~\cite{Nienhuis_Hilborst_Blote,Blote_Hilborst,Henley97,Zheng_Henley}, particularly when it is applied to periodic bipartite lattices that admit perfect matchings. On such lattices, the statistical mechanics of dimer configurations can be understood by mapping dimer coverings to configurations of an integer-valued `height' field on edges of the dual lattice. The hard-core constraint becomes a zero-divergence condition on this field --- i.e. a Gauss law --- allowing it to be re-expressed as the lattice curl of a scalar (vector) height variable in 2D (3D). The height mapping is most useful when entropically favoured dimer coverings correspond to locally flat height configurations. Dimer correlations can then be deduced using a coarse-grained free energy density for the height field, taking a local Gaussian form at long wavelengths. Monomers appear as vortex defects of the height field. In 2D, if the microscopic parameters correspond to the vorticity being irrelevant (as on the square and honeycomb lattices), the height model is in a \emph{rough} phase, implying critical (power law) dimer correlations and logarithmic confinement of monomers (i.e. the free energy cost of a pair of test monomers in an otherwise-perfect matching diverges logarithmically with their separation). When the vorticity is relevant (as on the square lattice with additional aligning interactions~\cite{Alet_etal}), the height model is in its \emph{flat} phase, with exponentially decaying connected correlations of dimers and linearly confined monomers. Since vortex defects are never relevant in 3D, dimer correlations are always algebraic, and monomers are deconfined. That is, a test pair can be separated to arbitrary distance with finite free energy cost. However, these arguments rely on both the existence of perfect matchings and the identification of locally flat height configurations with high-probability configurations. Neither is guaranteed for a generic bipartite graph.

Classical dimers have also been studied in settings with disorder, such as random regular graphs and Erd\H{o}s-Renyi random graphs~\cite{ErdosRenyi,AlbericiEA15,KarpSipser,Frieze86}, using approaches that are asymptotically exact in the thermodynamic limit~\cite{Zdeborova_Mezard,Zhou_Ou-Yang}. However, the absence of conventional spatial locality in these ensembles rules out any simple generalization of the notions of dimer correlations and monomer confinement.

\begin{figure*}
    \includegraphics[width=0.9\textwidth]{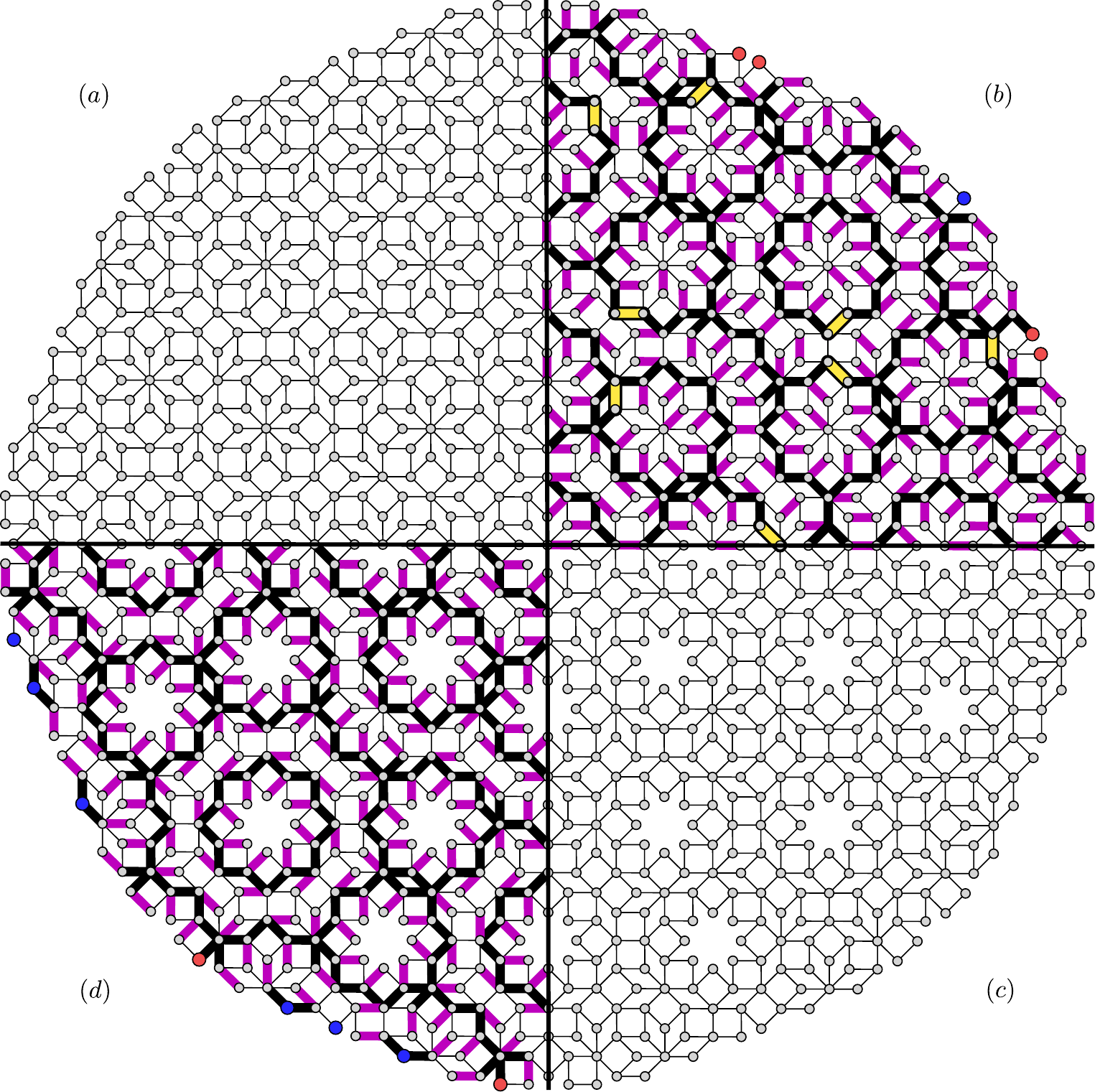}
    \caption{\emph{Clockwise:} \emph{(a)} A finite patch of the Ammann-Beenker (AB) tiling. Composed of copies of square and rhombus tiles, the
    (infinite) tiling covers the plane in an ordered fashion, yet never repeats periodically. \emph{(b)} A maximum matching (dimer covering) of a
patch of the \AB\ tiling. The tiling can be perfectly matched in the thermodynamic limit; on finite patches, an $O(1)$ \emph{number} of unmatched
vertices (monomers) generally appear, which can be moved to the boundary. Thick black lines indicate links which comprise overlapping pseudomembranes,
each of which collectively host one dimer. Yellow edges indicate a dimer on a pseudomembrane. \emph{(c)} A patch of the \ABB\ tiling, obtained from AB
by removing all 8-connected vertices. \emph{(d)} A maximum matching of a patch of the \ABB\ tiling. Pseudomembranes become membranes, hosting zero
dimers and leading to a decoupling of dimers residing on different \emph{ladders}.}
    \label{fig:ABwindow}
\end{figure*}

\begin{figure*}
    \includegraphics[width=0.49\textwidth]{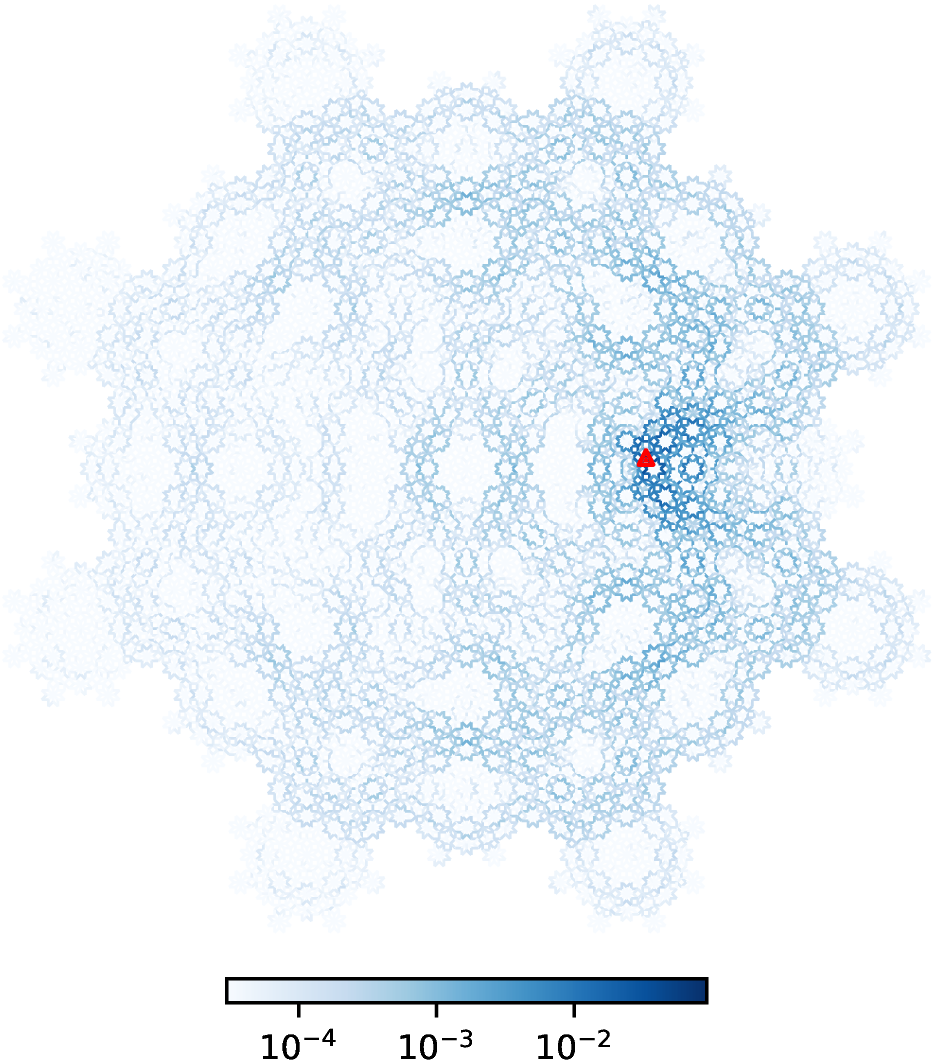}
    \includegraphics[width=0.49\textwidth]{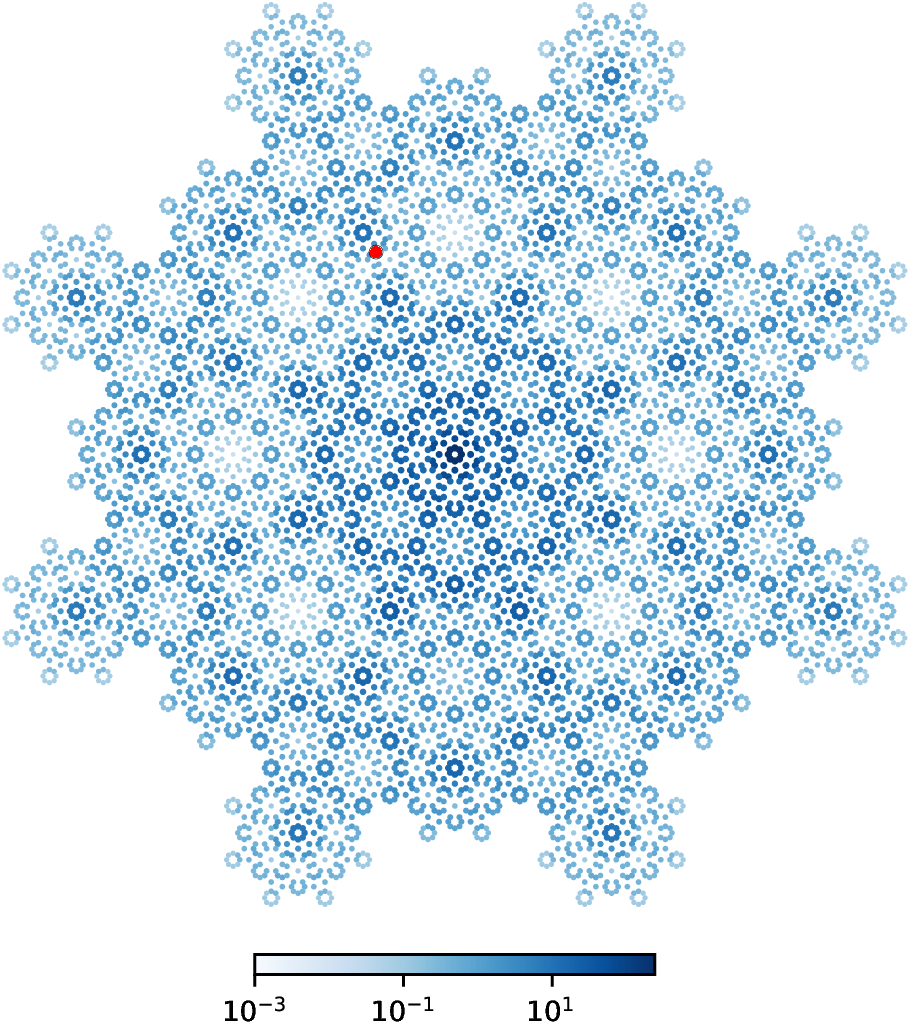}
\caption{\emph{Left:} The $e_j$-dependence of the connected dimer correlations, $|C(e_0,e_j)|$ (Eq.~\eqref{eq:dimcorr}). The source edge $e_0$, indicated by the solid red triangle, connects an $8_2$-vertex to a nearby ladder. The resulting correlations are typical of slowly decaying examples, and resemble a power law. (See Sec.~\ref{sec:numerics}). \emph{Right:} Density plot of the monomer correlation function $Z(x,y)$, for the $8_4$-vertex. $Z(x,y)$ is displayed as a function of $y$, with $x$ fixed at a vertex marked by a solid red dot. }  \label{fig:largecorrs}
\end{figure*}

Recent work has explored the problem of classical dimer models on Penrose tilings~\cite{Flicker_Simon_Parameswaran}. These are infinite tilings of the plane constructed from two types of tiles. The tiles fit together without gaps or defects, in such a way that no patch can be tessellated periodically to reproduce the pattern~\cite{Penrose74,GrunbaumShephard,BaakeGrimm}. Despite lacking the discrete translational symmetries of crystal lattices, they nevertheless display long-range order. For example, their Fourier transforms, which are tenfold rotationally symmetric, feature sharp Bragg peaks which can be labelled by a finite number of wave vectors. This latter condition defines the Penrose tiling to be \emph{quasiperiodic}~\cite{BaakeGrimm}. Penrose tilings came to prominence in the physics community with the discovery of \emph{quasicrystals}, real materials whose atoms are arranged quasiperiodically~\cite{Schechtman}. Considering the edges and vertices of the tiles as those of a bipartite graph, Penrose tilings do not admit perfect matchings despite having no {\it net} imbalance in their bipartite charge~\cite{Flicker_Simon_Parameswaran}. Instead, they have a finite density of monomers in the thermodynamic limit. The \emph{maximum matchings} on Penrose tilings, which contain the maximum number of dimers, have an unusually rich underlying structure, quite distinct from either periodic or random systems. In general maximum matchings, monomers can be thought of as moving via dimer re-arrangements. On Penrose tilings, monomers are always confined within regions bound by nested loops, or \emph{monomer membranes}. These membranes are composed of edges which do not host a dimer in any maximum matching. Each such region has an excess of vertices belonging to one or the other bipartite charge, and hosts a corresponding number of monomers. Adjacent regions have monomers of the opposite bipartite charge. The properties of these membranes and the regions that they enclose follow directly from the dimer constraint and the underlying symmetry of the tiling, and can hence be precisely determined. Ref.~\onlinecite{Biswas_etal} identified similar monomer-confining regions, separated either by membranes or perfectly matched regions, as components of the Dulmage-Mendelsohn decomposition of generic bipartite graphs~\cite{Dulmage_Mendelsohn,Pothen_Fan,HararyPlummer,Irving06}. This was used to investigate phase transitions of such monomer-confining regions in ensembles of periodic lattices with random vertex dilution, such as those used to model vacancy disorder in quantum magnets.

While both Refs.~\onlinecite{Flicker_Simon_Parameswaran} and \onlinecite{Biswas_etal} consider bipartite dimer models, the usual mapping to height models does not apply to quasiperiodic graphs, or graphs where vertices can have different co-ordination numbers. While a more general height mapping is possible in principle~\cite{Kenyon_review}, the resulting height functions typically do not lead to analytically tractable coarse-grained continuum free energy functionals. This is because there is no longer a simple relationship between the {\it local} configuration of the height field and the statistical weight of the {\it global} dimer covering. In any case, due to a sizeable density of monomers in the cases studied in Refs.~\onlinecite{Flicker_Simon_Parameswaran} and \onlinecite{Biswas_etal}, connected correlation functions of dimers are short-ranged and more or less unremarkable. In addition, monomer correlation functions are non-monotonic and strongly site-dependent, making it challenging to define a crisp notion of monomer confinement. These facts challenge the goal of a precise characterization of long-wavelength properties of dimer models in quasiperiodic environments.

\subsection{Results}

In this work, we meet this challenge in the setting of a distinct quasiperiodic dimer problem for which we can make a series of exact, and asymptotic,
statements. Specifically, we study classical dimers on the Ammann-Beenker (\AB) tiling, shown in Figures \ref{fig:ABwindow}a and ~\ref{fig:ABwindow}b. This tiling has been the
topic of much recent attention, with investigations in the context of magnetism~\cite{ThiemChalker15A,ThiemChalker15B,Koga},
superconductivity~\cite{AraujoAndrade}, critical eigenstates~\cite{NicolasEA17}, protected Majorana modes~\cite{VarjasEA19} and topological
insulators~\cite{ChenEA2020}. Like the Penrose tilings, \AB\ tilings exhibit discrete scale invariance: \emph{deflations} (vertex decimations followed
by rescaling lengths by the irrational silver ratio) map an \AB\ tiling to another \AB\ tiling. We prove that these tilings host perfect matchings in the thermodynamic limit, in contrast to the Penrose tilings investigated in Ref.~\cite{Flicker_Simon_Parameswaran}. Our proof makes use of the discrete scale invariance characteristic of quasicrystals: we find that at any given coarse-graining scale, special vertices that are left invariant by double deflations lead to the preservation of strong dimer-dimer correlations at the next scale. This suggests that certain regions associated with these vertices retain mutual dimer correlations. The preserved vertices have edge-co-ordination eight and we refer to them as `8-vertices'.

The success of this iterative construction of dimer coverings motivates us to consider an auxiliary problem on a related graph that we dub the
\ABB~tiling, Fig.~\ref{fig:ABwindow}c. This is obtained from the \AB~tiling by removing all the 8-vertices. The \ABB~tiling is also perfectly matched
and does not host monomers, and we show that it decouples into perfectly matched one-dimensional regions that we call \emph{ladders}, separated by
membranes. These membranes, like those in Refs.~\onlinecite{Flicker_Simon_Parameswaran} and \onlinecite{Biswas_etal}, are composed of edges between
different ladders that do not host a dimer in any maximum matching. However, unlike the monomer-confining regions of
Refs.~\onlinecite{Flicker_Simon_Parameswaran} and \onlinecite{Biswas_etal}, they separate perfectly matched regions. This structure allows us to
systematically treat the matching problem on the \AB~tiling once the 8-vertices are reintroduced. We show that membranes in the \ABB~tiling become
\emph{pseudomembranes} in the full \AB~tiling, concentric with the 8-vertices: the edges belonging to every pseudomembrane now collectively host exactly one dimer, as the monomer reinstated on the central 8-vertex of the region `crosses' the membranes in order to be matched with monomers from other 8-vertices.

Each 8-vertex is surrounded by at least one pseudomembrane. Loosely speaking, a double deflation maps this 8-vertex and its surrounding pseudomembrane-bounded region into a single vertex at the next scale. Since each pseudomembrane is crossed by exactly one dimer, this procedure preserves an ``effective dimer-constraint" at each successive scale--- thereby leading to effective matching problems at each scale. This remarkable property provides a heuristic picture of the emergence of correlated dimer fluctuations over large distances. It also suggests that the \text{AB} tiling is an intriguing example of a lattice-level (rather than continuum) coarse-graining of a constrained system that faithfully imposes the constraint at each successive decimation scale~\footnote{We emphasize the deterministic nature, since real-space decimations can have an especially simple structure in {\it random} systems.}.

We show that the discrete scale invariance exhibited by perfect matchings leaves its imprint in both dimer and monomer correlations, as exhibited in Fig.~\ref{fig:largecorrs}. Investigating connected dimer correlations numerically, we show that, while they are strongly anisotropic and site-dependent, the connected correlation functions of certain dimers have slow decays, consistent with power laws, cutoff by a set of lengthscales going up to the system size. This is particularly striking given the absence of the continuum height description
which mandates power law correlations in periodic bipartite lattices~\cite{Zheng_Henley,Chalker_review}. We then turn to monomer correlations, where we observe a pattern of `charged' correlations set by the pseudomembranes, whereby a monomer is strongly (weakly) correlated to monomers within pseudomembranes centred on 8-vertices of opposite (equal) bipartite charge to the monomer. We further show, that in the angular direction around an 8-vertex, monomer-monomer correlations asymptote to finite constant values at large
separation --- that is, a pair of monomers are \emph{deconfined} with respect to annular separation. We argue that these behaviours arise as a
consequence of effective matching problems at each scale. Finally, we investigate the phase diagram of dimer models on both the \AB~tiling and \ABB~tiling in the presence of a classical aligning interaction resembling the Rokhsar-Kivelson potential term, as a step on the road towards a study of quasiperiodic quantum dimer models.

The remainder of this paper is organized as follows. We introduce the necessary background on dimer covers and \AB~tilings in
Sec.~\ref{sec:background}. In Sec.~\ref{sec:perfect_matching_proof} we prove that perfect matchings exist for \AB~tilings in the thermodynamic limit.
This construction leads to the introduction of the auxiliary \ABB~tiling, which is also perfectly matched. In Sec.~\ref{sec:memproof} we prove the
existence of membranes in the \ABB~tiling; in Sec.~\ref{sec:quasiproof} we demonstrate how these become pseudomembranes in the \AB~tiling, and in
Sec.~\ref{sec:emp}, we explain how the pseudomembranes lead to the structure of effective matchings at different scales across the tiling. Turning to
numerical results on the \AB~tiling, after outlining our choice of samples and boundary conditions in Sec.~\ref{sec:boundaryconditions}, we identify
connected dimer correlations consistent with power laws (Sec.~\ref{sec:dimercorrs}), and long range monomer correlations with regions of annular
deconfinement (Sec.~\ref{sec:monocorrs}). In Sec.~\ref{sec:aligning_ints} we investigate the effects of including a classical aligning
interaction in both the \AB\ and  \ABB~tilings. We provide concluding remarks in Sec.~\ref{sec:conclusions}.

%
\section{Background}
\label{sec:background}
%

%
\subsection{Dimer models and graph theory}
\label{sec:matching_review}

In this section we introduce the necessary terminology to discuss dimer coverings on graphs. The graphs of interest in this paper have two important properties. First, they are \textit{bipartite}, meaning the vertices can be partitioned into two mutually exclusive subsets, $\mathcal{U}$ and $\mathcal{V}$, so that every vertex in $\mathcal{U}$ ($\mathcal{V}$) only has edges to vertices in $\mathcal{V}$ ($\mathcal{U}$). If two vertices belong to the same subset, we will say they have equal (bipartite) \emph{charge}; otherwise, they are oppositely charged. Second, the graphs admit planar embeddings. We keep the geometry of the tiling, although strictly speaking only the graph topology is relevant to the matching problem.

\begin{figure}[h!]
    \includegraphics[width=\linewidth]{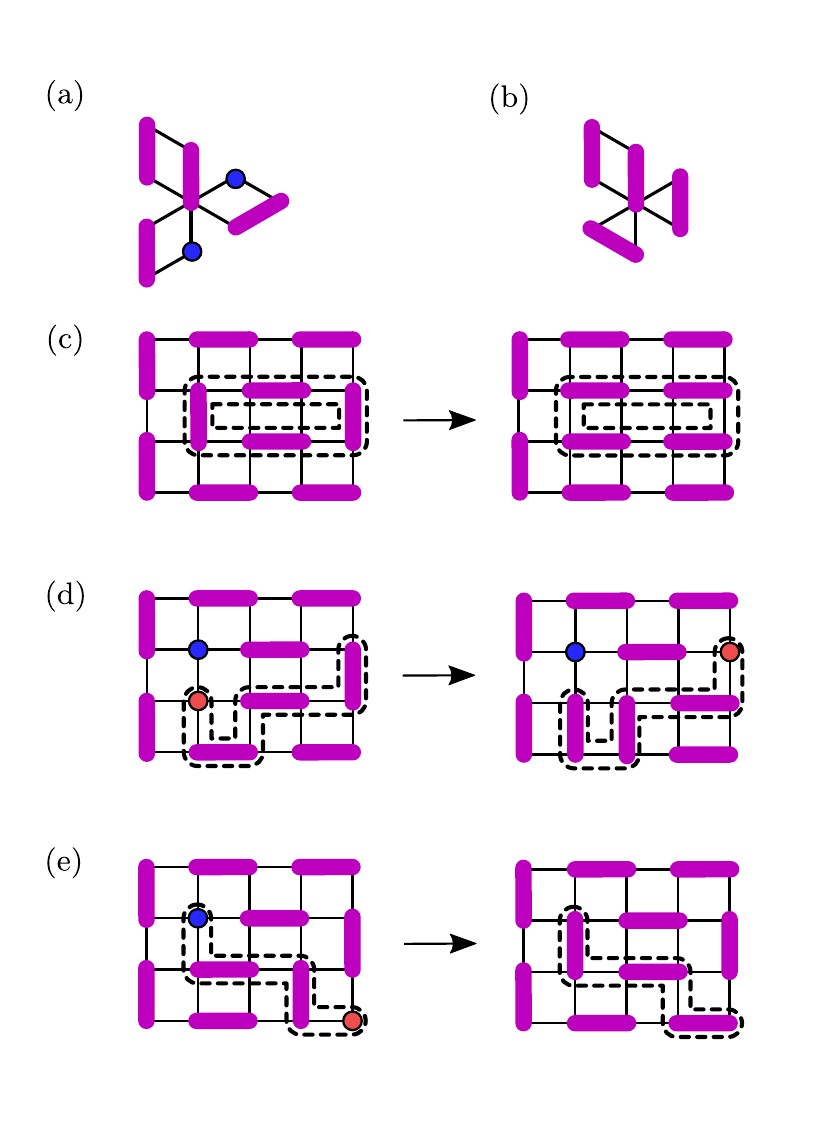}
    \caption{(a) A maximum matching which is not perfect. Unmatched vertices are marked with a blue circle. (b) A perfect matching. (c) An alternating cycle; augmenting  this cycle --- swapping covered and uncovered edges --- yields a new matching with the same number of dimers. (d) An even-length alternating path terminating at a monomer; upon augmenting this path, the monomer is transported to the other end of the path. (e) Augmenting an odd-length alternating path between monomers, or \emph{augmenting path}, annihilates the two monomers in favour of a dimer.}
  \label{fig:matchings}
\end{figure}

A \textit{matching} of a graph is a subset of edges such that no vertex is incident with more than one edge in the subset \cite{Gibbons}. A matching is equivalent to a dimer configuration, as edges in the matching can be covered by dimers, with no vertex touching more than one dimer. A vertex connected to an edge in the matching is said to be \textit{matched}; an unmatched vertex is called a \textit{monomer}. A \textit{perfect} matching is a matching with every vertex matched or, equivalently, a monomer-free dimer covering. Not every graph admits a perfect matching --- a simple counterexample is any graph with an odd number of vertices. A \textit{maximum} matching is a matching with the maximum number of dimers (minimum number of monomers): Fig.~\ref{fig:matchings}a-b shows examples of maximum and perfect matchings. If a graph admits a perfect matching, then any maximum matching is necessarily perfect. When considering the limit of infinite system size, as we will do here, a sensible definition of perfect matchings requires that the density of monomers vanishes, in this limit. This avoids the unnecessary complication of dealing with a small, $O(1)$ number of unpaired boundary vertices--- the exact number of such monomers depends on how exactly one chooses to terminate the system, and does not scale with the size of the graph. 

Usually a graph will have multiple maximum or perfect matchings and, from a statistical mechanics perspective, our interest is in this full configuration space, possibly up to certain constraints (e.g.\ we may only consider perfectly matched configurations, or penalise matchings with certain alignments of dimers). An \emph{alternating path} is a connected subset of edges along which edges are alternately covered and uncovered by dimers. Starting from any matching, switching which edges are covered/uncovered in a closed alternating path (\emph{alternating cycle}) results in another matching with the same number of dimers (see Fig.~\ref{fig:matchings}c). Switching those edges in an open odd-length alternating path changes the number of monomers in the matching  --- for example, two monomers can be created by removing one dimer. This process of switching covered and uncovered edges is known as \textit{augmenting} the path. All alternating cycles are of even length
(on a bipartite graph this statement is trivial since all cycles are of even length). 

On a bipartite graph, monomers are assigned a charge depending on which bipartite subgraph they reside on, and are able to `move' through the dimer vacuum along alternating paths: taking an even-length alternating path with one end terminating on one of the monomers, augmenting the path results in the monomer being translated to the opposite end of the path, while conserving the monomer's charge (see Fig.~\ref{fig:matchings}d). The creation of a monomer pair on a bipartite graph has a physical analogy in the excitation of a particle-antiparticle or defect pair above the vacuum state (viewed as one where all monomers are paired into neutral dimers). Given a maximum matching, all other maximum matchings in the configuration space can be reached by combinations of two basic `moves': (i) augmenting alternating cycles; and (ii) transporting monomers along alternating paths of even length. Detailed discussions on classical dimer coverings and their application to physics can be found in e.g. Refs.~\cite{Kasteleyn68, Baxter, Kenyon_review}.  

Monomers can be added to the matching by augmenting odd-length paths of dimers. The reverse process is also possible. Starting from an imperfect matching, if an odd-length alternating path can be found with end points terminating on two monomers (an \textit{augmenting path}), then this path can be augmented to annihilate the two monomers in favour of a dimer (see Fig.~\ref{fig:matchings}e). If augmenting paths can be found sequentially between \textit{all} remaining monomers of a matching, so that no monomer appears in more than one path, then the augmentation of these paths results in a perfect matching. We will use these facts to prove the Ammann-Beenker tilings can be perfectly matched.  

\subsection{Ammann-Beenker tilings}
\label{sec:abtiling}

Although we will use the language of graph theory in our discussion of \AB, the mathematical theory of quasicrystals has traditionally been developed
in the language of \emph{tilings}. A tiling is a filling of space with congruent shapes without overlaps or gaps~\cite{BaakeGrimm}. For clarity we
will use \emph{tiling} to refer to an infinite tiling of the plane; finite sections of a tiling we refer to as \emph{patches}. Periodic tilings can be
formed by repetition of a single tile, such as a square or hexagon. Conventional crystals can be seen as regular tilings of 3D space by an atomic unit cell; the underlying symmetries can be deduced from diffraction experiments, and yield the 230 crystallographic space groups~\cite{Landau}. However, the only rotational symmetries compatible with such \textit{periodic} long-range order (in 2D or 3D) are 2-, 3-, 4-, and 6-fold. When diffraction experiments on \ch{AlMn}-alloys revealed structures compatible with 5-fold rotational symmetry~\cite{Shechtman84}, crystallographic theory had to be extended to include \emph{quasi}crystals --- alloys with long-range atomic order but no translation symmetry. Quasicrystals can feature finite patches of 5-,
8-, 10- or 12-fold rotational symmetry~\cite{Levine84, Levine86, SocolarSteinhardt86}.

\begin{figure}[t]
    \includegraphics[width = \linewidth]{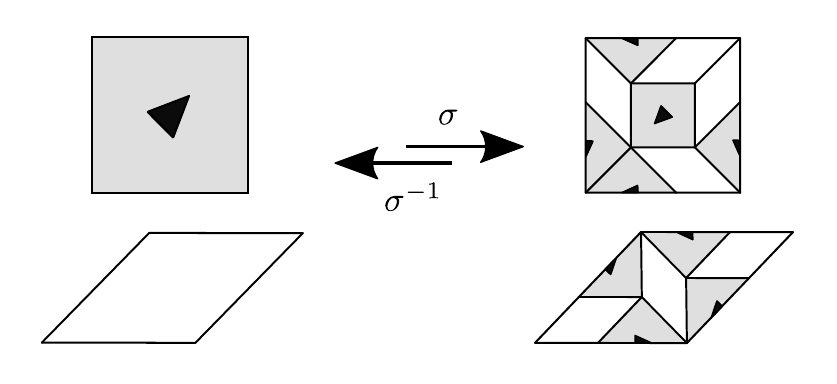}
   \caption{The inflation rule $\sigma$, in which a tile is decomposed into smaller tiles, followed by a rescaling of all lengths by a factor of the
   silver ratio $\delta_S = 1+\sqrt{2}$ (which we do not show here). The inverse $\sigma^{-1}$, deflation, is uniquely specified. An arbitrarily large
   \AB~tiling can be generated from a small patch by repeated inflations. (Note the inflation of the square tile breaks the rotation symmetry so we must
   keep track of orientation, as captured by the triangular motif shown. We will usually leave the markings absent in future figures for clarity.)}
    \label{fig:inflrules}
\end{figure}

\begin{figure*}    
    \includegraphics[width=\textwidth]{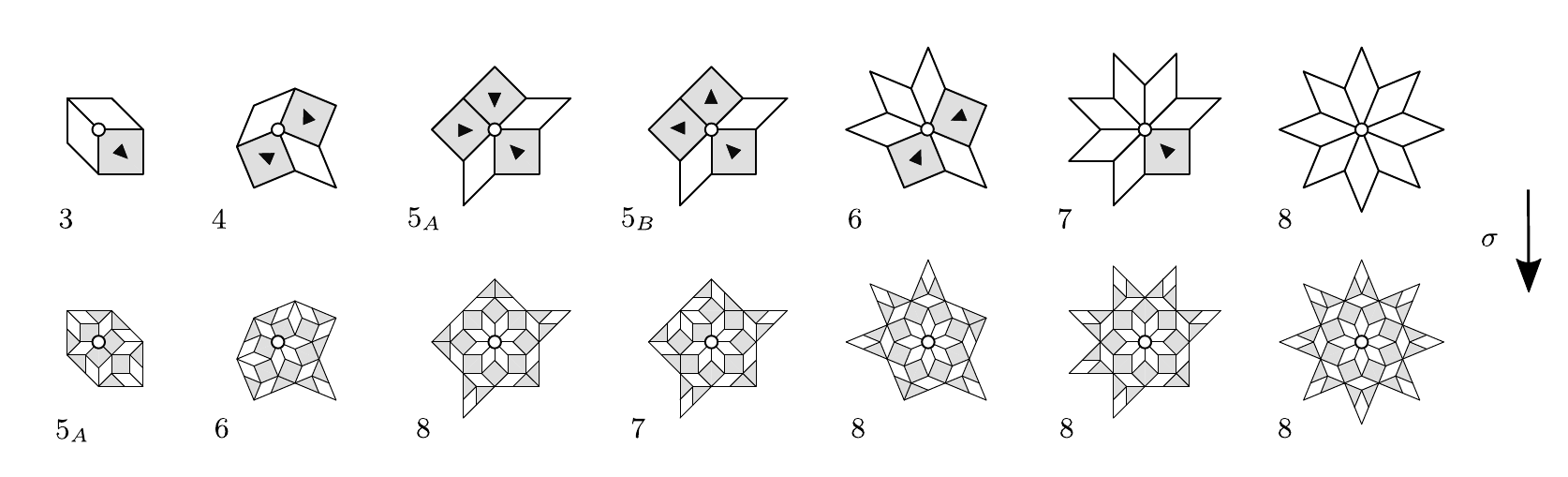}
    \caption{The seven vertex configurations, and their action under inflation (Fig.~\ref{fig:inflrules}). All vertices map to $8$-vertices under at most two inflations.}
    \label{fig:vertexconfigs}
\end{figure*}

Quasiperiodic tilings are a class of planar tilings without translation symmetry, capable of accounting for the symmetries observed in quasicrystals which are forbidden in periodic tilings~\cite{BaakeGrimm, Grimm19, GrunbaumShephard}. They are distinguished from the more general set of aperiodic tilings by the fact that their diffraction patterns can be indexed by a finite number of wavevectors~\cite{BaakeGrimm}. The most famous family of quasiperiodic tilings is due to Penrose~\cite{Penrose74}. Penrose tilings have local patches of 5-fold symmetry (and can have at most one true 5-fold rotational centre). The \textit{Ammann-Beenker tilings}~\cite{Beenker, GrunbaumShephard} have 8-fold symmetry, and together with the decagonal (10-fold) and dodecagonal (12-fold) tilings, they make up the `Penrose-like' tilings~\cite{BoyleSteinhardt16B}. The remarkable fact that the four symmetries displayed by these tilings account for the symmetries of all physical quasicrystals (i.e.~those observable in experiment) was explained by Levitov based on arguments of energetic stability~\cite{Levitov88}. The Penrose-like tilings have therefore seen extensive study in connection to physics.

Classical dimers on Penrose tilings were studied in previous work~\cite{Flicker_Simon_Parameswaran}; here we focus on the Ammann-Beenker (\AB) tiling,
a patch of which is shown in Fig.~\ref{fig:ABwindow}a. The \AB\ tiling is constructed from copies of two inequivalent tiles: a square, and a rhombus with angles $\pi/4$ and $3\pi/4$. Both tiles have edges of unit length. No shifted copy of this tiling can be exactly overlaid with the original. Many mathematical properties of \AB~tilings are well understood and can be found in e.g.~\cite{BaakeGrimm}. Here we briefly mention those relevant to the following discussion.

The Ammann-Beenker tilings (plural) constitute an uncountable set of LI (\emph{locally indistinguishable} or \emph{locally isomorphic}) tilings. Two tilings are LI if any \textit{finite} patch of one can be found in the other. In this way, all \AB\ tilings `look the same' from a local perspective, distinguished only in their global structure. In fact, every finite patch recurs with positive frequency in all tilings. This property accounts for the long-range order of the quasiperiodic tilings. This structure can be compared to that of the periodic tilings, where the long-range order arises from the repeated unit cell. Results that we discuss in the following sections apply in generality to the entire set of \AB~tilings. 

Any \AB~tiling $\mathcal{T}$ can be generated by several methods~\cite{Senechal,BaakeGrimm,GrunbaumShephard}. Square and rhombus tiles can have their edges decorated by \emph{matching rules} which force the tiles to fit together quasiperiodically; an \AB~tiling can be created as a slice through a higher dimensional periodic lattice (the \emph{cut-and-project} technique); or the tilings can be generated via an \textit{inflation} procedure. This inflation method allows us to discuss the scale symmetry of the tilings, and is our primary workhorse in this paper. 

Starting from a finite seed patch $\mathcal{T}_0$ (e.g.~a single square tile) an \textit{inflation rule} $\sigma$ is repeatedly applied to grow the patch as $\mathcal{T}_n = \sigma^n(\mathcal{T}_0)$, with the number of tiles growing exponentially under inflation. $\sigma$ consists of two steps: \textit{decomposition}, where every tile is divided into smaller tiles as shown in Fig.~\ref{fig:inflrules}, followed by \textit{rescaling}, where the decomposed tiling is scaled so that the new tiling is formed from exact copies of the original tiles. The scaling factor is the silver ratio $\delta_S$, defined by
\begin{equation}
  \delta_S^2 = 2\delta_S+1
  \label{eq:delta_S_def}
\end{equation}
and equal to
\begin{equation}
  \delta_S = 1+\sqrt{2}.
  \label{eq:delta_S}
\end{equation}
The Penrose tiling instead has as its scale factor the golden ratio $\varphi = \frac{1+\sqrt{5}}{2}$. Note that it is the \emph{length} (rather than area) of the tile edges that scales by $\delta_S$ under decomposition: this can be seen geometrically in Fig.~\ref{fig:inflrules}, with each edge divided into a rhombus edge and the square diagonal. The area of each tile scales by $\delta_S^2$ under decomposition. In the following, if we say one tiling is larger than another by some power of the silver ratio, we are always referring to the respective lengths of their tile edges. 

Patches of arbitrarily large size can be generated by the inflation process. The tiling $\mathcal{T}$ is recovered after an infinite number of inflations. The inverse of the inflation rule, $\sigma^{-1}$, \textit{deflation}, is also uniquely specified on a tiling, consisting of \textit{composition} --- reconstruction of the larger tiles from smaller tiles --- and rescaling by the inverse of the silver ratio. Acting with $\sigma$ or $\sigma^{-1}$ on $\mathcal{T}$ returns another tiling in the same LI class.

Every possible configuration of tiles around a vertex in \AB~is identical to one of the seven vertex configurations shown in Fig.~\ref{fig:vertexconfigs} (up to rotations). The 5-vertices ($5_A$ and $5_B$) yield two distinct results upon inflation depending on the orientation of the adjacent square plaquettes. All other vertices can be uniquely identified by their co-ordination number (valence, in graph theory nomenclature). Under $\sigma$, each vertex configuration is mapped to a different configuration, e.g.\ $4 \rightarrow 6$, with the exception of the 8-vertex, which inflates to another 8-vertex. These inflations are also shown in Fig.~\ref{fig:vertexconfigs}. 

\begin{figure}[t]
    \includegraphics[width=\linewidth]{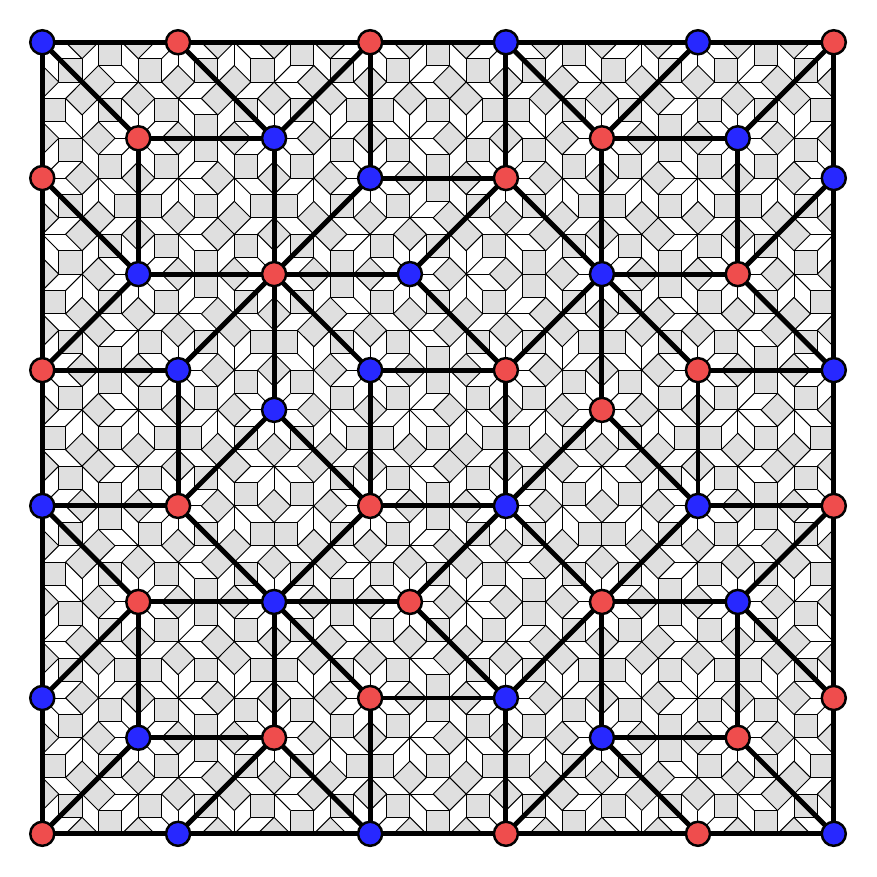}
    \caption{The scale symmetry of the \AB~tiling: the $8$-vertices of an \AB~tiling are positioned to sit at the vertices of another locally
    indistinguishable \AB~tiling, with lengths increased by a factor of $\delta_S^2$, where $\delta_S=1+\sqrt{2}$ is the silver ratio. 8-vertices are coloured by their bipartite charge.}
    \label{fig:scalesym}
\end{figure}

\begin{figure}[t]
    \includegraphics[width=\linewidth]{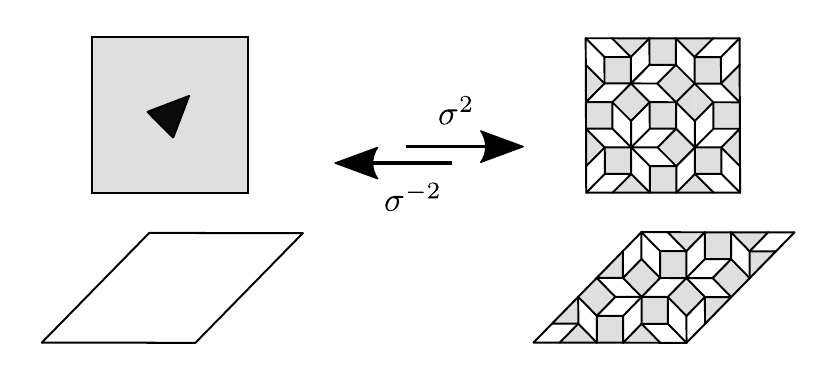}
    \caption{The $\sigma^2$ (in/de)flation rules, which map vertices directly to 8-vertices in the completed tiling. Here we leave off triangular motifs in the inflation for clarity.} 
    \label{fig:inflsquared}
\end{figure}

The scale symmetry of the Ammann-Beenker tilings arises from this inflation structure, and specifically from the invertibility of $\sigma$. One expression of this symmetry is that all 8-vertices of $\mathcal{T}$ are positioned at the vertices of the tiling obtained by twice-composing $\mathcal{T}$, i.e.\ the twice-deflated tiling $\mathcal{T}_{-2} \equiv \sigma^{-2}(\mathcal{T})$ with lengths rescaled by $\delta_S^2$. This follows from noticing that every vertex is mapped to an 8-vertex under two inflations, as can be checked from the vertex inflations in Fig.~\ref{fig:vertexconfigs}. By the inverse, deflating twice maps the 8-vertices of $\mathcal{T}$ to the vertices of $\mathcal{T}_{-2}$. This scale symmetry is more clearly shown in Fig.~\ref{fig:scalesym}, which shows a patch of $\mathcal{T}$ overlaid with the scaled $\mathcal{T}_{-2}$. The 8-vertices of $\mathcal{T}$ are coloured by their bipartite charge, and are seen to coincide with the vertices of the scaled $\mathcal{T}_{-2}$. We will rely on this symmetry in the rest of the paper. Anticipating this 8-vertex $\leftrightarrow$ vertex mapping, we introduce $\sigma^2$ inflation tiles, shown in Fig.~\ref{fig:inflsquared}. Compared to the basic inflation tiles (Fig.~\ref{fig:inflrules}), the $\sigma^2$ tiles have the property that their vertices become $8$-vertices in the inflated tiling.

\begin{figure}
    \includegraphics[width=\linewidth]{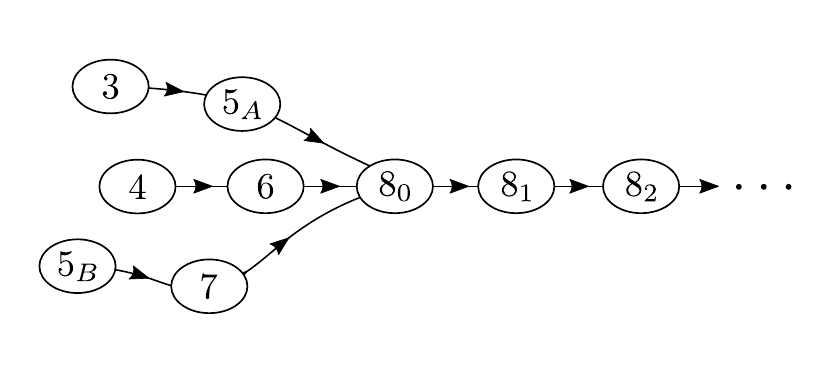}
    \caption{The flow of vertices under successive inflations. The ordering of $8$-vertices is defined so that an order-zero 8-vertex, $8_0$, deflates once to a $5_A$-, 6-, or 7-vertex, and an order-$n$ 8-vertex, $8_n$, deflates to $8_{n-1}$.}
    \label{fig:inflflow}
\end{figure}

This scale hierarchy continues: under $\sigma^2$, the 8-vertices of $\mathcal{T}$ inflate to 8-vertices, and these twice-inflated 8-vertices are positioned to sit at the vertices of a tiling with lengths scaled by $\delta_S^4$. With this in mind, it is convenient to define an ordering to the $8$-vertices. An $8$-vertex is \textit{order-zero} if it maps to a 7-, $5_A$-, or 6-vertex under a single deflation; \textit{order-one} if it deflates to an order-zero 8-vertex; and so on. This flow of vertices is represented in Fig.~\ref{fig:inflflow}. We denote an order-$n$ $8$-vertex an $8_n$-vertex. An $8_n$-vertex is therefore the $n^\textrm{th}$ inflation of an order-0 8-vertex, $8_0$. We find that $8_n$-vertices of $\mathcal{T}$ sit at the vertices of an \AB\ tiling, LI to $\mathcal{T}$, with lengths scaled by a factor of $\delta_S^{(n+2)}$. Since \emph{all} vertices of $\mathcal{T}$ are mapped to order-$n$ or higher 8-vertices under $\sigma^{(n+2)}$, 8-vertices of order-$n$ and higher can also be viewed as those vertices of $\mathcal{T}$ that are preserved under $n+2$ \textit{deflations}. We will denote the $n$-times deflated tiling of $\mathcal{T}$ as $\mathcal{T}_{-n}$ (generally we will not distinguish between deflations and decompositions, unless the distinction is important).

\begin{figure}[t]
    \includegraphics[width=0.65\linewidth]{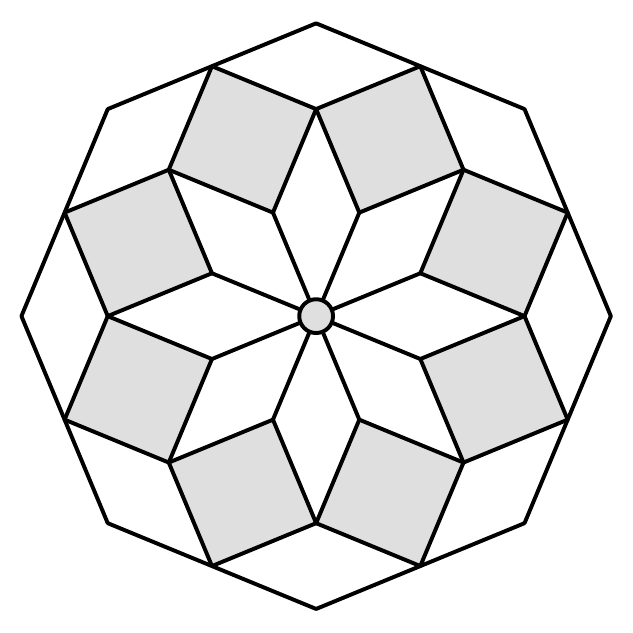}
\caption{The local empire of an $8_0$-vertex, the simply connected set of tiles that always appear around an $8_0$-vertex in the tiling. The local empire has $\mathcal{D}_8$ symmetry.}
  \label{fig:eightemp}
\end{figure}

To conclude this section, we discuss the 8-fold symmetry that defines the Ammann-Beenker tilings. First, as mentioned previously, the Bragg spectrum
(Fourier transform of the lattice) has a discrete 8-fold rotational symmetry ($D_8$ in Sch\"{o}nflies notation)~\cite{Baake2002}. Second, and most importantly to us, the symmetry shows up in the structure of tiles surrounding an 8-vertex. Every vertex configuration has a set of tiles which always appears around the vertex wherever it is placed in the tiling. Such a motif is known as the vertex \emph{empire}~\cite{GrunbaumShephard,EffingerDean}. Vertex empires are generally not simply connected; the set of empire tiles simply connected to a vertex is called the \emph{local empire} of the vertex. The local empire of the $8_0$-vertex is shown in Fig.~\ref{fig:eightemp}, and has $D_8$ symmetry. The inflation $\sigma$ preserves this symmetry, and so the radius of symmetry of an $8_n$-vertex is a factor of $\delta_S$ larger than for an $8_{n-1}$-vertex. The local empire of an $8_n$-vertex (which we will refer to as an $8_n$-empire) in fact includes that of an $8_{n-1}$-vertex within it. According to the definition of the LI-class of tilings, every finite region can be found with positive frequency across the tiling. It follows that $D_8$-symmetric 8-empires of arbitrarily large size can be found across the tiling, with a frequency corresponding to the frequency of the respective 8-vertices.

%
\section{Perfect matchings on the Ammann-Beenker tilings}
\label{sec:perfect_matching_proof}
%

We now show that Ammann-Beenker tilings can be perfectly matched. We work in the thermodynamic limit, where the tiling covers the infinite plane and
boundary effects can be ignored. Taking the thermodynamic limit is a delicate procedure in quasiperiodic systems since one must necessarily work with
open boundary conditions and bound their contribution as the system size increases. By `perfect matchings' in the thermodynamic limit, we mean those
with zero monomer density. While efficient algorithms exist to determine the maximum matching of finite graphs~\cite{HopcroftKarp73,DuffEA12}, showing that an infinite graph (the relevant case in the thermodynamic limit) admits a perfect matching is in general non-trivial. Quasiperiodic Penrose tilings cannot be perfectly matched: though the infinite tiling is charge neutral, monomers are confined within regions with an excess of one or the other type of bipartite
charge. Such confinement emerges as a direct consequence of the quasiperiodic geometry: monomers cannot annihilate those of the opposite charge as
they cannot cross so-called \textit{monomer membranes}. Membranes are sets of edges which cannot be covered by dimers in any maximum matching. Equivalently, no augmenting paths exist between monomers on opposite sides of a membrane. Deleting these edges disconnects the graph. 

The perfect matching problem on \AB\ is of a similarly non-trivial nature. We note that the matching problem on \textit{finite} patches of the \AB\
tiling depends sensitively on the patch considered. Clearly a patch with a net imbalance of bipartite charge never admits a perfect matching. However,
the
\AB~tilings have vanishing net charge density in the thermodynamic limit, because the average vertex connectivity is the same for the two
bipartite subsets. Less trivially, it is also possible to find charge-neutral patches with unpaired monomers. An example is the patch generated by inflating the basis
square four times: exactly two monomers remain in the maximum matching, with opposite charges. This is akin to the Penrose problem where membranes
stop the monomers from matching. In contrast to that case, where the \emph{density} of monomers approaches $\sim 10\%$, the observed
\textit{number} of unpaired monomers on \AB~remains of order unity as the tiling is grown. We have verified these statements using the above
graph-theoretic methods to compute maximum matchings on a large sample of patches generated via inflation. While similar membrane structures to those
in Penrose tilings exist on \AB\ (as we will show in Sec.~\ref{sec:mem_quasimem}), in the latter case they do not frustrate the perfect matching, as
we will show below. On finite patches, the presence of monomers arises from defects at the boundary which migrate into the bulk, and no finite density persists in the thermodynamic limit. 

The strategy of our proof that the AB tilings can be perfectly matched is as follows. Consider the perfect matching problem on an \AB~tiling $\mathcal{T}$. We first match up all vertices of $\mathcal{T}$ except for the 8-vertices (recall that these are the vertices which remain under two deflations). We then show that the problem of matching the remaining 8-vertices maps to a perfect matching problem on the twice-deflated tiling $\mathcal{T}_{-2}$. On this tiling, we match all but the 8-vertices again (this matches all $8_0$- and $8_1$-vertices on the original $\mathcal{T}$). We then use the scale symmetry of $\mathcal{T}$ to iterate this procedure and obtain an upper bound on the monomer density of the \AB~tiling after $n$ deflations. This density vanishes exponentially as $n\to \infty$, corresponding to taking the thermodynamic limit.

Our motivation for separating out the 8-vertices comes from the requirement for a bipartite graph to be charge-neutral in order to admit a perfect matching. This is true of the \AB~tiling in the thermodynamic limit. Note, however, that any $D_8$ local empire has an excess charge of at least one. This follows since every $D_8$ local empire has an 8-vertex at its centre (by definition), and its eight-fold rotational symmetry mandates that the total number of vertices within the region must be $8m+1$, for some integer $m$. Since this number is odd, there must be an excess of bipartite charge. The same argument shows that any symmetric region centred on 8-vertices cannot be perfectly matched without placing at least one dimer outside of the region. Since such regions exist at all scales across the tiling, we can always find arbitrarily large regions that cannot be perfectly matched internally. This structure of coupled $D_8$ regions complicates the problem. By first removing the 8-vertices at the centres of these regions, we can match the rest of the tiling systematically before progressively reintroducing and matching the remaining monomers. 

\subsection{Proof of Perfect Matching}

\begin{figure}[!ht]
    \includegraphics[width=\linewidth]{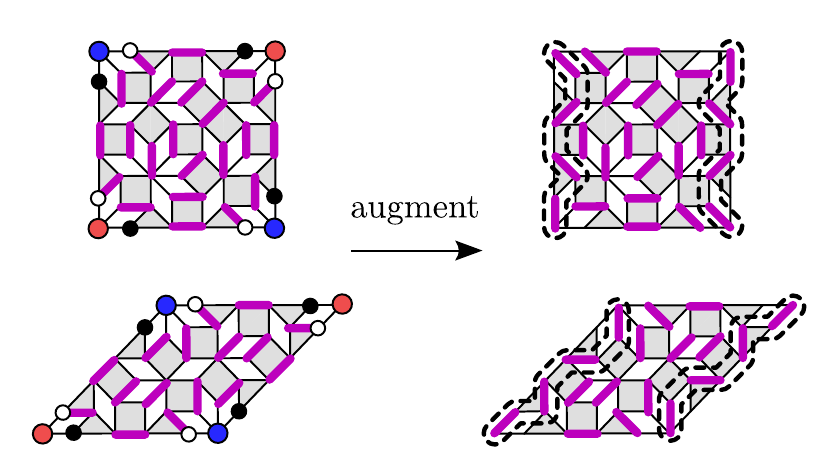}
    \caption{(\emph{Left}) The dimer inflation tiles: deflating with the squared tiles in Fig.~\ref{fig:inflsquared},
    $\mathcal{T}\rightarrow\mathcal{T}_{-2}$, and then re-inflating with these dimer-decorated tiles places dimers on $\mathcal{T}$. This matches all vertices of $\mathcal{T}$ except the $8$-vertices (marked with red and blue circles denoting bipartite charge). Unmatched vertices on tile boundaries (black) overlap with matched vertices (white) in the completed tiling. (\emph{Right}) Augmenting paths always exist between neighbouring 8-vertices (the other choice of charge-neutral pairing follows from the symmetry of the dimers).}
  \label{fig:dimerinflrules}
\end{figure}

\begin{enumerate}
    \item Let $\mathcal{T}$ be an \AB\ tiling not covered by dimers. Let $\mathcal{T}_{-2}$ be the tiling obtained by deflating twice on $\mathcal{T}$. Replacing the tiles of $\mathcal{T}_{-2}$ by the \emph{dimer inflation tiles} (DITs) in Fig.~\ref{fig:dimerinflrules}, we obtain a matching of $\mathcal{T}$, such that only the 8-vertices remain unmatched.

    The last statement follows from noting that (a) no dimers touch the corner vertices of the DITs, which map to the 8-vertices of $\mathcal{T}$ (see Sec. \ref{sec:abtiling}); (b) all vertices in the interior of the DITs are matched; and (c) all remaining unmatched vertices on the boundary of the DITs (large, black) overlap consistently with matched vertices on adjacent tile boundaries (large, white). No double covering occurs.

    The density of unpaired monomers on $\mathcal{T}$ after application of the DITs gives an upper bound on the true monomer density $\rho$ (i.e.\ the density remaining in any maximum matching of $\mathcal{T}$ ). Therefore, 
    \begin{equation} 
      \rho \leq \nu_{8v},
      \label{eq:monomerdensity1}
    \end{equation}
    where $\nu_{8v} = \delta_S^{-4} \sim 0.03$ is the density of 8-vertices on the Ammann-Beenker tiling~\cite{BaakeGrimm}. 

    \item Next, we allow the monomers to move around the tiling via augmenting paths. We define 8-vertices (monomers) to be \textit{deflate neighbours} on $\mathcal{T}$ if they become true nearest-neighbours on $\mathcal{T}_{-2}$. An augmenting path always exists between two deflate-neighbouring monomers. This is clear from the right of Fig.~\ref{fig:dimerinflrules}, since deflate-neighbouring monomers sit on adjacent corner vertices of either one of the DITs.

    The bound on $\rho$ will be reduced by augmentation of these paths, annihilating two deflate-neighbouring monomers to give a dimer. Each augmentation removes the monomers from future possible pairings, but otherwise does not affect the possibility of annihilations between other monomers, since augmenting paths between disjoint pairs of deflate-neighbouring monomers can be chosen so as not to intersect. The paths in Fig.~\ref{fig:dimerinflrules} are non-intersecting. 

    Augmentation of a complete (non-intersecting) set of paths between deflate-neighbouring monomers results in a perfect matching of $\mathcal{T}$. According to our definition of deflate-neighbours, a complete set of augmenting paths is given by a perfect matching of $\mathcal{T}_{-2}$, via the identification of \textit{dimers} on $\mathcal{T}_{-2}$ with \textit{augmenting paths} on $\mathcal{T}$. Fig.~\ref{fig:scalesym} shows that augmenting paths between deflate-neighbouring 8-vertices indeed has the structure of a matching problem at the next scale.

    \item A perfect matching of $\mathcal{T}_{-2}$ is evidently no easier to obtain than for $\mathcal{T}$. However, by application of the DITs to the tiling $\mathcal{T}_{-4}$, we match all vertices of $\mathcal{T}_{-2}$ except the 8-vertices. Augmenting the corresponding paths on $\mathcal{T}$ annihilates all monomers on $8_0$- and $8_1$-vertices of $\mathcal{T}$. The remaining unmatched vertices correspond to the $8_{n>1}$-vertices, and the bound on $\rho$ is reduced by another factor of $\nu_{8v}$:
            \begin{equation} 
              \rho \leq  \nu_{8v}^2.
            \end{equation} 

    \item We now apply this procedure to all scales of the \AB~tiling. We call a monomer on an $8_n$-vertex an order-$n$ monomer. Finding augmenting paths between all order-$n$ and order-$(n+1)$ monomers of $\mathcal{T}$ corresponds to finding augmenting paths between all order-$(n-2)$ and order-$(n-1)$ monomers of the $\mathcal{T}_{-2}$ tiling, and by induction, to a matching of the non-8-vertices of the $\mathcal{T}_{-(n+2)}$ tiling (here we identify order-$n$ for $n<0$ with the non-8-vertices according to Fig.~\ref{fig:inflflow}). Each matching of the non-$8$-vertices of $\mathcal{T}_{-(n+2)}$ is performed via the DITs, reducing the bound on $\rho$ by a factor of $\nu_{8v}$. Since the infinite tiling $\mathcal{T}$ can be deflated an arbitrary number of times (and $\nu_{8v}<1$), the density of monomers must tend to zero in the thermodynamic limit:
    \begin{equation} 
      \rho \leq  \nu_{8v}^{2n} ,\hspace{1cm} \lim_{n\rightarrow\infty}\rho \rightarrow 0,
    \end{equation}
    and the tiling admits a perfect matching $\square$.

\end{enumerate}

We point out finally that the above proof does not guarantee a vanishing \emph{number} of monomers on the tiling. In fact, it is easy to construct counterexamples: the infinite tiling obtained by repeated inflation of the 8-fold symmetric patch in Fig.~\ref{fig:eightemp} is one. This tiling must have an excess charge of at least one monomer, which can be seen as the charge residing on the `order infinity' 8-vertex at the centre of the tiling. It will never be matched under a finite number of the above inflation matching rules. These monomers are artefacts of how the boundary of the graph is terminated (for the above example, we can imagine adding a small number of vertices to the graph boundary which break the symmetry and allow the monomer to be matched). Crucially, the number of such monomers is $O(1)$, \textit{i.e.} it does not scale with the size of the graph. Consequently, the density of monomers vanishes on all vertices in the thermodynamic limit, and all observables of interest remain unchanged by the specific choice of boundary. 
We emphasize that an $O(1)$ number of monomers on the infinite graph is not a peculiarity of our problem; a similar number of monomers appears for periodic graphs with certain choices of boundary conditions (\textit{e.g.}, an $L_x \times L_y$ square lattice with with both $L_x$ and $L_y$ odd always hosts one monomer).

%
\section{Membranes, Ladders and Effective Matching Problems} 
\label{sec:mem_quasimem}
%

In proving the existence of perfect matchings on the Ammann-Beenker tilings, 8-vertices of $\mathcal{T}$ were matched (by augmenting paths) if the corresponding sites shared a dimer on the deflated tiling $\mathcal{T}_{-2n}$. Since the 8-vertices sit at the vertices of tilings with lengths rescaled by powers of the silver ratio, $\delta_S^{2n}$, a natural question is whether an effective description emerges in terms of matchings at larger lengthscales. In this section, we will show how the 8-vertices of the \AB\ tiling are surrounded by concentric structures of sets of edges collectively hosting exactly one dimer. We call these monomer `pseudomembranes'. Each pseudomembrane, therefore, imposes an effective dimer constraint at a larger scale, with the pseudomembrane bounding a region acting as an effective vertex (`effective unit'), connected to the rest of the graph by exactly one dimer. This effective matching description, in turn, has far-reaching consequences for the system's correlations, which we explore in Sec.~\ref{sec:numerics}.

Monomer \emph{membranes} were introduced in Refs.~\onlinecite{Flicker_Simon_Parameswaran} and \onlinecite{Biswas_etal}, as sets of edges hosting {\it
zero} dimers. In Penrose tilings membranes separate unmatched regions with an excess bipartite charge. The complete \AB\ tiling does not host
membranes. However, we identify similar structures of sets of edges --- the pseudomembranes --- which host exactly \emph{one} dimer in any perfect
matching. Like membranes, pseudomembranes capture how certain aspects of the quasiperiodic graph structure are encoded in the set of perfect
matchings. The properties of membranes and pseudomembranes can be understood in terms of the Dulmage-Mendelsohn decomposition of the graph and its
`fine' generalization as discussed in Appendix~\ref{app:memreview}. Here, we provide their construction and establish their properties. We do this by
first introducing an auxiliary tiling, `\ABB', obtained from \AB\ by deleting all 8-vertices. \ABB\ admits exact membranes, which separate perfectly matched quasi-1D regions (`ladders'). This is in contrast to Refs.~\onlinecite{Flicker_Simon_Parameswaran} and \onlinecite{Biswas_etal} where membranes confine monomers. We next demonstrate how the exact membranes become pseudomembranes on \AB\ with the reintroduction of the 8-vertices. The pseudomembranes lead to the natural identification of perfectly matched regions of $D_8$ symmetry (`$\mathcal{H}$ regions') within the local empires of 8-vertices, which play an important role in the monomer correlations discussed in Sec.~\ref{sec:numerics}. To conclude the section, we comment on the effective matching problem. 

\begin{figure}[t]
  \centering
  \includegraphics[width=\linewidth]{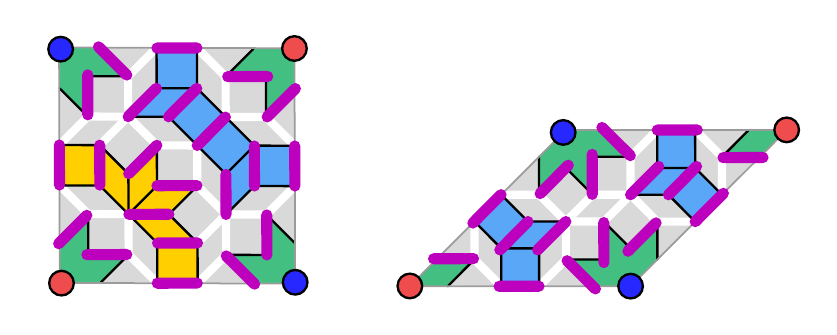}
  \caption{The \ABB\ ladder tiles: dimer inflation tiles of Fig.~\ref{fig:dimerinflrules} with edges incident to $8$-vertices removed. The removed $8$-vertices are marked with solid blue and red circles, the colours denoting the two
  bipartite charges. The yellow and blue shaded segments join up to form \emph{ladders} which host dimers in a maximum matching of \ABB. The edges in
  the unshaded regions comprise \emph{membranes}, which never host a dimer in any maximum matching of \ABB. The edges enclosing the green shaded
  region form \emph{stars} of $16$ edges around the absent $8$-vertices, and can be perfectly matched in the bulk of \ABB.}
  \label{fig:abbtiles}
\end{figure}

\subsection{Membranes and ladders on the \ABB~tiling}
\label{sec:memproof}
\begin{figure*}[t]
  \centering
  \includegraphics[width=0.95\textwidth]{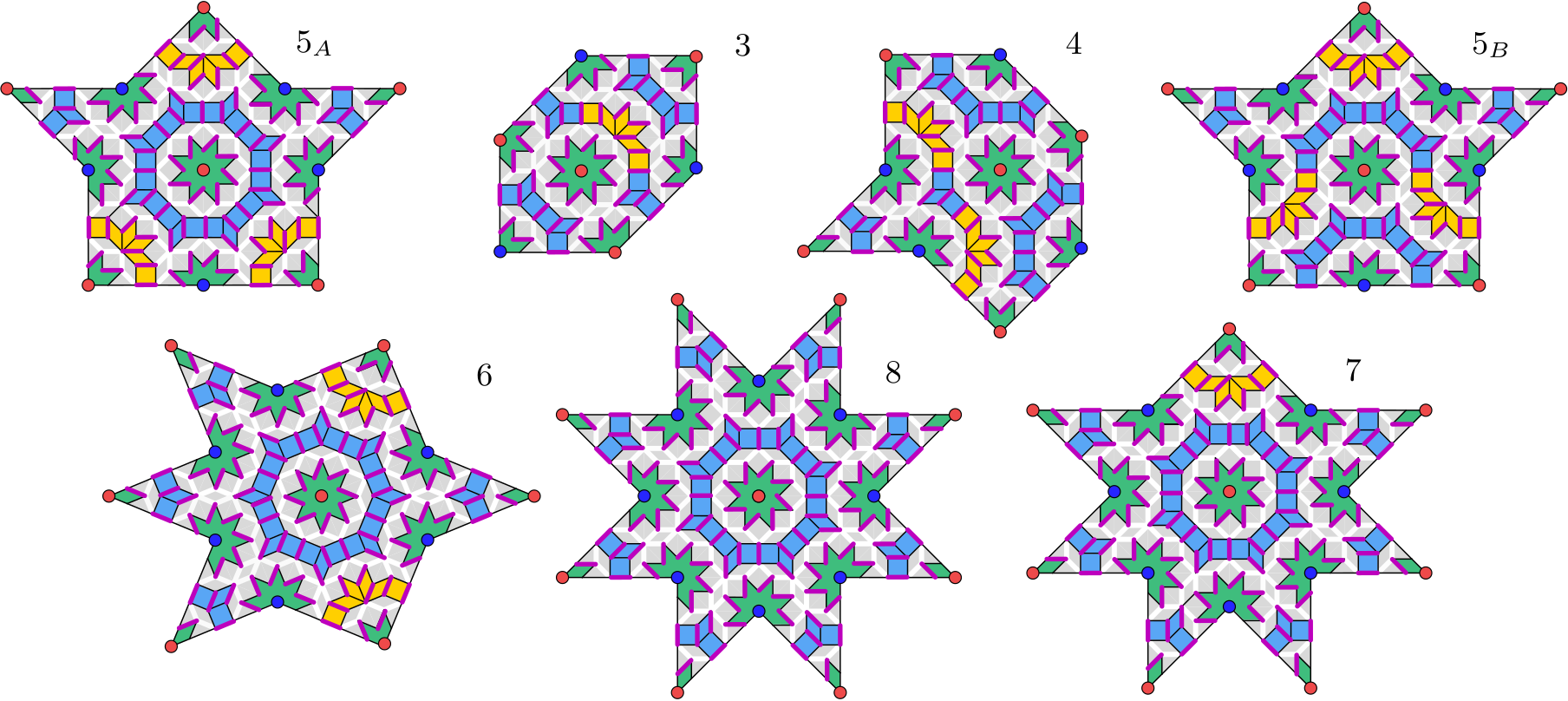}
  \caption{All vertex configurations of the ladder-decorated tiles of Fig.~\ref{fig:abbtiles}. This shows all possible ways in which ladder segments can join up to form ladders in the \ABB~tiling. The segments shaded with yellow and light blue in Fig.~\ref{fig:abbtiles} match up to form ladders, whereas the green segments match up to form stars in the bulk of the \ABB~tiling.}
  \label{fig:abbvertices}
\end{figure*}

The important point to take from Sec.~\ref{sec:perfect_matching_proof} is that at each (double) deflation step, only the augmenting paths between two
8-vertices survive as a dimer at the next scale. Crucially, all information that correlates different inflation scales involves the $8$-vertices. By
removing these vertices from \AB\ (equivalently localizing monomers on all 8-vertices), we obtain the \ABB\ tiling. \ABB\ has particularly simple correlations and provides insight into the more complicated statistics of \AB. 

The \ABB\ tiling can be perfectly matched. This follows trivially from Step 1 of the Perfect Matching proof, since the DITs (Fig.~\ref{fig:dimerinflrules}) match all vertices except for the 8-vertices. We will use a decorated version of the DITs, Fig.~\ref{fig:abbtiles}, to emphasise structures that emerge when the tiles are placed in the full \ABB\ tiling. In Fig.~\ref{fig:abbtiles} the edges connecting the $8$-vertices (solid blue and red circles) to the rest of the graph are removed. The blue- and yellow-shaded plaquettes join to form the regions we call \emph{ladders}. The colour codes relate to the inflation properties of the ladders (see Appendix~\ref{app:ladder_infl}). The edges enclosing the green-shaded regions close to form loops of 16 edges, centred on the missing 8-vertices, which we call \emph{stars} (or \emph{order}-0 ladders). To see how individual ladder segments match up to form ladders in the graph, it is useful to refer back to the vertex configurations. They are shown for the \ABB\ tiles in Fig.~\ref{fig:abbvertices} (compare Fig.~\ref{fig:vertexconfigs} for the undecorated tiles). The continuation of blue and yellow regions into ladders, as well as the formation of closed stars around the green regions, is apparent. 

\begin{figure}[t]
    \includegraphics[width=\linewidth]{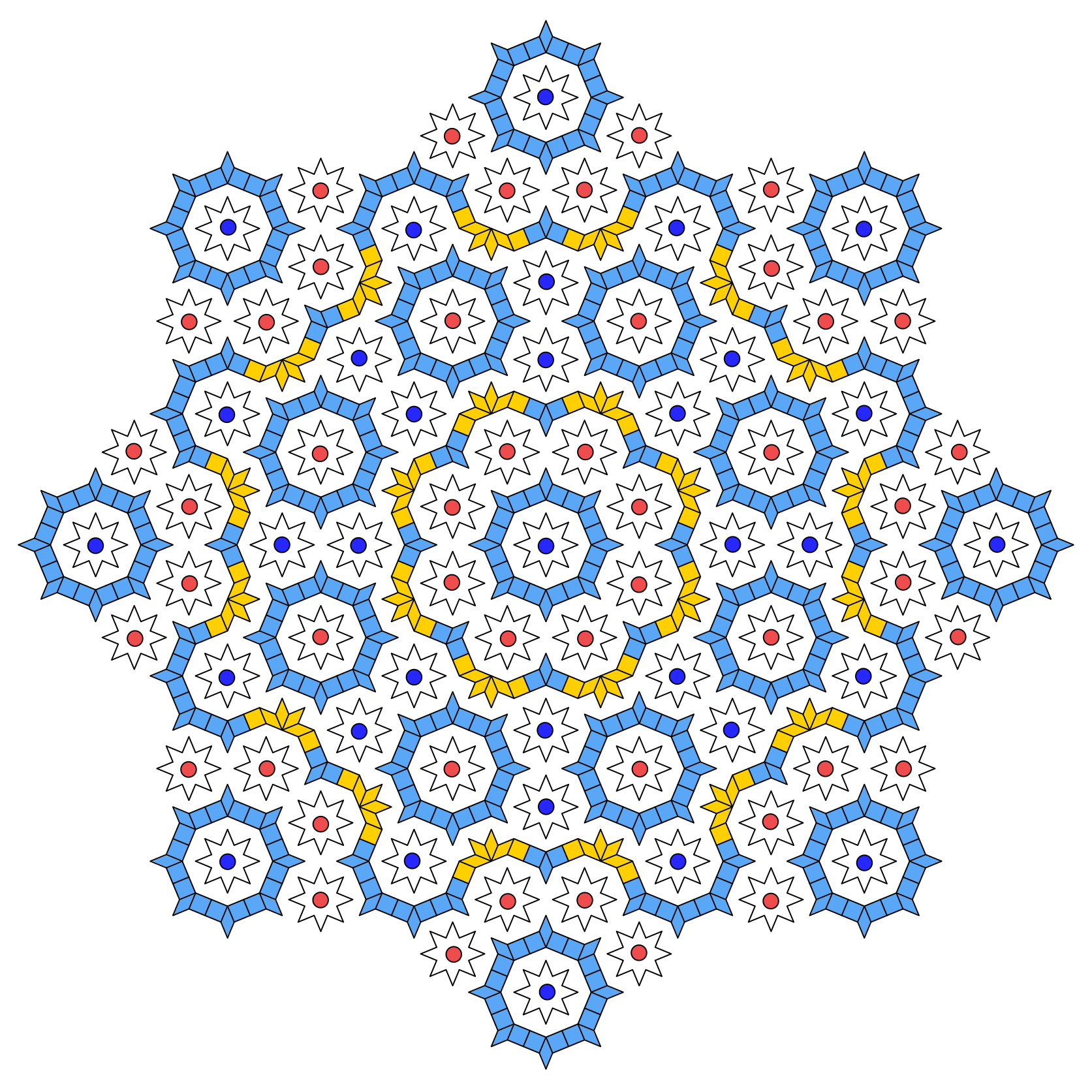}
    \caption{A finite patch of the \ABB~tiling, in terms of stars and ladders. An $8_n$-vertex is the centre of concentric ladders of orders $0\leq
    m\leq n$, which close with $D_8$ symmetry. White space represents impermeable membranes separating ladders.}
    \label{fig:laddersandstars}
\end{figure}

We denote the closed all-blue ladder appearing in the $5_A$-, $6$-, $7$- and $8$-vertex configurations of Fig.~\ref{fig:abbvertices} the $L_1$ ladder. Higher order ladders, $L_n$, are constructed by $n-1$ inflations $\sigma$ ( Sec.~\ref{sec:abtiling}) of $L_1$. The star is obtained by a single deflation of $L_1$, and so we denote the star a zero-order ladder, $L_0$. A detailed discussion of this inflation structure is provided in Appendix~\ref{app:ladder_infl}. From the $D_8$ symmetry of the $L_0$ and $L_1$ ladders, the fact that \emph{all} ladders that close\footnote{The reader may note that unclosed system-spanning ladders are not ruled out by the ladder continuations of Fig.~\ref{fig:abbvertices}. Such ladders in fact exist, but are necessarily of `order infinity' in the inflation scheme: they therefore occur with zero frequency on the infinite tiling and will not play a part in the remaining discussion.} do so with $D_8$ symmetry can also be proven. In general, an order-$n$ 8-vertex is surrounded by concentric $D_8$ symmetric ladders of orders 0-through-$n$, which justifies the terminology. This structure can be seen in Fig.~\ref{fig:laddersandstars}, where we display a finite patch of the \ABB\ tiling, with ladders of orders up to $L_3$ surrounding the correspondingly ordered 8-vertices. 

In every perfect matching, stars and ladders form perfectly matched regions, separated by exact membranes which do not host
any dimers. Fig.~\ref{fig:abbtiles} has dimers residing only on edges that belong to the stars and ladders. We will show that all vertices in each star (ladder) must be matched to vertices in the same star (ladder) in all perfect matchings. Therefore, any edges external to the stars and ladders (thick white edges in Fig.~\ref{fig:abbtiles} and Fig.~\ref{fig:abbvertices}) are never covered by dimers in any perfect matching and hence form exact membranes. 

The proof follows from two observations. First, stars are closed loops of $16$ vertices, with eight vertices in each bipartite subset (we denote the two subsets $\mathcal{U}$ and $\mathcal{V}$) and no vertices in the interior (due to the removal of the $8$-vertices in constructing \ABB). For each star, all `exterior' vertices with edges to the rest of the graph (i.e.~the `points' of the star) are of the same charge. The `interior' vertices alternate with these and therefore are all of the opposite charge. Since they cannot match with any other vertex, we must match each interior vertex with an adjacent exterior vertex. Therefore each star hosts eight dimers forming one of two possible alternating paths around the star, leaving no exterior vertices unmatched.

Ladder segments in the tiles of Fig.~\ref{fig:abbtiles} always contain the same number of vertices of each charge. This property is inherited by any section of a ladder built up from these segments. Fig.~\ref{fig:abbvertices} shows the vertex configurations for the ladder-decorated tiles, allowing us to read off all the ways in which different ladder segments can match up to form ladders. For each ladder section in Fig.~\ref{fig:abbvertices}, all vertices with edges to other ladder segments are of the same charge, say $\mathcal{U}$. While $\mathcal{V}$-vertices do have edges connecting them to stars, those edges can never host dimers in a perfect matching, as we have already shown that the vertices in a star are always matched to vertices within the star. Since there are an equal number of vertices of each charge in any ladder, this immediately implies that all $\mathcal{U}$-vertices in a ladder must match to $\mathcal{V}$-vertices in the same ladder in a perfect matching to avoid a contradiction. Thus, edges outside ladders and stars constitute membranes which never host a dimer in any perfect matching of the \ABB~tiling.

This splitting of the \ABB\ tiling into ladder regions separated by membranes means that no statistical correlation can occur between distinct ladders. The no-dimer constraint implies membranes act as impermeable boundaries for each ladder subsystem, and the dimer partition function of the entire 2D tiling factorises into a product of partition functions for lower-dimensional quasiperiodic ladders. Thus, the notoriously difficult problem of enumerating dimer configurations on a graph is simplified on \ABB. We use this fact in Appendix \ref{app:transfer_matrices} to calculate an asymptotically exact result for the free energy of dimers on \ABB, via transfer matrices. These properties also provide insight into the more complex correlations of \AB, as we will return to in Sec.~\ref{sec:monocorrs}. The ladder regions and membranes in any perfectly matched patch of \ABB can be determined algorithmically by using the Dulmage-Mendelsohn decomposition described in Appendix~\ref{app:memreview}. 

\subsection{Pseudomembranes on the \AB~tiling}
\label{sec:quasiproof}

Armed with our results on \ABB, we now return to the full \AB~tiling, and show that the restoration of the deleted 8-vertices transforms the membranes of \ABB~to pseudomembranes on \AB. As we defined earlier, a pseudomembrane is a connected set of edges which collectively host exactly one dimer between them in any maximum matching. Pseudomembranes satisfy two additional properties: (i) deleting all the edges in a pseudomembrane disconnects the graph into two components; (ii) pseudomembranes close with $D_8$ symmetry, and are centred on $8$-vertices.

\begin{figure}[t]
  \includegraphics[width=0.4\linewidth]{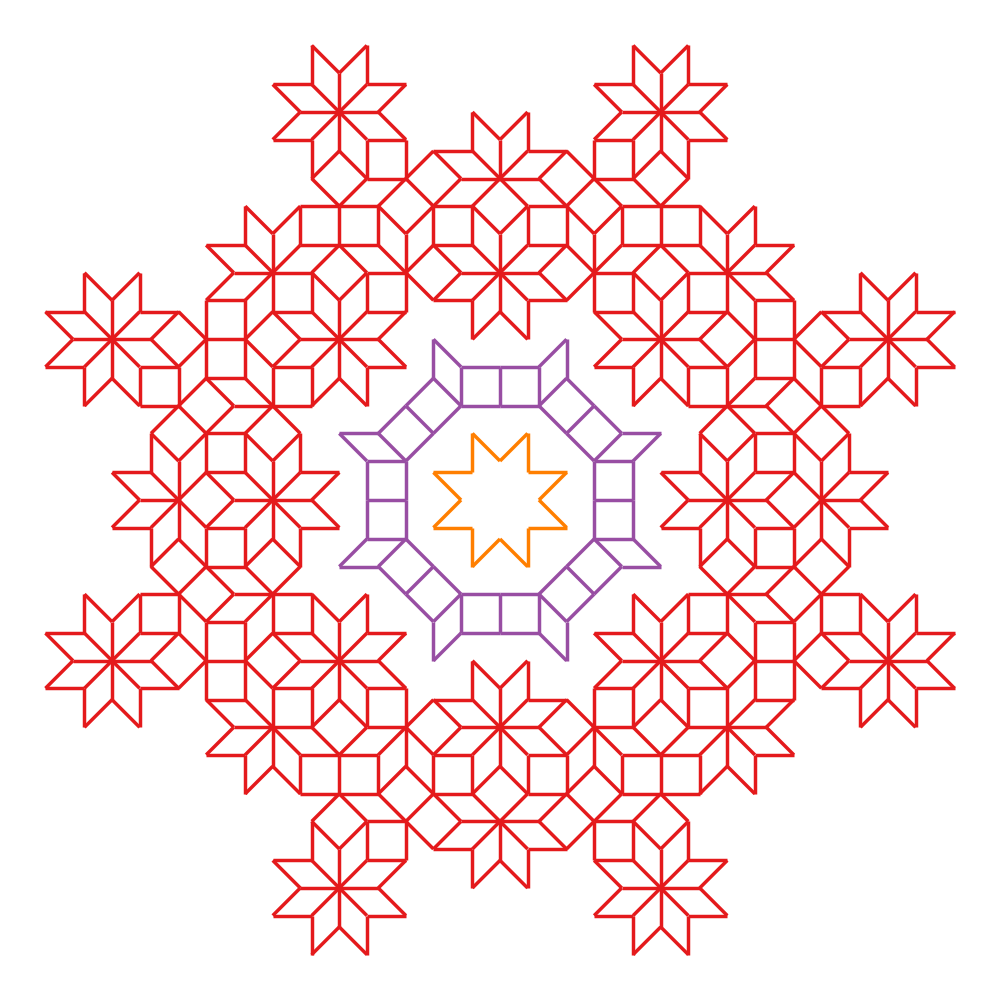}
  \centering
  \includegraphics[width=0.8\linewidth]{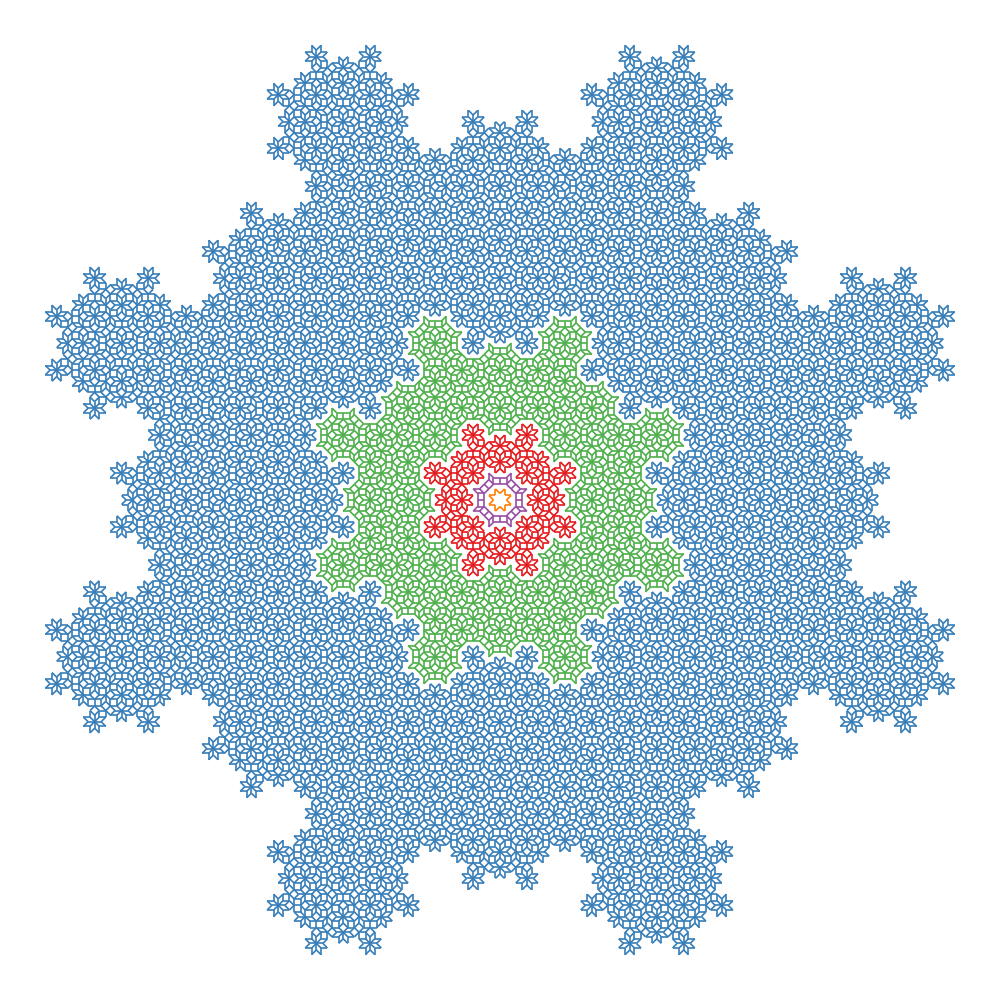}
  \caption{Membranes in perfectly matched \emph{punctured $8_{2n}$-empires}, which separate perfectly matched $\mathcal{H}_i$ regions demarcated with
  different colours. An $8_n$-empire is the local empire of an $8_n$-vertex. A punctured $8_n$-empire is an $8_n$-empire without the central
  $8$-vertex. On adding the $8$-vertex back, each membrane becomes a \emph{pseudomembrane} which can host a single dimer on its edges. \emph{Top}:
  membranes in a punctured $8_2$-empire, separating regions $\mathcal{H}_0$ (gold star), $\mathcal{H}_1$ (purple) and $\mathcal{H}_2$ (red).
  \emph{Bottom}: membranes in a punctured $8_4$-empire, separating regions $\mathcal{H}_{0-4}$. The $8_4$-empire contains a copy of the $8_2$-empire.
  Small components of the empires outside the largest $\mathcal{H}_i$ regions are not shown.}
  \label{fig:pseudomembranes}
\end{figure}

Anticipating the second point, we consider local empires of $8_n$-vertices, which we refer to as $8_n$-empires. An $8_n$-empire is generated by inflating the local empire of an $8_0$-vertex (Fig.~\ref{fig:eightemp}) $n$ times. Since the $8_n$-empire cannot be perfectly matched (due to the charge imbalance), we first remove the central $8$-vertex from the $8_n$-empire, yielding what we term a \emph{punctured $8_{n}$-empire}. Generally, in any matching of the punctured empire, small components of the tiling lying outside the outermost $\mathcal{H}_i$ will remain unmatched, on account of the finite patch boundary. Choosing boundary conditions to exclude these outer vertices, the remaining region can be perfectly matched, and this is what we will take an $8_n$-empire to mean in the rest of section. With this caveat, the punctured $8_n$-empire is perfectly matched, and splits into certain perfectly matched $D_8$-symmetric annular subregions, which we label as $\mathcal{H}_i$. In all perfect matchings of a punctured $8_n$-empire , vertices in each subregion $\mathcal{H}_i$ can only match to other vertices within $\mathcal{H}_i$. The $\mathcal{H}_i$ are separated from each other by exact membranes concentric with the central 8-vertex\footnote{These $\mathcal{H}$ regions are the perfectly matched components of the Dulmage-Mendelsohn decomposition reviewed in Appendix~\ref{app:memreview}}. The zeroth subregion is the star, $\mathcal{H}_0\equiv L_0$, and the first subregion is the first-order ladder, $\mathcal{H}_1\equiv L_1$. Higher order subregions correspond to connected components of \AB\ formed from lower order 8-empires, closed ladders, and the membrane links in between. As explained in Appendix~\ref{app:proof_punctured}, \emph{even} subregions are essentially inflations of the star $L_0$, while \emph{odd} subregions are inflations of $L_1$. We therefore refer to $\mathcal{H}_i$ as `star-like' for $i$ even, and `ladder-like' for $i$ odd.  These statements generalise the result that the stars and ladders of \ABB\, separated by membranes, can be perfectly matched. 

 The proof of these statements, along with a more detailed discussion of the inflation structure of the $\mathcal{H}$ regions, can be found in Appendix~\ref{app:proof_punctured}. Essentially, we first adapt the methods of Sec.~\ref{sec:perfect_matching_proof} to use the inflation structure of $\mathcal{H}$ regions and prove that each $\mathcal{H}_i$ subregion can be perfectly matched. We then follow it up with a proof that in all perfect matchings vertices in a region $\mathcal{H}_i$ match only to other vertices within the same region $\mathcal{H}_i$.   We display the membranes and $\mathcal{H}_i$ subregions for punctured $8_2$ and $8_4$-empires in Fig.~\ref{fig:pseudomembranes}.

If the central $8$-vertex is reinstated to this perfect matching it will host a monomer. Moving this monomer out of any region enclosed by a membrane of the punctured empire converts the membranes to pseudomembranes. This follows since moving the monomer requires the augmentation of alternating paths of even length which terminate on the monomer. Since such paths traverse a membrane that originally enclosed the monomer, they will place a single dimer across the former membrane. Since only a single monomer was added to the perfect matching, it follows that there can be no more than one such dimer in any pseudomembrane. 

One might wonder if a monomer can recross the membrane in the opposite direction and thereby place another dimer across the membrane; this is ruled
out by some simple observations about the perfectly matched regions. Any two perfectly matched regions separated by a membrane have the property that
all edges in the membrane connect vertices of the same bipartite charge in one region to vertices of the opposite bipartite charge in the other region
(Appendix \ref{app:proof_punctured}). If the central $8$-vertex is a $\mathcal{U}$-vertex, this implies that each perfectly matched component of a punctured $8_n$-empire has only $\mathcal{V}$-vertices on its inner boundary and $\mathcal{U}$-vertices on its outer boundary. When a monomer is introduced by reinstating the central $8$-vertex, the monomer recrossing a membrane would imply the existence of a perfectly matched region such that its inner boundary is crossed twice by the monomer and its outer boundary is not crossed by the monomer. This matches two $\mathcal{V}$-vertices of that region with vertices outside the region, leaving an excess of two $\mathcal{U}$-vertices in the rest of the region. It is therefore not possible in a maximum matching. 

Note that the results of the previous paragraph also hold for $8_n$-empires with generic boundary conditions. Punctured $8_n$-empires admit a perfect matching only when the unmatched components outside the outermost $\mathcal{H}$ region are excluded, otherwise they typically host a few monomers at their boundaries. While such monomers have a vanishing density in the thermodynamic limit, our construction of pseudomembranes outlined above does not carry over. However, when such an $8_{n}$-empire is embedded in a larger system, the monomers at the boundaries are annihilated with other monomers in a maximum matching (recall that \AB~tilings host perfect matchings in the thermodynamic limit), and pseudomembranes appear.

In general, an $8_n$-empire hosts $n+2$ pseudomembranes concentric with the $8$-vertex at the empire's centre (including the membrane enclosing the outermost $\mathcal{H}$ subregion and the eight edges incident to the $8$-vertex). The $8_n$-empire also hosts smaller $8_m$-empires, and all these empires host their own pseudomembranes, concentric with each empire's central 8-vertex. These $8$-vertices are necessarily of order $m<n$. Taken together, the edges that make up the pseudomembranes of \AB\ are those that form the membranes of the \ABB~tiling, as described in Sec.~\ref{sec:memproof}. 

\subsection{The effective matching problem}
\label{sec:emp}

The full \AB\ tiling exhibits a rich hierarchical structure of nested pseudomembranes. Each pseudomembrane bounds a region (set of edges and vertices) which acts as an \emph{effective unit} in the matching problem, such that only one dimer connects the region to the rest of the graph. Motivated by this, we define an $8_n$-unit to be the region bounded by the $n+2$-th pseudomembrane from the centre of the local empire of any ($m>n$) $8_m$-vertex. Note the $8_n$-unit coincides with the definition of the $8_n$-empire, minus the components outside and including the outermost pseudomembrane. 

In the sense of Sec.~\ref{sec:perfect_matching_proof}, the perfect matching problem can be seen as the problem of matching up $8_1$-units, \emph{mediated} by the background of perfectly matched ladder regions. This means that each $8_1$-unit matches to a ladder (made up of non-8-vertices) with a single dimer (placed on the pseudomembrane). The edges of this ladder in turn match to other $8_1$-units. Similarly, based on the scale symmetry of the \AB~tilings, the perfect matchings can be described by the matching up of $8_n$-units mediated by \emph{effective ladders}. Here we define an $n$-th order effective ladder as a ladder built from $8_{n-1}$- and $8_{n-2}$-units (so that the effective ladder deflates to one of the ladders of Sec.~\ref{sec:memproof} on the $\mathcal{T}_{-2n}$ tiling). We dub these `effective matching problems'.

\begin{figure*}
\setlength\fboxsep{0pt} \setlength\fboxrule{0.0pt} \fbox{\includegraphics[width=18cm]{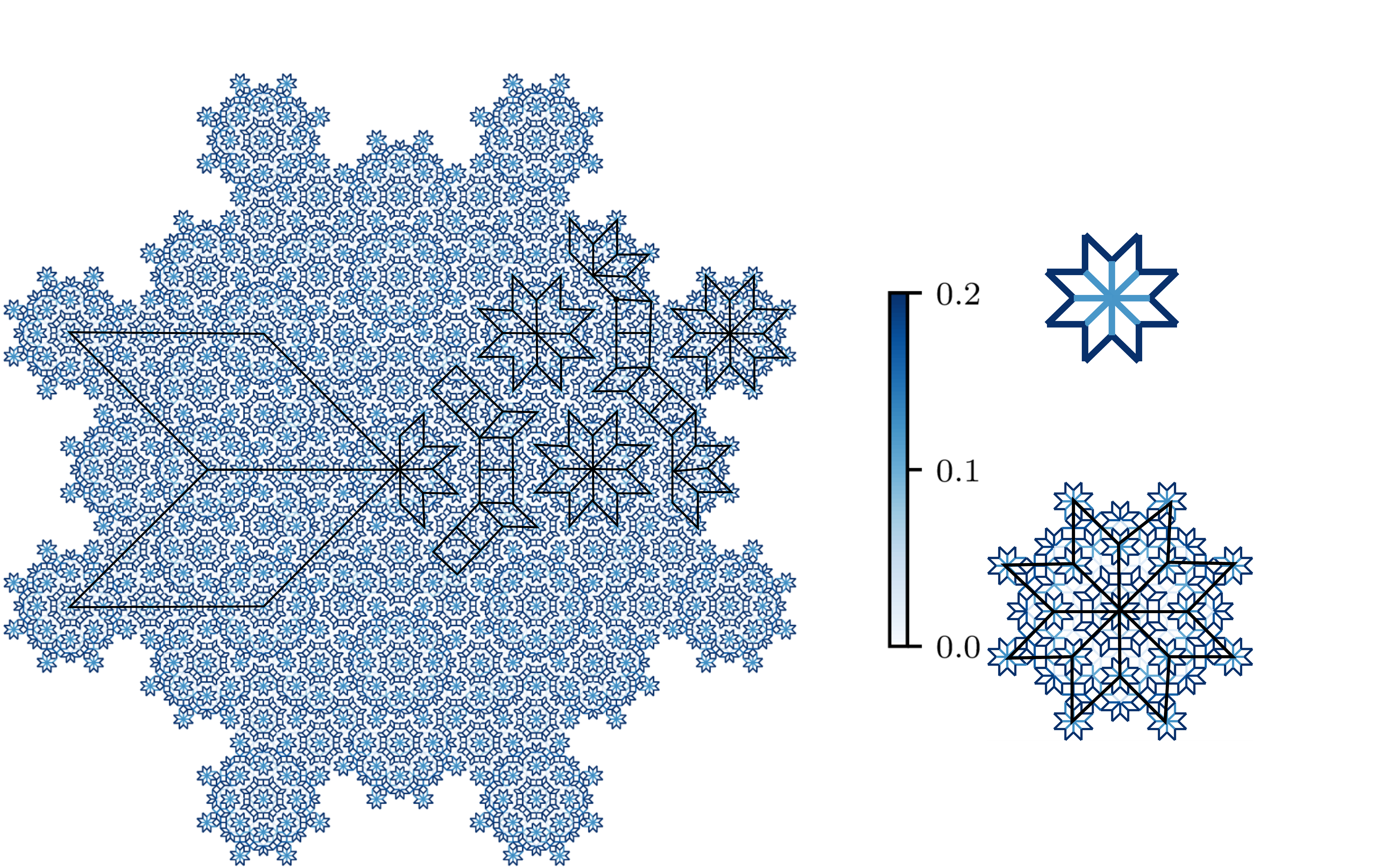}}
\caption{Left: Dimer occupation densities on an $8_4$-unit, a finite $D_8$-symmetric patch of \AB~tiling. The region is bounded by the $6$-th
pseudomembrane. The lighter colours indicate a rich nested structure of pseudomembranes associated with different $8$-vertices and their local
empires. The black lines emphasize the scale symmetry of the lattice (and consequently maximum matchings): they denote sections of the larger
\AB-tilings ($\mathcal{T}_{-2}$ and $\mathcal{T}_{-4}$) composed of the $8_2$ and $8_0$-units. Each $8_n$-vertex is surrounded by a pseudomembrane,
and hence connected to the rest of the graph by at most one dimer. They sit at the vertices of a larger \AB-tiling ($\mathcal{T}_{-n}$), and act as
effective units. Right: An $8_0$-unit (top) and an $8_2$-unit (bottom), with edges coloured to indicate dimer-occupation densities.}
\label{fig:ddensities}
\end{figure*}

In Fig.~\ref{fig:ddensities}, we display dimer occupation densities for the $8_4$-, $8_2$- and $8_0$-units to highlight the structures discussed in this section. The data were obtained using Monte Carlo methods, as discussed in the next section. The pseudomembranes appear as rings of edges with near-zero dimer density, concentric with the centres of local $D_8$ symmetry. For the $8_4$-unit we have highlighted the structures of the $2$nd order effective ladders. The matching of the $8_4$-unit can therefore be seen from multiple scales: as the matching of vertices on the basic $\mathcal{T}$ graph; as the effective matching of $8_0$-vertices, mediated by the ladders of the \ABB\ tiling; or of the effective matching of $8_2$-units, mediated by the $2$\textsuperscript{nd}-order effective ladders. The scale symmetry in the matching problem --- where effective units are matched at each scale of the tiling --- hints at the possibility for non-trivial signatures of the scale invariance in the dimer and monomer correlations. We now proceed to explore this question numerically.

%
\section{Numerical results on the \AB~tiling}
\label{sec:numerics}
%

In this section, we explore the correlations of the \AB\ tiling via a numerical study of the classical dimer model on finite patches of the \AB~tiling using the \textit{directed-loop algorithm}. Originally introduced to efficiently sample space-time configurations using quantum Monte Carlo algorithms~\cite{Evertz_Lana_Marcu,Syljuasen_Sandvik,Syljuasen}, this method has been adapted to sample the configuration space of classical dimer models and to access their monomer correlations~\cite{Sandvik_Moessner,Alet_etal}. The algorithm involves introducing two monomer defects into a maximum matching and transporting one of the monomers around an alternating cycle until annihilating it with its partner monomer (or with any other monomer, in graphs not admitting perfect matchings). Since detailed balance is satisfied for these intermediate configurations with two monomers, the corresponding partition function can be sampled. This gives us access to monomer correlations without additional computational effort. For completeness, we review the algorithm in Appendix~\ref{app:numericalmethods}.

\subsection{Choice of samples and boundary conditions}
\label{sec:boundaryconditions}

Since quasiperiodic systems are not translationally invariant and do not admit periodic boundary conditions, some care must be taken in choosing appropriate finite patches for our numerical studies of the \AB~tiling. We consider finite patches with exact $D_8$ symmetry. We anticipate that understanding matching problems on such patches can yield results representative of an arbitrary finite patch of \AB~tilings, since those are characterized by an effective matching problem of $D_8$ empires. Additionally, every finite patch of tiling is a part of a larger $D_8$ empire of the infinite tiling. 

For our largest simulations we consider an $8_4$-unit. We use the $8_4$-unit rather than the $8_4$-empire as the components of the empire outside the largest pseudomembrane generate significant boundary effects. The central $8$-unit has $6$ concentric pseudomembranes around it. As explained in Sec.~\ref{sec:quasiproof}, in any perfect matching of the \AB~tiling, precisely one dimer straddles the pseudomembrane to correlate the dimer configurations inside and outside the region it encircles. Hence, it is reasonable to expect that calculations of observables defined entirely inside a region enclosed by a pseudomembrane will not be severely affected by the rest of the tiling even in finite patches. With these modifications, the largest \AB~patch we consider contains $15473$ vertices and $30136$ edges. In some cases, we consider a smaller patch consisting of the $8_2$-unit. This sample contains $481$ vertices and $872$ edges. Below we present dimer and monomer correlations for these finite patches. Note that all maximum matchings of these patches host a single monomer with the charge of the central $8$-vertex. If the finite patch were extended to the infinite tiling, this monomer would annihilate with an oppositely charged monomer in another region, creating the single dimer crossing the pseudomembrane. With this in mind, we also consider monomer correlations on the charge-neutral punctured tilings, obtained from the above patches by removing the central vertex. We will show that the two cases (entire and punctured tilings) demonstrate qualitatively different behaviours for the monomer correlations. 

\subsection{Dimer correlations}
\label{sec:dimercorrs}

The dimer occupation densities are displayed in Fig.~\ref{fig:ddensities}. Motivated by the structure of effective matchings and the possibility of non-trivial long range correlations, we investigate the connected correlations $C(e_i,e_j)$ of dimers on edges $e_1$ and $e_2$, defined to be
\begin{equation}
C(e_i,e_j)= \langle n(e_i) n(e_j) \rangle - \langle n(e_i) \rangle \langle n(e_j)\rangle,
\label{eq:dimcorr}
\end{equation}
where $n(e_i)=1$ if the edge $e_i$ hosts a dimer, and $n(e_i)=0$ otherwise. For the system sizes under consideration we find that connected dimer correlations do not decay exponentially, as they would for either long-range ordered or disordered dimer covers. Characterizing $C(e_i,e_j)$ is complicated owing to its lack of translational invariance, and its high degree of inhomogeneity. Fig.~\ref{fig:largecorrs} shows the dimer-correlation function $C(e_0,e_j)$ for an edge $e_0$ which connects a $4$-vertex to a ladder (formed of $2$-vertices).

\begin{figure}[t]
\includegraphics[width=0.9\columnwidth]{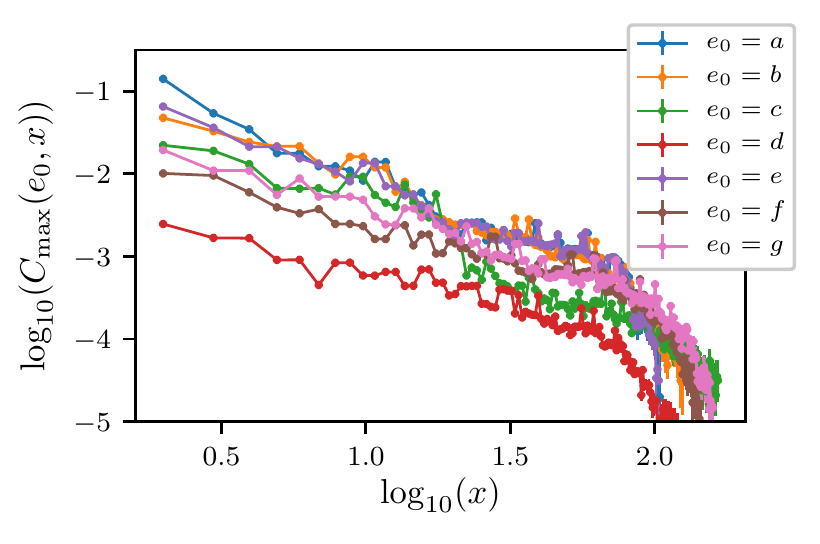}
\caption{Power law connected correlations of dimers for a $D_8$-symmetric sample (an $8_4$-unit). For several edges $e_0$, $C_{\mathrm max}(e_0,x)$,
the maximum absolute value of dimer correlations at a graph distance of $x$ edges from $e_0$ has a slow decay consistent with a power law. The apparent dip
at $\log_{10}(x)\sim2$ corresponds to effects of the sample boundary. We have displayed several choices of $e_0$ in this figure: $a$, $b$ and $c$ are
edges between $8_0$-units and $L_2$ ladders; $d$ is a edge between an $8_0$-unit and an $L_3$ ladder; $e$ is a edge on an $L_1$ ladder; $f$ connects
an $8_2$-unit to $L_4$; $g$ is a edge between an $8_2$-unit and an $L_1$ ladder. A large-scale plot of the tiling showing all the different source edges for this plot is provided in Appendix~\ref{app:legend}.}
\label{fig:ddcorrsplaw}
\end{figure}

\begin{figure}[t]
\includegraphics[width=0.9\columnwidth]{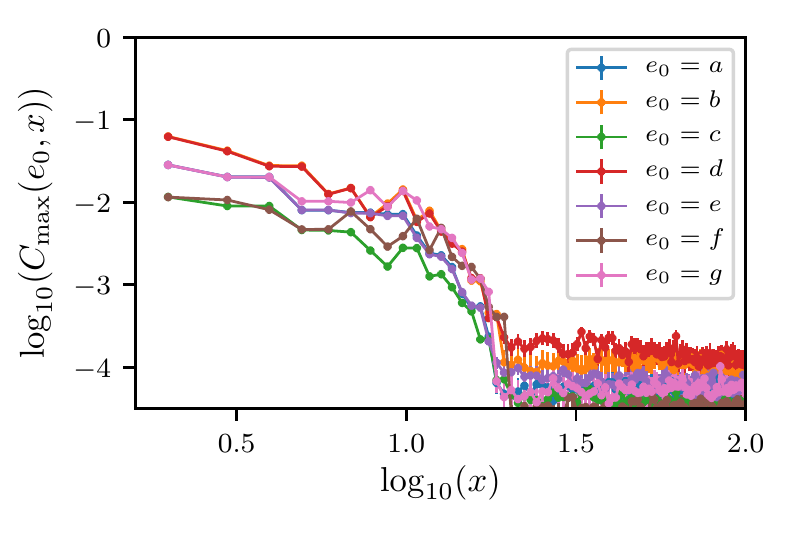}
\caption{Connected correlations of dimers for a $D_8$-symmetric sample (an $8_4$-unit). In contrast to Fig.~\ref{fig:ddcorrsplaw}, for several edges
$e_0$, $C_{\mathrm max}(e_0,x)$, the maximum absolute value of dimer-correlations at a graph distance of $x$-edges from $e_0$, is sizeable only inside
the $8_2$-unit in which $e_0$ lies. In this figure, we have chosen $e_0$ to be edges $a-h$, all connecting an $8_0$-unit to an $L_1$ ladder, inside a
larger $8_2$-unit. A large-scale plot of the tiling showing all the different source edges for this plot is provided in Appendix~\ref{app:legend}.}
\label{fig:ddcorrsbounded}
\end{figure}

To characterize the decay of correlations, we calculate $C_{\max}(e_0,x)$: the maximum value of $|C(e_0,e_j)|$ such that the edge $e_j$ has a graph
distance of $x$ edges from $e_0$. We display this quantity, computed for seven different choices of $e_0$, in Fig.~\ref{fig:ddcorrsplaw}. We see a
slow decay across the system, consistent with power law asymptotic behaviour, cut-off only by the boundaries of the sample. We conjecture that this slow power law is a manifestation of the
description in terms of the effective matching problems, at all scales, in terms of $8_n$-vertices, mediated by the effective ladders formed of
$8_{n-2}$- and $8_{n-1}-$units as described in Sec.~\ref{sec:emp}. We emphasize that these power-law-like correlations are neither homogeneous nor
translationally invariant, and are unrelated to the familiar power laws appearing in bipartite lattices with a continuum Gaussian action. Not all
edges have dimer correlations which decay slowly across the whole system. In general, different dimer correlation functions decay slowly until a
cutoff lengthscale, after which they rapidly fall off. The cutoff lengthscales are set by the pseudomembranes. To gain an understanding of this, we
look at the dimer correlations within the basic star surrounding the 8-vertices. For a dimer on one of the eight central membrane links, connected
correlations outside the star are identically zero.\footnote{This can be seen by checking that for each choice of dimer connecting the central
8-vertex to the star, and the dimer connecting the star to the rest of the tiling, there is exactly one possible dimer configuration on the star
itself.} Combined with the discussion of effective matching problems, this suggests that links on and inside the $n$-th pseudomembrane have dimer
correlations which are effectively cut off by the $(n+1)$-th pseudomembrane. In a system of linear size $L$, we therefore expect slowly decaying correlations over all scales, cut off by lengthscales in the set $L, L/\delta_S^2, L/\delta_S^4 \ldots$ We confirmed this numerically. For example, inside an $8_2$-unit, the edges connecting $8_0$-units to $L_1$ ladders (\textit{i.e.}, inside the $3$-rd pseudomembrane) do not have significant connected correlations outside the $8_2$-unit (bounded by the $4$th pseudomembrane). Such correlations are displayed in Fig.~\ref{fig:ddcorrsbounded}.  We remark that the correlations outside the $8_2$-unit are significantly reduced, but unlike an elementary star, they are not exactly zero; the analyis of the elementary star presented above merely provides us with a guiding principle to understand the cut-off lengthscales in the decay of dimer correlations for larger regions.

\subsection{Monomer correlations}
\label{sec:monocorrs}

Next we turn to monomer correlations. For bipartite graphs with perfect matchings, the monomer correlation function is just the partition function
$Z(x,y)$ over dimer configurations, up to normalisation, with two oppositely charged monomers situated on vertices $x$ and $y$. As noted in the section introduction, the symmetric patches have an extra vertex in the bipartite subset of the central 8-vertex. Without loss of generality we denote this subset $\mathcal{U}$. To probe the physics associated with a single monomer defect pair in maximum matchings, we first introduce an auxiliary partition function $Z'(x_1,x_2,y)$, having two $\mathcal{U}$-monomers on the vertices $x_1$ and $x_2$ and a single $\mathcal{V}$-monomer at $y$. We then define the two-particle correlation function as 
\begin{equation}\label{eq:mmcorr_def}
    Z(x,y) = \sum_{x_1,x_2} Z'(x_1,x_2,y)(\delta_{x_1,x}+\delta_{x_2,x}),
\end{equation} 
which calculates the pair correlation of oppositely charged monomers in a gas of three monomers.

To understand the monomer correlations, it is useful to consider dynamical processes like those in Fig.~\ref{fig:matchings}d applied to the three-monomer gas, where the monomers `random-walk' across even-length alternating paths by augmenting them.  As explained in Appendix.~\ref{app:numericalmethods}, such processes can be designed to respect detailed balance in the configuration space with three monomers (thereby leading to a Monte Carlo scheme to computationally study such monomer correlations). Consequently, they provide valuable insight into ensemble averages in the three-monomer gas like the pair correlation functions. In the rest of this section, we will use the term `random-walk' to mean such processes.

Without loss of generality, we consider the movement of $\mathcal{V}$-monomers while the $\mathcal{U}$-monomers are held fixed. Fig.~\ref{fig:largecorrs} shows a typical profile for $Z(x,y)$, with $x$ fixed and $y$ allowed to vary. While the structure of correlations  $Z(x,y)$ varies considerably with the $\mathcal{U}$-monomer position $x$, we observe two distinct features which hold across all values of $x$: 
\begin{itemize}
  \item $Z(x,y)$ is strongly peaked for $y$ within the first few pseudomembranes near the central $8$-vertex, taking its maximum value on the eight
      $\mathcal{V}$ vertices neighbouring the central $8$-vertex. Minima reside on the eight highest order $\mathcal{V}$-charged 8-vertices.
      Additional \emph{local} maxima (minima) occur for $y$ around (on) the locations of $\mathcal{U}$-charged ($\mathcal{V}$-charged) 8-vertices. Sites around (on) higher order $\mathcal{U}$-charged ($\mathcal{V}$-charged) 8-vertices are associated with stronger maxima (minima) of $Z(x,y)$.
    \item $Z(x,y)$ has comparable support around the entire angular range of the region, and is not confined in the vicinity of the $\mathcal{U}$-monomer.
\end{itemize}

The first point describes a `charged' attraction between the free $\mathcal{V}$-monomer and the tiling's 8-vertices, with the strength of the interaction set by the order of the 8-vertex. By charged, we mean that the $\mathcal{V}$-monomer has large correlations around the $\mathcal{U}$-charged 8-vertices (the central vertex being the highest order $\mathcal{U}$-charge in the region), but small correlations on and around $\mathcal{V}$-charged 8-vertices. This can be understood from the properties of the (pseudo) membranes. It is simplest to first understand the effect for exact membranes around the central vertex (i.e.~on the punctured tiling with the central vertex removed). The $\mathcal{V}$-monomer acts as a random walker, and begins its walk in some $\mathcal{H}_i$ region bounded by the two membranes $\mathcal{M}_i$ and $\mathcal{M}_{i+1}$. No alternating path exists that crosses $\mathcal{M}_{i+1}$ and also has an end point terminating on the monomer. This follows from the arguments in Appendix~\ref{app:proof_punctured}. Hence, the monomer's walk cannot cross the $\mathcal{M}_{i+1}$ membrane. Heading inward, however, the monomer may cross $\mathcal{M}_i$ at any point $x_i$ on the membrane, but may then only return to $\mathcal{H}_i$ by re-crosssing $\mathcal{M}_i$ at the \emph{same} point. It may further cross the next membrane, $\mathcal{M}_{i-1}$ at any $x_{i-1}$, but cannot come back unless it can return to the same $x_{i-1}$. For large membranes, the monomer is effectively `trapped' on the inward side (like a lobster in a pot). The net effect for high-order 8-vertices is thus a large probability for the monomer to be found near the centre. 

\begin{figure}
    \centering
    \includegraphics[width=0.5\linewidth]{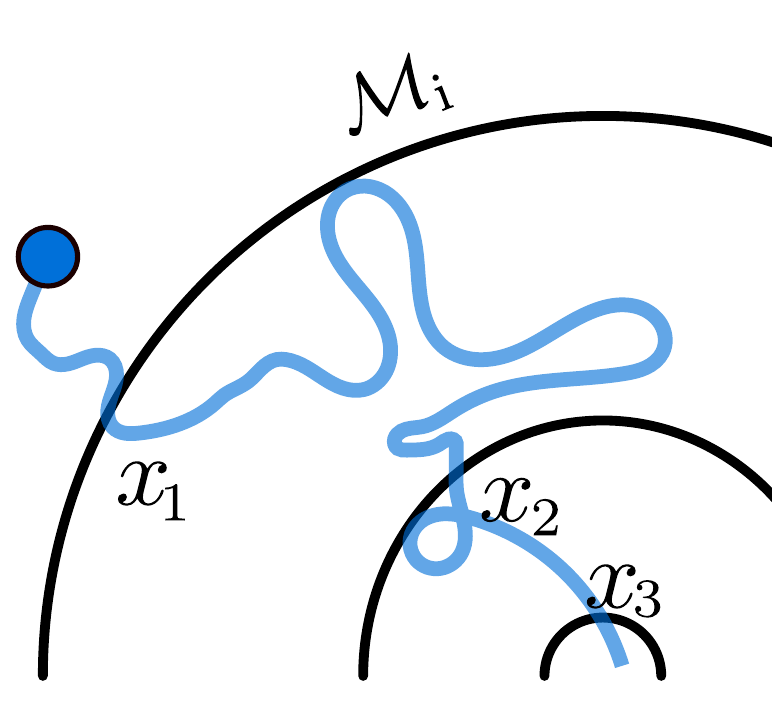}
    \caption{A random walking $\mathcal{V}$-monomer feels a global attraction to the central $\mathcal{U}$-charged 8-vertex on account of the
    membranes. The monomer, moving inward, can cross a membrane $\mathcal{M}_i$ at any point $x_i$, but then may only re-cross the membrane at the same point.}
    \label{fig:membranetrap}
\end{figure}

The reinstatement of the central vertex places a single dimer on each pseudomembrane $\mathcal{M}_i$ at a position $x'_i$. For large patches, with the averaging described in Eq.~\ref{eq:mmcorr_def}, a $\mathcal{U}$ monomer is likely to be found far from the centre, and so almost all pseudomembranes will host a dimer. This dimer provides an additional path for the $\mathcal{V}$-charged monomer to escape the trapping by pseudomembranes. That is, once a $\mathcal{V}$ monomer crosses inward from $\mathcal{H}_i$ to $\mathcal{H}_{i-1}$ through the pseudomembrane $\mathcal{M}_i$ at $x_i$, it can re-cross either at $x_i$ or $x'_i$. However, for large membranes, this does not significantly affect the arguments for trapping of monomers presented earlier. The existence of local maxima and minima of monomer correlations across the tiling's 8-vertices follows from the same reasoning applied to the pseudomembranes of all 8-vertices. This directly implies a high probability for the monomer to be found near the $\mathcal{U}$-charged 8-vertices. For the $\mathcal{V}$-charged 8-vertices, the effect is reversed. Now the monomer may only cross inward at the single point where a dimer exists on the pseudomembrane, but it may cross outward at any point along the pseudomembrane. Thus, a $\mathcal{V}$ monomer is trapped by the $\mathcal{U}$-charged pseudomembranes, and blocked by the $\mathcal{V}$-charged pseudomembranes. From the same argument, it is easy to see that for a fixed position of the $V$-charged monomer $y$, a random-walking $\mathcal{U}$ monomer is trapped by $\mathcal{V}$-charged pseudomembrane and blocked by $\mathcal{U}$-charged pseudomembranes. This implies that $Z(x,y)$, for fixed $y$, has local maxima (minima) for $x$ around (on) $\mathcal{V}$-charged ($\mathcal{U}$-charged) $8$-vertices. 

Higher order $8$-vertices, having more concentric pseudomembranes, result in stronger maxima/minima on sites on or around them. The configuration with
largest pair correlation (consequently smallest free energy) corresponds to the $\mathcal{V}$-monomer around the central $8$-vertex and the 
$\mathcal{U}$-monomer  around the highest order $\mathcal{V}$-charged 8-vertex. The separation between the central $\mathcal{U}$-charged 8-vertex and
the highest order $\mathcal{V}$-vertex scales with the size of the sample, and this implies that monomers have the
lowest free energy when separated over very large distances. Thus, monomers are radially `anti-confined', in striking contrast to familiar cases in periodic lattices where monomers are typically either confined or deconfined.

\begin{figure}
    \centering
    \includegraphics[width=0.9\linewidth]{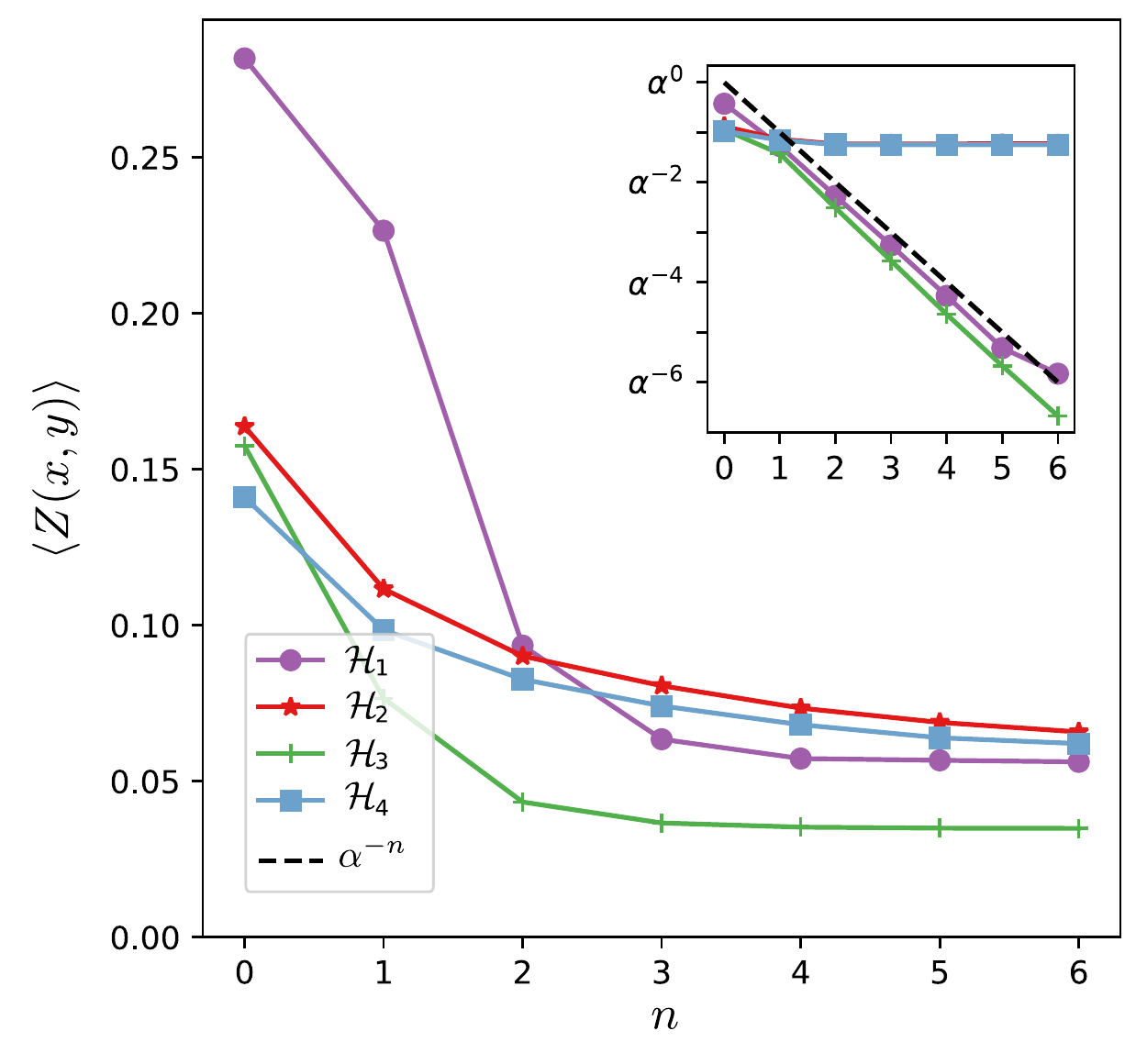}
    \caption{Averaged correlation function for a $\mathcal{U}$-monomer at position $x$ and a $\mathcal{V}$-monomer at $y$, both within
    a given $\mathcal{H}_i$ region. $\langle Z(x,y) \rangle$ is averaged over $x$ in the sector $\mathcal{S}_0$ and over $y$ in $\mathcal{S}_n$. On
the extended graph with 13 sectors (see main text), the furthest sector from $\mathcal{S}_0$ corresponds to $n=6$. For every $\mathcal{H}_i$ region,
$\langle Z(x,y) \rangle$ decays to a \emph{finite} constant at large $n$, signifying annular deconfinement of the monomers. \emph{Inset:} on the
punctured tiling with the central vertex removed, star-like ($i$ even) $\mathcal{H}_i$ regions remain deconfined, while ladder-like ($i$ odd) are
\emph{confined}, decaying exponentially to zero at large $n$. The log scale on the inset $y$-axis is set by $\alpha=\frac{1}{2+\sqrt{3}}$}.
\label{fig:mmcorr}
\end{figure}

The above mechanism effectively describes the behaviour of correlations in the radial direction (towards the central vertex, and other 8-vertices as a
sub-leading effect). With this in mind, we ask how the correlation behaves at large angular distances (around the central vertex) within a given
$\mathcal{H}_i$ region. $\mathcal{H}_i$ can be partitioned into eight identical sectors (`wedges'), each spanning an interior angle $2\pi/8$, which we
label by $\mathcal{S}_{n}$ for $n=0\ldots7$. We fix $x$ in $\mathcal{S}_0$ and ask how $Z(x,y)$ behaves for $y$ in $\mathcal{S}_n$ and large $n$ (by
`large', we mean the distance in sectors from $\mathcal{S}_0$). On the Ammann-Beenker tiling the largest $n$ we have access is $n=4$. However, we can
artificially increase $n$ by inserting additional symmetric wedges into the basic \AB\ tiling. This allows us to probe the large-$n$ limit without
changing the basic physics. In Fig.~\ref{fig:mmcorr} we display the results for an effective $8_4$-unit extended to 13 sectors; this patch is large
enough to see clear asymptotic behaviour while remaining computationally accessible. For each of the five $\mathcal{H}_i$ regions in the effective
$8_4$-unit, we plot $Z(x,y)$ averaged over all $x$ in $\mathcal{S}_0$ and all $y$ in $\mathcal{S}_n$, for $n \leq 6$. In each region we observe a
distinct plateau to a \emph{non-zero} constant at large $n$. This implies that, within a given $\mathcal{H}_i$ region, the free energy cost to
separate a monomer along the annular direction does not increase with distance . We dub this effect `annular deconfinement'. Along with the charged
attraction to 8-vertices explained above, this completes our characterisation of the structure of monomer correlations in the \AB\ tiling.

It is interesting to contrast this behaviour to the monomer-monomer correlations for the punctured tiling, the inset to Fig.~\ref{fig:mmcorr}. Here we
see that the correlations for the \emph{ladder}-like regions ($i$ odd) $\mathcal{H}_1$ and $\mathcal{H}_3$ now decay exponentially with the sector
distance $n$, whereas those of the star-like ($i$ even) regions again plateau to a finite constant. In fact, a transfer matrix calculation for the
$\mathcal{H}_1$ region requires that at large $n$, the leading behaviour of the correlation function goes as 
\begin{equation}\label{eq:confined_leading}
    Z_1(x,y) \propto \bigg(\frac{1}{2+\sqrt{3}}\bigg)^{n}.
\end{equation}
The effective matching problems (recall all odd-$i$ $\mathcal{H}_i$ form ladders of effective $8_{i-2}$-units) suggest that analogous behaviour should
emerge for the $\mathcal{H}_3$ region, and all other ladder-like region. This statement is supported by the near-identical exponential fit of
$\mathcal{H}_3$ in the inset of Fig.~\ref{fig:mmcorr}. However, once the central 8-vertex is reinstated, this large-$n$ `confinement' of the
ladder-like regions on the punctured tilings is fragile to the deconfinement of the star-like regions. On the punctured tilings the $\mathcal{H}_i$
regions are bounded by two exact membranes $\mathcal{M}_i$ and $\mathcal{M}_{i+1}$, and monomer correlations within the region $\mathcal{H}_i$ can be
described by considering the $\mathcal{V}$-monomer to be confined within the $\mathcal{H}_{i}$ region. Note that the $\mathcal{V}$-monomer can indeed cross $\mathcal{M}_i$ to escape to the region $\mathcal{H}_{i-1}$, but since it can only cross back through the same point, its excursion outside $\mathcal{H}_i$ does not contribute to monomer correlations within $\mathcal{H}_i$. The reinstatement of the central vertex places a dimer on each of the membranes, meaning the monomer can now take paths that cross one of the pseudomembranes at one point $x$, and re-cross at another point $x'$. For a monomer starting and ending its path in a ladder-like region, these paths provide a mechanism to deconfinement: the monomer takes a path which crosses over to a neighbouring star-like region, where it is annularly deconfined, only crossing back over near the end of the path. On the entire tiling, the monomer can therefore find  deconfining paths between all points starting and ending in the same $\mathcal{H}_i$ region.


\subsection{Aligning Interactions}
\label{sec:aligning_ints}

\begin{figure}[t]
  \includegraphics[width=\linewidth]{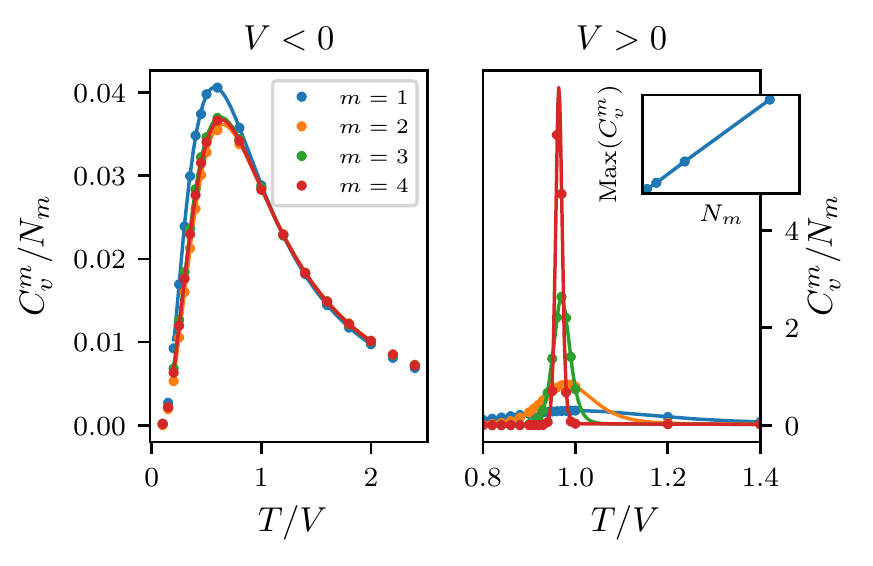}
  \caption{The intensive specific heat, $C^m_v/N_m$, of an $m^\textrm{th}$-order ladder with aligning interactions, as a function of the temperature $T$. Lines denote transfer matrix calculations, whereas the points represent data obtained from Monte Carlo simulations. \emph{Left}: When $V<0$, the intensive specific heat has a broad feature independent of the ladder size, associated with a crossover to columnar states. \emph{Right}: When $V>0$, the ladders exhibit a sharp feature in the intensive specific heat, which scales with the number of edges in a ladder (shown in inset). }
  \label{fig:ladd_cv}
\end{figure}

\begin{figure}[t]
  \includegraphics[width=\linewidth]{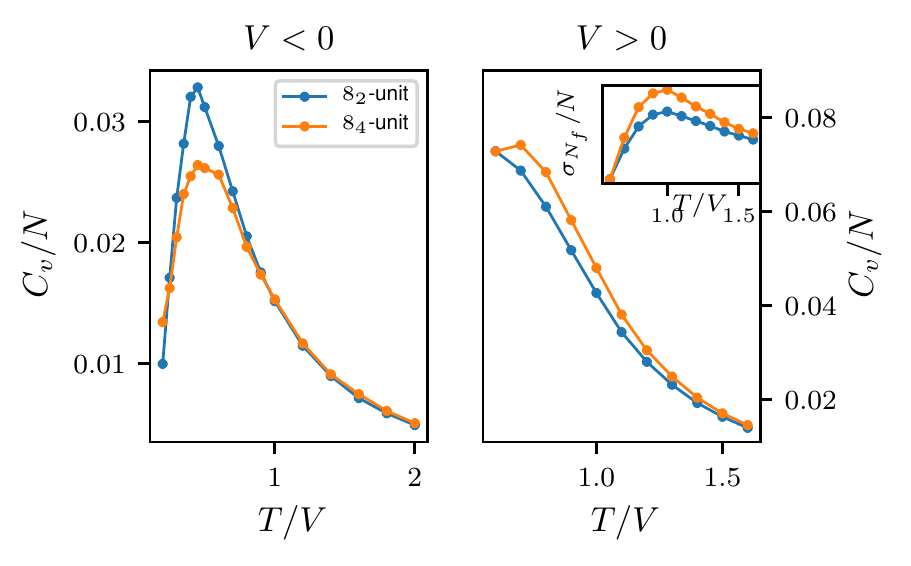}
  \caption{The intensive specific heat, $C_v/N$ of finite \AB~patches with aligning interactions, as a function of the temperature $T$.  \emph{Left}: When $V<0$, the crossover to columnar states on the ladders is associated with a broad feature which does not scale with the size of the tiling. \emph{Right}: For $V>0$, the sharp peak seen in the case of the ladders in Fig.~\ref{fig:ladd_cv} is significantly broadened here. \emph{Inset}: $\sigma_{N_f}/N$ (Eq.~\eqref{eq:sigmanf}), indicating fluctuations in the number of flippable plaquettes, has a broad feature which indicates a crossover to staggered states on the ladders.}
  \label{fig:ab_cv}
\end{figure}

Finally, we briefly discuss the role of aligning interactions. To explore the possibility of ordered dimer phases and their associated transitions in the
absence of periodic lattice symmetries, we consider a classical model defined by the energy function 
\begin{equation}
  \includegraphics[width=0.8\columnwidth]{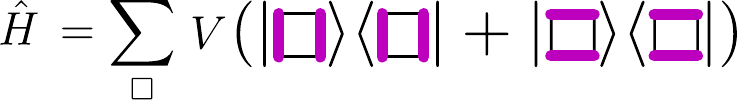}
  \label{eq:RK}
\end{equation}
where the sum is taken over all plaquettes $\square$ (the four edges comprising a square or rhombus tile of \AB). The potential $V$ either favours
(negative sign) or disfavours (positive sign) the presence
of `flippable' plaquettes which host a pair of parallel dimers. Our energy function can be understood as the classical limit of the quantum
Rokhsar-Kivelson Hamiltonian~\cite{RokhsarKivelson88}, with the kinetic term set to zero.  Working entirely within the manifold of maximum matchings,
the partition function is weighted according to
\begin{align}
Z=\sum_{\mathcal{C}} e^{-V N_F(\mathcal{C})/T},
\label{eq:aligning_ints}
\end{align}
where $N_F(\mathcal{C})$ counts the number of flippable plaquettes in a dimer configuration $\mathcal{C}$. $T$ is a temperature that we introduce for
convenience. Only the combination $V/T$ has physical significance. Regardless of sign, a non-zero value stabilizes ordered phases which break lattice
symmetries.  

To orient the discussion, first consider the effect of aligning interactions on the ladders which make up the \ABB~tiling. Negative $V$ favours the
maximally flippable columnar states on the ladders, while positive $V$ favours the two staggered states, which are in topologically distinct sectors.
That is, each staggered state cannot be reached from any other state by a sequence of local plaquette flips. These ladders can be treated exactly
using the transfer matrix approach, as described in App.~\ref{app:transfer_matrices}.


We calculate the intensive specific heat for an $m^\textrm{th}$-order ladder with $N_m$ edges (this is obtained by taking derivatives of the free
energy in the usual way). We show the result in Fig.~\ref{fig:ladd_cv}. For $V<0$, the intensive specific heat exhibits a broad peak which does not
scale with the size of the ladder. This feature corresponds to the loss of entropy incurred in the crossover to the columnar state. Each ladder has a
macroscopically degenerate number of columnar ground states --- in the language of App.~\ref{app:ladder_infl}, each $A$ segment contributes two states while each $B$ segment contributes three. When $V>0$, the intensive specific heat has a sharp feature at $T/V\approx 1.0$, signalling the onset of staggered states in the ladders. As shown in the inset of Fig.~\ref{fig:ladd_cv}, $C^m_V/N_m \sim N_m$. This conventionally signals a first-order phase transition, though the latter term should be used cautiously since one can only take a particular sequence of $N_m$. Note, however, that this does \textit{not} drive a first-order transition in the \ABB~tiling, since any thermodynamic singularity is suppressed by the exponential decay in the density of ladders with their order $m$.

We now compute the intensive specific heat $c_v = C_v/N$ of \AB~patches with $N$ edges (Fig.~\ref{fig:ab_cv}) using Monte Carlo simulations. As in the case of the \ABB~ladders, for $V<0$, $c_v$ has a broad feature at $T/V\approx 0.5$ which corresponds to the onset of columnar-like configurations on the ladders. On \AB, dimer correlations are no longer restricted to ladders, but since the ladders host most of the flippable plaquettes, tuning $V$ enhances the role of the ladders even in this case. However, when $V>0$, the sharp feature associated with the transition to the staggered states at $T/V\approx 1$ in the \ABB~ladders (Fig.~\ref{fig:ladd_cv}) is no longer present in the \AB~patches.  

We calculate the standard deviation $\sigma_{N_F}$ of the number of flippable plaquettes $N_F$ as
\begin{align}
  \sigma_{N_F}= (\langle N_F^2 \rangle - \langle N_F\rangle ^2).
  \label{eq:sigmanf}
\end{align}
The results are displayed in the inset of Fig.~\ref{fig:ab_cv}. We see that $\sigma_{N_F}/N$ exhibits a broad feature at $T\approx 1$, which suggests that the dimers settle into configurations resembling staggered states on the ladders \textit{without} undergoing a phase transition.

\begin{figure}[!ht]
  \includegraphics[width=5cm]{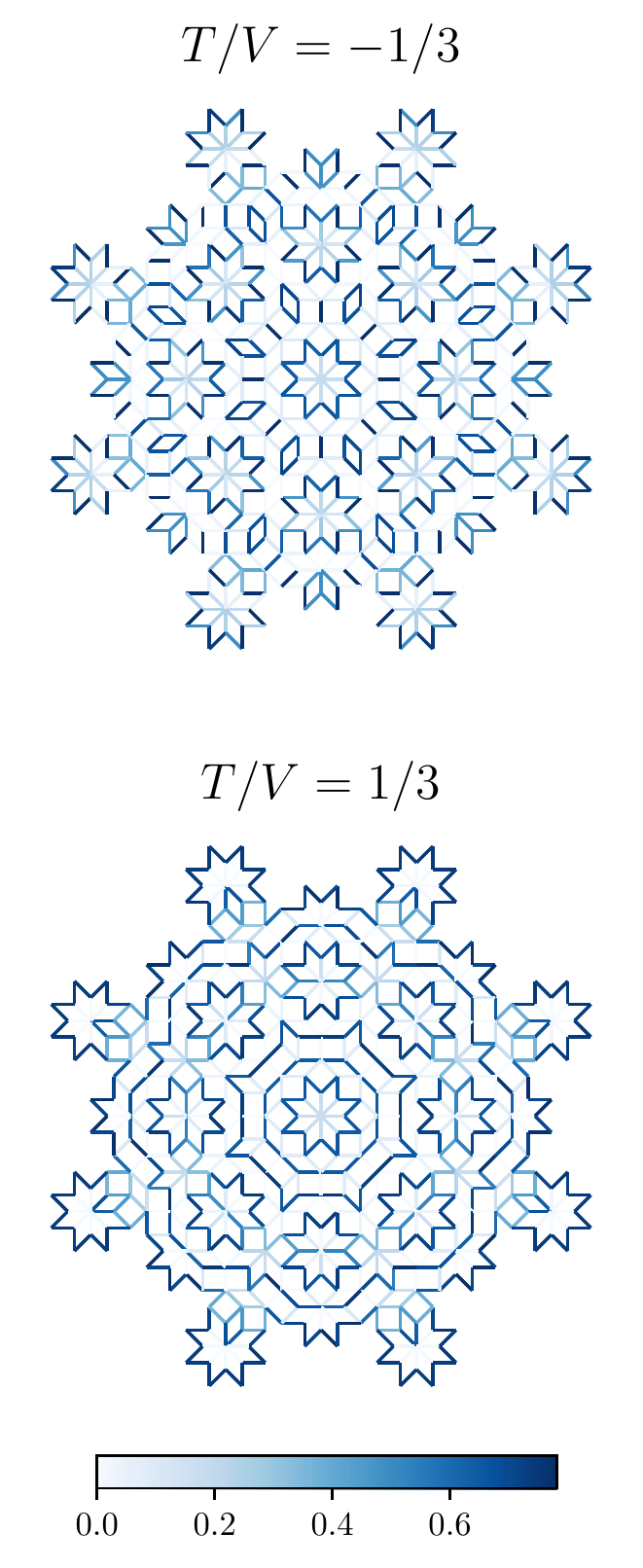}
  \caption{Dimer densities on an \AB~patch ($8_2$-vertex) with aligning interactions. \emph{Top}: at $T/V=-1/3$, the dimer densities reveal a strong tendency to align along the legs of the ladders, indicating that staggered configurations on the ladders are favoured. \emph{Bottom}: at $T/V=1/3$, the dimers tend to align along the rungs of the ladders, indicating that columnar states on the ladders are favoured.}
  \label{fig:orders}
\end{figure}

Fig.~\ref{fig:orders} shows the dimer densities of an \AB~patch in the presence of aligning interactions of both signs. We see that at large negative values of $V$ the dimers align along the rungs of the ladders, while at large positive values of $V$ the dimers align along the legs of the ladders. This suggests that typical dimer configurations of the full \AB~tiling in these limits can be described in terms of the cartoons of columnar and staggered states on the ladders. 

\section{Conclusions}
\label{sec:conclusions}
%

We have demonstrated that classical dimers on the \AB~tiling admit perfect matchings in the thermodynamic limit, with a rich structure linked to the interplay of constraints with quasiperiodicity. A crucial feature of our analysis is the identification of collections of edges whose dimer content can be strictly bounded from above in any maximum matching. Previous work on the Penrose tiling identified exact membranes, sets of edges that host zero dimers in maximum matchings, separating regions hosting monomers with distinct bipartite charge~\cite{Flicker_Simon_Parameswaran}. The present work extends this analysis in two ways. First, we identified exact membranes on the auxiliary quasiperiodic \ABB~tiling obtained by deleting $8$-vertices from the \AB~tiling. These membranes can be understood in terms of the fine Dulmage-Mendelsohn decomposition of a bipartite graph, reviewed in Appendix~\ref{app:memreview}. Second, returning to the full \AB~tiling by replacing the $8$-vertices, we found that membranes become \emph{pseudomembranes}, sets of edges that collectively host precisely one dimer in any perfect matching. 

The pseudomembrane structure of \AB~implies the existence of `effective matching problems' at every scale of the quasiperiodic tiling. Effective
units bounded by pseudomembranes are matched together in the same way as the basic graph vertices, with a hard-core dimer constraint imposed by the
pseudomembranes. This scale invariance leads to long-ranged, anisotropic, and site-dependent connected dimer correlations with a series of cut-off scales, which we found evidence for
using classical Monte Carlo simulations. Monomer statistics are likewise dominated by the tiling's pseudomembranes. In contrast to familiar cases, e.g.~monomer
correlations on the periodic square tiling, where a pair of oppositely charged monomers are confined by the pair
separation, on the AB tiling a neighbouring pair of test-monomers reach their lowest free energy for very large separations (of the order of the
system size). Along with this radial anti-confinement, we also observe annular deconfinement  around the 8-vertices. Our results on dimer and monomer correlations
place the behaviour of the dimer model on AB tilings in sharp contrast to familiar examples of dimer models of bipartite periodic graphs, where both dimer correlations (either algebraic or short-range) and test-monomer correlations (confining) are translationally invariant and isotropic, like the coarse-grained continuum height action which describes these systems. From the correlations discussed above, it is unlikely that the dimer model on AB tilings have such a description. Putting the connection between the effective matching problems and the underlying statistics on a firm quantitative basis is an important future direction.

The existence of perfect dimer matchings on the \AB~tiling in the thermodynamic limit is also distinguished from the situation in other
two-dimensional quasicrystals~\cite{Flicker_Simon_Parameswaran}. There are six minimal quasicrystals in two dimensions as identified in the
classification scheme of Ref.~\onlinecite{BoyleSteinhardt16B}. All vertices of the Penrose tiling belong to regions of one or other excess bipartite
charge. The oppositely charged regions are separated by exact membranes. These are the prototypes for the \AB~pseudomembranes in the present study. In
some sense, however, it is the lack of perfectly matched regions which makes the Penrose tiling unique, just as it is the absence of unmatched regions
which makes \AB~unique. The remaining four 2D quasicrystals contain regions with an excess of one or the other bipartite charge alongside perfectly
matched regions. The division between different regions is always formed by sets of edges which cannot host dimers in maximum matchings. 
Other graph decorations of the tiles can change these results. In the tiling literature such decorations are said to give `mutually locally derivable' (MLD) tilings~\cite{BaakeGrimm}. For example, it is possible to define a tiling MLD to the rhombic Penrose tiling which can be perfectly matched~\cite{Flicker_Simon_Parameswaran}. Such special cases aside, however, the remaining minimal quasicrystals, while still quasiperiodic and long-range ordered, begin to approach the generic results found in random or disordered bipartite graphs with two-dimensional embeddings~\cite{Biswas_etal,Dulmage_Mendelsohn,Pothen_Fan}. It is in this context that the results in the present study truly stand out.


Turning to the more thorny issue of quantum fluctuations, the existence of columnar and staggered states for opposing signs of the alignment potential
suggests that upon including dimer resonance moves, the phase diagram of quantum dimers on quasicrystals will resemble the Rokshar-Kivelson picture
for periodic lattices~\cite{RokhsarKivelson88,Moessner_Raman}. Therefore we anticipate the existence of at least a point in the quantum dimer phase
diagram, between the columnar and staggered limits, where equal-time dimer correlations resemble those of the classical models studied in this work.
However, whether this point requires fine tuning, or instead represents the properties of a robust phase of matter, is a more delicate question.
Historically, studies of dimer models on periodic bipartite lattices have made use of the height representation~\cite{Henley04,MoessnerSondhiFradkin};
in the quantum case, the height action must be supplemented by instanton contributions linked to the integer-valued nature of the height field.
Instantons destroy the long-range quantum dimer correlations in two spatial dimensions, rendering them exponentially short-ranged. In essence, this is
Polyakov's argument for the absence of a deconfined phase of $U(1)$ quantum lattice gauge theories in three spacetime dimensions~\cite{Polyakov1977}.
This means that the power law correlations of the RK model require the fine-tuning characteristic of a multicritical point rather than the stability
of a robust phase. The situation is less clear in the present case since, as we have noted above, there does not seem to be an obvious local height
action that characterizes maximally probable dimer configurations on quasicrystals. In fact, the anisotropic and site-dependent nature of dimer and monomer correlations strongly suggest the absence of any local, isotropic and translationally invariant action of coarse-grained fields which describe our system.  Therefore, Polyakov's arguments do not apply directly, and as we have seen, the notions of confinement and deconfinement become richer and subtler. Consequently there is a possibility that long-wavelength dimer correlations persist in the quantum problem. Verifying this is challenging, since in order to make precise statements one must approach the thermodynamic limit via a discrete sequence of inflations, and we are forced by quasiperiodicity to work with open boundary conditions. This means that the relevant computational cost likely becomes prohibitive before finite-size effects have been suppressed. In light of this, further study of the classical problem to determine if there is a convenient, possibly non-local, characterization of maximally probable dimer configurations seems warranted, as this might open a route to an analytical treatment. 

A more numerically tractable direction is to explore dimer models on the $D_8$-symmetric ladders introduced in the context of the \ABB~tiling in App.~\ref{app:transfer_matrices}. Recall that each of the eight symmetry-related segments of the ladder forms a system which tends to quasiperiodicity in the thermodynamic limit in its own right. This presents an interesting avenue for investigating quasiperiodic quantum dimer ladders. Periodic ladders, and closely associated frustrated spin ladders, have long been studied to shed light on the phases and transitions of quantum dimers~\cite{ChepigaMila,ChepigaMila19,VekuaHonecker,HikiharaStarykh,HungEA06,HijiiEA03}. We expect studies of quasiperiodic ladders to be similarly fruitful.

On a given bipartite graph there is a well-established mapping between monomers in dimer matchings, and localised electronic zero-energy modes in tight-binding models~\cite{Longuet-Higgins,WeikEA,Biswas_etal}. Here, the number of monomers hosted in a maximum matching is associated with the number of zero modes, while the monomer-confining regions are associated with wave functions whose support is confined within a compact subgraph. The computed monomer densities and geometry of monomer-confining regions on the Penrose tiling~\cite{Flicker_Simon_Parameswaran} are consistent with the density of zero modes and the nature of confined states obtained from investigations of hopping problems on the Penrose tiling~\cite{KohmotoSutherland,KogaTsunetsugu,Day-RobertsEA}. Ref.~\onlinecite{Koga} recently computed a finite  density of confined zero modes on the \AB~tiling; naively, this appears to be in conflict with our results that demonstrate that the \AB~tiling can be perfectly matched with vanishing monomer density in the thermodynamic limit. This apparent contradiction may be resolved by noting that the zero modes obtained in Ref.~\cite{Koga} are `fragile', in the sense that they move away from zero energy upon the introduction of arbitrarily weak disorder in the hopping matrix elements. In contrast, the Penrose tiling hosts `strong' zero modes that survive to any disorder strength. Formally, the monomer density computed in the dimer cover problem exactly equals the density of strong zero modes, but is not linked to the density of fragile zero modes. The fragility of the \AB~zero modes may be explicitly verified by computing the spectrum of the random-hopping problem: for any nonzero randomness, exactly one zero mode from Ref.~\cite{Koga} survives (the strong mode associated with the unavoidable central monomer on 8-fold symmetric  patches), with all the remaining modes moving to finite energy.

Finally, we comment on a relationship between the ideas explored in this paper and fracton phases of matter~\cite{Chamon2005,bravyi2009no,Haah2011,GlassyFractonDynamics,Pretko17,NandkishoreHermele18, PretkoChenYou}. The latter are usually defined on translationally invariant lattices, and blend topological features with sensitivity to geometry. Type-I fracton phases host quasiparticle excitations which cannot move individually, but which can combine into pairs or quadruplets to move along lines or planes. There is a passing resemblance to the membranes in \ABB, in which the minimum excitation out of a perfect matching would be the deletion of a single dimer, creating a monomer-antimonomer pair. Membranes restrict each individual monomer to move on a subset of vertices of the same bipartite charge. But the pair together is free to move anywhere, with one or other monomer `opening a door' through a membrane which the other closes. However, a key feature of mobile fracton pairs is that their separation is fixed, while the monomer pair has no such constraint. Very recent work~\cite{Surowka} has extended the duality between fractons and elasticity theory~\cite{PretkoRadzihovsky18} to quasicrystalline systems, but as yet it is unclear whether this has direct implications for the results presented here. In type-II fracton phases the excitations can only move along fractal subsets of the full system. Loosely speaking, they are an instance where a complicated set of gauge constraints on a simple lattice leads to emergent low-energy behaviour in which the natural gauge-charged objects are fractals. Contrast this with the present example, where a conventional Gauss-law-like structure (imposed by the dimer constraint) leads to a setting where `gauge lines' are themselves subject to fractal --- and fractally distributed --- barriers. It seems therefore that the study of dimer models on quasicrystals presents an intriguing counterpoint to fractons. In future, it will be interesting to explore whether properties such as unusual topological robustness at finite temperature and glassy dynamics characteristic of fractonic phases also emerge in the quasiperiodic dimer setting.

\begin{acknowledgements}
We thank Kedar Damle, Paul Fendley, and Shivaji Sondhi for insightful discussions. J.L.~and F.F.~acknowledge the advice, support, and encouragement of Uwe Grimm throughout this work, which we dedicate to his memory. We acknowledge support from the European Research Council under the European Union Horizon 2020 Research and Innovation Programme via Grant Agreement No. 804213-TMCS (S.A.P., S.B.), and EPSRC Grant EP/S020527/1 (S.A.P., S.H.S.). F.~F.~acknowledges support from the Astor Junior Research Fellowship of New College, Oxford. J.~L.~acknowledges support from the Ogden Trust. Statement of compliance with EPSRC policy framework on research data: This publication is theoretical work that does not require supporting research data.
\end{acknowledgements}

\begin{appendix}

%
\section{Membranes in perfectly matched bipartite graphs}
\label{app:memreview}
%

Here, we provide a graph-theoretic picture of how perfect matchings on a bipartite graph  decouples into perfectly matched regions separated by membranes. These regions are such, that in all perfect matchings, vertices in each region are matched to sites within the same region. Further, edges constituting membranes which separate these regions are never matched in any perfect matching. It turns out that these regions are precisely the components of the `fine' version of the Dulmage-Mendelsohn decomposition of bipartite graph theory\cite{Pothen_Fan}, well known in computer science and computational graph theory~\cite{Dulmage_Mendelsohn, Pothen_Fan,irving06,HararyPlummer}. As we will describe below, the literature also provides us with computationally efficient methods to determine these regions.

We note that in general bipartite graphs a maximum matching is not necessarily perfect. In general, such maximum matchings decouple into regions which host monomers in addition to perfectly matched regions and membranes~\cite{Flicker_Simon_Parameswaran, Biswas_etal}. These regions also turn out to be components of a more general Dulmage-Mendelsohn decomposition, involving monomer carrying regions. However, this does not directly concern us in the context of \AB~ tilings, and we will focus on the special case of perfectly matched graphs. 

Consider a bipartite graph $G$ with a perfect matching, with the bipartite subsets denoted by $\mathcal{U}$ and $\mathcal{V}$. Given a perfect matching $\mathcal{M}$, any other perfect matching $\mathcal{N}$ can be reached by flipping the dimer-occupancy states on  some set of alternating cycles on the graph. One can see this by first constructing the symmetric difference $M \oplus N$. $M \oplus N$ consists of edges which are matched in either $\mathcal{M}$ or $\mathcal{N}$, but not both. It is easy to see that $M \oplus N$ decomposes into cycles. By construction, these cycles are alternating cycles in $\mathcal{M}$. Starting with $\mathcal{M}$, flipping the dimer-occupancies on these alternating cycles gives us $\mathcal{N}$. The preceding arguments bring alternating cycles to the centre stage in our discussions on the finer structure of perfect matchings of bipartite graphs. Loosely speaking, the perfectly matched regions are the ones into which alternating cycles localize, while the membranes are edges through which no alternating cycles pass (such that they are not matched in any perfect matching). 

Let us formalize this intuition to define these regions precisely. First, consider an equivalence relation on the vertices of a graph. Given a maximum matching $\mathcal{M}$, two vertices $v_a$ and $v_b$ are related if there is an alternating loop going through both $v_a$ and $v_b$. This is an equivalence relation on the vertices, and therefore it divides up the vertices into equivalence classes $V_i$. The vertices in the class $V_i$ along with the edges which have both vertices in $V_i$ define a subgraph $G_i$. The $G_i$s  precisely describe the regions we are interested in--- all vertices in $G_i$ are perfectly matched within $G_i$ for all perfect matchings of the graph $G$. Edges which do not belong to any of these $G_i$ constitute the membranes--- these edges are not a part of any alternating cycle, and consequently do not host a dimer in any perfect matching.

We now turn to the algorithmic determination  of these regions~\cite{Pothen_Fan}. It is convenient  to construct an auxiliary directed graph $G_d$. To construct $G_d$, given a perfect matching $\mathcal{M}$ in  the graph $G$, we first direct all unmatched edges from $\mathcal{U}$-vertices to the $\mathcal{V}$-vertices. Then all matched edges are collapsed to a single vertex, labelled by the $\mathcal{U}$-site. This specifies the directed graph $G_d$, with vertices labelled by $\mathcal{U}$-vertices. By construction, alternating cycles of $G$ are now directed cycles of $G_d$. This recasting frames the problem in terms of a standard one: determination of ``strongly connected components" of a directed graph. Two vertices $v_a$ and $v_b$ of a directed graph are strongly connected if there exist a directed path from $v_a$ to $v_b$, and another directed path from $v_b$ to $v_a$. Strong connection defines an equivalence relation, and the corresponding equivalence classes define the strongly connected components. Standard algorithms, such as ones by Tarjan~\cite{Tarjan} and Kosaraju~\cite{sharir} can efficiently solve the problem of determination of strongly connected components of a bipartite graph.

Finally, we briefly outline an alternative, but illuminating, view of the problem in terms of block-triangular factorisation (BTF) of matrices, which was the initial context in which many of the above ideas were introduced~\cite{Pothen_Fan}.
A bipartite graph can be represented by a matrix $G$ with rows denoting one bipartite subset $\mathcal{U}$, and columns denoting the
other subset $\mathcal{V}$.  An edge between an $\mathcal{U}$-vertex $i$ and a $\mathcal{V}$-vertex $j$ corresponds to a nonzero value of 
the element $G_{ij}$. For our purposes, the actual numerical value of $G_{ij}$ is not important, but it may encode other information like the weights of edges in a weighted matching problem. Readers may be more familiar with the graph adjacency matrix $\mathcal{A}$, a square matrix labelled by vertices of the graph, and whose nonzero entries correspond to edges. In terms of $G$, the adjacency matrix is
\begin{equation}
\mathcal{A} = \left( \begin{array}{cc} 0 & G\\ G^T &0 \end{array}\right). 
\end{equation}

The decoupling of the graph into perfectly matched regions and membranes can also be viewed as a block-triangular factorisation (BTF) of the matrix $G$, achieved by a permutation of its rows and columns. Such a BTF is of the general form

\begin{align}
\label{fine_2}
G= \left(
 \begin{matrix}
   G_1    & M_{12} &\ldots&M_{1n}\\
   0    & G_{2} &\ldots&M_{2n}\\
 \vdots & \vdots &\ddots &\vdots\\
0 & 0 &0 &G_n\\
\end{matrix}
\right
).
\end{align}
The diagonal blocks $G_i$ are square matrices which denote the perfectly matched regions, while the off-diagonal blocks $M_{ij}$  denote the edges which make up membranes. This can be easily seen. Consider the first diagonal block $G_1$. First note that the $\uu$-vertices in $G_1$ must match to the $\vv$-vertices in $G_1$ because we know that $G$ is perfectly matched and the $\vv$-vertices in $G_1$ have no neighbours outside of the $\uu$-sites in $G_1$. This implies that the edges in the blocks $M_{11} \cdots M_{1n}$ are never matched in a perfect matching, \textit{i.e.,} they constitute membranes. With the $\uu$-vertices in $G_1$ used up, the $\vv$-vertices in $G_2$ must all match to $\uu$-vertices in $G_2$, implying that the edges in $M_{23}\cdots M_{2n}$ constitute membranes. This argument can be continued to show that sites in the regions $G_i$ are perfectly matched to sites in the same region, while edges in $M_{ij}$ are never matched in any perfect matching. Of course, the BTF ultimately depends on the graph, and it might happen that no such BTF is possible for certain graphs, or more precisely the BTF corresponds to a single diagonal block encompassing the whole graph.

\section{Inflation structure of ladders on the \ABB\ tiling}
\label{app:ladder_infl}

\begin{figure}
    \includegraphics[width=\linewidth]{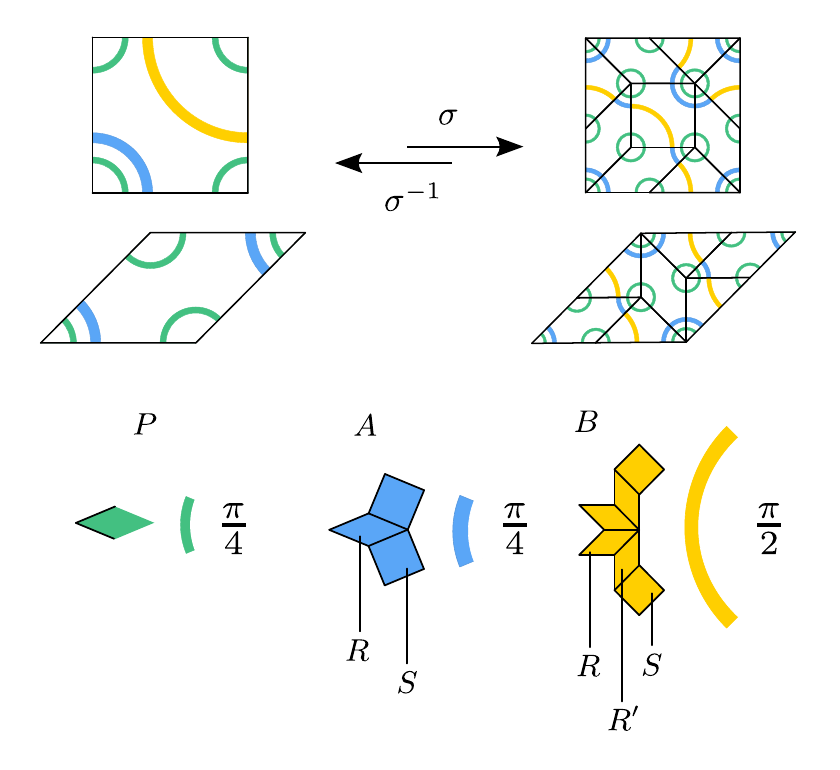}
    \caption{The inflation of the ladder tiles in Fig.~\ref{fig:abbtiles}. We distinguish three distinct units that build the stars and ladders of the \ABB~tiling: $P$, $A$, and
    $B$. White space separating segments represents impermeable membranes. Denoting the three different tile environments that appear in any segment
as $S$, $R$, and $R'$, we have $A=SRS$, $B=SR'R^2R'S$. The
segments are marked on the tiles by coloured arcs of internal angles $\frac{\pi}{4}$, $\frac{\pi}{4}$, $\frac{\pi}{2}$ for segments $P$, $A$, $B$ respectively, which retain the symmetry of the original segment.}
    \label{fig:ladderinflate}
\end{figure}

As explained in Sec.~\ref{sec:mem_quasimem}, the structure of \ABB~is composed of stars, ladders and membranes. The \AB\ inflation rule $\sigma$
implies an inflation rule, displayed in Fig.~\ref{fig:ladderinflate}, for the ladder tiles of Fig.~\ref{fig:abbtiles} (technically, to perform
inflation on \ABB, the inflation is performed on the \AB~tiling and then the $8$-vertices are once again removed, mapping back to \ABB). Rather than
work at the level of the tiles, we distinguish three repeating units: the basic 2-edge unit of the star, which we label $P$ (shaded green); a ladder
segment with the structure $SRS$ (square-rhombus-square), which we label as $A$ (shaded light blue); and a ladder segment with the structure
$SR'RRR'S$, which we label $B$ (shaded yellow). Each unit connects at its two end-points with other units, forming longer segments. We note the basic
rhombus tile can appear in two distinct environments ($R$ and $R'$): the $R$ rhombi appear in the $A$ segment or in the centre of the $B$ segment, and
have a single vertex which only has edges to two other vertices in the ladder; the $R'$ rhombi appear sandwiched between $S$ and $R$ in $B$. (Viewing
each ladder as a graph, the $R'$ environment is equivalent to the $S$ environment in the way in which it connects to neighbouring tiles, and this identification is used in defining the transfer matrices
for enumerating matchings, Eq.~\ref{eq:A&Bmatrices}.) We denote a tile or segment $T$ repeated consecutively $n$ times as $T^n$.

From Fig.~\ref{fig:ladderinflate} we can read off the inflation of the segments:

\begin{equation}
  P \rightarrow A, \hspace{0.3cm} A \rightarrow {B}^{1/2}A{B}^{1/2}, \hspace{0.3cm} B \rightarrow {B}^{1/2}A^4{B}^{1/2}.
\end{equation}
Although a one-to-one identification creates fractional-power $B$ segments, two of these always join to form a whole segment: $(B^{1/2})^2 = B$. We are free to shift our inflation rule by $B^{1/2}$ to the right, obtaining the more convenient form

\begin{equation}
 \xi\ \hspace{.5cm} :\hspace{.5cm} P \rightarrow A, \hspace{.5cm} A \rightarrow AB, \hspace{.5cm} B \rightarrow A^4B.
 \label{eq:xiinflation}
\end{equation}
A 1D inflation rule $\xi$ over the three-letter\footnote{Since $P$ never occurs after the first inflation the alphabet is effectively two-letter.} `alphabet' $(P, A, B)$ is then obtained if we further define
\begin{equation}
  \xi(MN) = \xi(M) \xi(N)
  \label{eq:homomorphism}
\end{equation}
for any `words' $M,N$. Due to the invertibility of $\sigma$, the inverse of $\xi$ must also exist: we denote it $\xi^{-1}$. The 2D inflation $\sigma$
is therefore reduced to an effective 1D inflation $\xi$ on the ladders forming the \ABB~tiling.

The $P$ segments always close into isolated stars with $D_8$ symmetry, i.e.~$P^8$ (with periodic boundary conditions). Each ladder either closes in a
loop or spans the entire tiling --- this follows by noting from the configurations in Fig.~\ref{fig:abbvertices} that ladders cannot branch or
terminate in the tiling. Additionally, if a ladder closes, it does so with $D_8$ symmetry. To see this, observe that the inflation of the ladder tiles
(Fig.~\ref{fig:ladderinflate}) preserves the symmetry of the segments. Then, given a closed ladder loop, repeated deflation must eventually result in
a loop containing $P$ segments, since $\xi^{-1}$ only destroys $P$s. It follows that ladders can only close with $D_8$ symmetry.

The possibility of a system-spanning ladder on the infinite tiling is evident by considering the case of repeated inflation of any single segment; the infinite ladder is obtained in the limit. The possibility of \emph{multiple} system spanning ladders is more interesting. Similar curves to those in Fig.~\ref{fig:ladderinflate} can be used to
decorate the tiles of the Penrose tiling; in that case, Penrose and Conway have independently shown that at most two system-spanning curves can exist
in a tiling~\cite{Gardner}. We expect that a similar statement holds for the Ammann-Beenker tiling. Such infinite ladders are very rare in any case
--- from the LI property of \AB, any finite patch of tiling can be taken to lie inside a closed ladder of large enough extent, wherein all ladders
must close. The existence of these rare ladders will have negligible effect on the matching problem, and in the main text we assume all ladders are closed with $D_8$ symmetry. 

With the terminology of Sec.~\ref{sec:memproof}, we define the star, $P^8$, to be the order-0 ladder, set $A_0 \equiv P$, and henceforth phrase our discussion in terms of ladders of order $n\geq 0$ according to the inflation hierarchy. With the $D_8$ symmetry constraint, and assuming periodic boundary conditions, we obtain the structure of the order-$n$ ladder as 
\begin{equation}
 L_n = (\xi^n(A_0))^8 \equiv A_n^8.
 \label{eq:ladderstruc}
\end{equation}
Thus the order-1 ladder corresponds to $A^8$, the order-2 ladder to $(AB)^8$, the order-3 ladder to $(ABA^4B)^8$, and so on.

The order-$n$ ladders
are fractal structures in the limit $n\rightarrow\infty$, and their fractal dimension can be derived from the preceding inflation rules. Denoting the
number of $A$ ($B$) segments in an order-$n$ ladder as $\alpha _n$ ($\beta _n$), the growth under $\xi$ is specified by the inflation matrix, 
\begin{equation}\label{eq:inflationmatrix}
    \begin{pmatrix} \alpha_{n+1} \\ \beta_{n+1} \end{pmatrix} = \begin{pmatrix} 1 & 4 \\ 1 & 1 \end{pmatrix} \begin{pmatrix} \alpha_{n} \\ \beta_{n}
\end{pmatrix}.
\end{equation}
The largest eigenvalue of this matrix is 3, and so the number of both $A$ and $B$ segments in a ladder grows by 3 under inflation, in the above
limit. It would appear to follow that the length of a ladder likewise increases by a factor of 3 under inflation. However, there is a subtlety here in the fact that the orientations of $A$ and $B$ segments are \emph{reversed} under a single
inflation; this is easily resolved by considering the action under $\xi^2$. Then the length of a ladder increases by a factor of 9, whereas all edge lengths
are scaled by the silver ratio $\delta_S^2$. The (box-counting) fractal dimension $d_F$ of the ladders is then found by setting $(\delta_S^2)^{d_F} =
9$,
from which
\begin{equation}\label{fractal_dim}
    d_F = \frac{1}{\log_3 \delta_S} \approx 1.246\ldots
\end{equation}
This is similar to the case on the Penrose tiling, where the membranes were found to have fractal dimension $1/\log_2
\varphi$~\cite{Flicker_Simon_Parameswaran}, where $\varphi$ is the golden ratio. 

The statement that an $8_n$-vertex is the centre of concentric
ladders of orders $0\leq m\leq n$ follows from the fact that all $8_0$-vertices are enclosed by stars, and an order-$n$ ladder inflates from a star.
Consequently, the number of ladders at a given order $n$ is in one-to-one correspondence with the number of $8$-vertices of order $m \geq n$. This in turn
can be computed by counting the number of {\it all} $8$-vertices on the tiling obtained by deflating $n$ times. Given a tiling with $N$ vertices, for
$N$ large, the $n$-fold deflated tiling has $\sim N/\delta_S^{2n}$ vertices, a fraction $ 1/\delta_S^4$ of which are $8$-vertices. These statements become exact as $N\to\infty$. Combining these results, we see that the number $\mathcal{N}_{n}$ of order-$n$ ladders on an $N$-site tiling is given by 
\begin{equation}
  \mathcal{N}_{n}  \equiv \nu_n N \hspace{0.5cm} \text{with} \hspace{0.5cm} \nu_n \to \frac{1}{\delta_S^{4+2n}} \text{ as } {N\to \infty}. \label{eq:Nn}
\end{equation}
Here, $\nu_n$ is the frequency of order-$n$ ladders, equal to the sum of frequencies of $8_m$-vertices for $m\geq n$ (so $\nu_0 \equiv \nu_{v8}$ in the
notation of section \ref{sec:perfect_matching_proof}). Note that $N$ refers here to the number of vertices on the `parent' \AB~tiling; the number of vertices on the \ABB~tiling is given by eliminating the $N/\delta_S^4$ 8-vertices, with

\begin{equation} 
  \lim_{N\to\infty} N_* = N(1-\delta_S^{-4}). 
  \label{eq:Nstar}
\end{equation}

\section{Statistical mechanics of the \ABB\ tiling: transfer matrix calculations}
\label{app:transfer_matrices}

%

We consider the statistical mechanics of perfect matchings of the \ABB\ tiling, which, as noted in Sec.~\ref{sec:memproof}, is particularly simple on
account of the membrane constraints. This allows us to derive an exact asymptotic form for the free energy of perfect matchings on the \ABB\ tiling, and compute e.g.\ monomer
correlations on the ladders.

The most general partition function for the dimer
problem on a graph $G$ is a weighted sum over all dimer configurations $\mathcal{C}$, 
\begin{equation}
    Z = \sum_{\mathcal{C}}w(\mathcal{C}).
    \label{eq:partition_function}
\end{equation}
Restricting to equally weighted perfect matchings (i.e.\ $w=1$ if $\mathcal{C}$ is a perfect matching, otherwise $w=0$), $Z$ simply counts perfect
matchings. Given a perfect matching of $G$, all other perfect matchings can be obtained by augmenting alternating cycles
(Sec.~\ref{sec:matching_review}). On \ABB\, an alternating cycle cannot intersect a membrane. Therefore, the enumeration of
configurations on \ABB\ is obtained as the product over the enumerations on each ladder. That is, the partition function factorizes over ladders, with
each ladder defining an independent 1D dimer problem. All ladders of order $n$ have the identical partition function $Z_n$, and so the partition function in the thermodynamic limit, $Z_*$, is simply 
\begin{equation}
Z_* =\prod_{n=0}^\infty Z_n^{\mathcal{N}_n}.
\label{eq:Z*}
\end{equation}
where $\mathcal{N}_n$ is the number of order-$n$ ladders on the tiling, given in Appendix~\ref{app:ladder_infl}. The one dimensional nature of the
stars and ladders suggests that the partition functions $Z_n$ can be efficiently calculated in terms of transfer matrices. To outline our approach
for the \ABB~ladders, we first consider the simpler problem of enumerating coverings of a closed periodic ladder
consisting only of square tiles.

The basic repeating unit of the ladder can be directly lifted to write down the transfer matrix in graphical notation:
\begin{figure}[h!]
\centering
\begin{tikzpicture}
    \draw (-0.75,0.5) node[left] {$i$} -- (0.75,0.5)  node[right] {$i'$};
    \draw (-0.75,-0.5) node[left] {$j$} -- (0.75,-0.5) node[right] {$j'$};
    \draw (0,-0.5) -- node[right] {$\alpha$} (0,0.5);
    \filldraw [black] (0,-0.5) circle (2pt);
    \filldraw [black] (0,0.5) circle (2pt);
\end{tikzpicture}
\end{figure}

\noindent which simply assigns an index to each edge of the unit. Each index runs over the two values $(0,1)$, representing the edge in an uncovered state and a
state with a dimer. The transfer matrix entry corresponding to a given set of indices $(iji'j'\alpha|$ is assigned a weight $w=1$ if the
corresponding state obeys the hard-core dimer constraint and perfectly matches the two black vertices, otherwise we set $w=0$. For example, $(10010|$ is a valid dimer
configuration, but $(11001|$ is not. The internal index $\alpha$ is then summed over, and the left (right) indices are combined into a single index $(ij|$
($(i'j'|$) to enable us to write down the transfer matrix $\mathcal{P}_{iji'j'} \equiv (ij|i'j')$, or
\begin{equation} 
    \label{eq:transfer1}
    \mathcal{P} = \begin{pmatrix} 1&1&0&0\\1&0&0&0\\0&0&0&1\\0&0&1&0\end{pmatrix}.
\end{equation}
Note that we have chosen to order the indices as $(00|,\ (11|,\ (10|,\ (01|$ to take advantage of the block diagonal nature
of $\mathcal{P}$. This splitting reflects a more fundamental property of the configuration space of perfect matchings, wherein the space is further subdivided
into topological sectors: while all configurations are reachable from one another via alternating cycles, only configurations in the same sector are connected via
\emph{local} cycles; configurations in distinct topological sectors can only be connected via cycles which wind around the entire system (in one of the directions with
periodic boundary conditions). The periodic square ladder has three possible sectors, $s=0, \pm1$. The $s=\pm1$ sectors contain only one configuration each. These are the `staggered' configurations with dimers alternating between the top and bottom legs of
the ladder. These are captured by the lower block of $\mathcal{P}$. All other configurations belong to the $s=0$ sector and are enumerated by
the upper block matrix.  For a closed periodic ladder consisting of $M$ plaquettes, the partition function is then\footnote{The same result appears in models of chains of Rydberg atoms~\cite{Lesanovsky,
Turner18}, owing to a mapping between the dimer constraint and the `Rydberg blockade' constraint, which forbids two neighbouring atoms to be in
simultaneously excited states.} 
\begin{equation}
    \tilde{Z}_M = \Tr(\mathcal{P}^M) = F_{M+1}+F_{M-1}+2,
  \label{eq:periodicladder}
\end{equation}
with $F_M$ the $M$-th Fibonacci number, and $F_1 = F_2 = 1$. The term $+2$ is the staggered contribution.

The aperiodic ladders defined by Eq.~\eqref{eq:ladderstruc} are treated in an analogous fashion. Clearly (\ref{eq:transfer1}) still defines the transfer
matrix between two square tiles. We similarly introduce the matrix for an $A$-section rhombus (left) and the $B$-section central ($R$-type) rhombi (right) as 
\begin{figure}[h!]
\begin{tikzpicture}
    \draw (0,0.92388) -- (0.38268,0);
    \draw (0,0.92388) -- (-0.38268,0);
    \draw (0,-0.92388) -- (0.38268,0);
    \draw (0,-0.92388) -- (-0.38268,0);
    \draw (0.38268,0) -- (1.13268,0) node[right] {$i'$};
    \draw (-1.13268,0) node[left] {$i$} -- (-0.38268,0) ;
    \draw (0,-0.92388) -- (0.75,-0.92388) node[right] {$j'$} ;
    \draw (-0.75,-0.92388) node[left] {$j$} -- (0,-0.92388) ;
    \filldraw [black] (0,0.92388) circle (2pt);
    \filldraw [black] (0.38268,0) circle (2pt);
    \filldraw [black] (0,-0.92388) circle (2pt);
    \filldraw [black] (-0.38268,0) circle (2pt);
\end{tikzpicture}
  \begin{tikzpicture}
    \draw (0,0) -- (0.70712,0.70712);
    \draw (0,0) -- (-0.70712,0.70712);
    \draw (0,0) -- (0,-1);
    \draw (0,-1) -- (-0.70712,-0.29290);
    \draw (0,-1) -- (+0.70712,-0.29290);
    \draw (0.70712,0.70712) -- (+0.70712,-0.29290);
    \draw (-0.70712,0.70712) -- (-0.70712,-0.29290);
    \draw (0,-1) -- (0.75,-1) node[right] {$j'$};
    \draw (-0.75,-1) node[left] {$j$} -- (0,-1);
    \draw (+0.70712,-0.29290) -- (+1.45712,-0.29290) node[right] {$i'$};
    \draw (-1.45712,-0.29290) node[left] {$i$} -- (-0.70712,-0.29290);

    \filldraw [black] (0,0) circle (2pt);
    \filldraw [black] (0.70712,0.70712) circle (2pt);
    \filldraw [black] (-0.70712,0.70712) circle (2pt);
    \filldraw [black] (0,-1) circle (2pt);
    \filldraw [black] (+0.70712,-0.29290) circle (2pt);
    \filldraw [black] (-0.70712,-0.29290) circle (2pt);
\end{tikzpicture}
\end{figure}

\noindent with the internal states of the rhombi again summed over and only the external legs $(iji'j')$ `free'. These are written explicitly as
\begin{equation}
    \mathcal{Q} = \begin{pmatrix} 2&1&0&0 \\ 1&0&0&0 \\ 0&0&0&1 \\ 0&0&1&0 \end{pmatrix}, \hspace{1cm} \mathcal{R} = \begin{pmatrix} 3&1&0&0 \\ 1&0&0&0 \\ 0&0&0&1 \\ 0&0&1&0 \end{pmatrix} \hspace{.5cm}.
\end{equation}
It is seen that each rhombus simply adds a state for which none of the free legs hosts a dimer. Again the ladders decouple into three topological
sectors, with two staggered configurations entirely analogously to the periodic ladders. 

The three matrices $\mathcal{P},\ \mathcal{Q},\ \mathcal{R}$ allow for exact enumeration of the ladder configurations. Unfortunately, they do not commute and so cannot be
simultaneously diagonalised --- consequently, each $Z_n$ calculation requires multiplication of an ever-longer (exponentially growing) string of
matrices which tends to quasiperiodicity in the thermodynamic limit. The inflation rules defined on the ladders, however 
(\ref{eq:xiinflation}), allow the traces to be computed efficiently (approximately linearly in $n$) to arbitrary finite order. 

We take advantage of the block diagonal nature of the matrices to keep only the $s=0$ block. The staggered sector contributes two configurations to
each ladder partition function. Then we introduce two matrices for the basic $A$ and $B$ sections of the ladders:
    \begin{align}\label{eq:A&Bmatrices}
    \mathcal{A}_1 \equiv \mathcal{P}'\mathcal{Q}' = &\begin{pmatrix} 3&1 \\ 2&1 \end{pmatrix} \hspace{.5cm} \nonumber \\
    \mathcal{B}_1 \equiv \mathcal{P}'^2\mathcal{R}'\mathcal{P}' = &\begin{pmatrix} 9&7 \\ 5&4 \end{pmatrix}.
\end{align}
(Primed matrices represent the upper block of the respective matrix.) Here we used the fact that the transfer matrix for the $R'$ rhombi appearing in
the $B$ section is the same as the square's transfer matrix, $\mathcal{P}$ (or $\mathcal{P}'$). The inflation rule $\xi$ in (\ref{eq:xiinflation}) then carries over straightforwardly to the generalized transfer matrices $\mathcal{A}_n$ and $\mathcal{B}_n$ via 
\begin{equation}
  \mathcal{A}_{n+1} = \mathcal{A}_n \mathcal{B}_n, \hspace{.5cm} \mathcal{B}_{n+1} = \mathcal{A}_n^4\mathcal{B}_n.
  \label{eq:matmul}
\end{equation}
The partition function of the order-$n$ ladder in the $s=0$ sector is thus
\begin{equation}
  Z_n^{(0)} =  \Tr\left[\mathcal{A}^8_n\right].
  \label{eq:Ztrace}
\end{equation}
which effectively solves (\ref{eq:Z*}) as the $\mathcal{A}_n$ can be computed recursively. We note that $Z_0$, the partition function of the stars, is
equal to two: these two configurations can be seen as analogues to the ladder's two staggered states (the only alternating cycle connecting the two
configurations winds all the way around the star). 

The free energy density per edge of the \ABB\ tiling in the thermodynamic limit is given in terms of the ladder partition functions as 
\begin{align}
    f_* = \lim_{N\to\infty} -\frac{\ln Z_*}{N_{E*}} = - \frac{1}{6(1+4\delta_S)} \sum_{n=0}^\infty \frac{\ln Z_n}{\delta_S^{2n}}
\end{align}
or with $(\ref{eq:Ztrace})$,
    \begin{equation}
    f_* = -\frac{1}{6(1+4\delta_S)}\Bigg[\ln 2 +\sum_{n=1}^\infty\frac{\ln\big(\Tr\left[\mathcal{A}^8_n\right]+2\big)}{\delta_S^{2n}}\Bigg].
    \label{eq:fstar}
\end{equation}
where we have used Eq.~\eqref{eq:Nn} and Eq.~\eqref{eq:Nstar}, and simplified powers of $\delta_S$ using Eq.~\eqref{eq:delta_S_def}. Here $N_{E^*}$ is
the number of edges on the \ABB\ tiling in the thermodynamic limit, calculated in terms of the number of vertices $N$ of the \AB\
tiling as $N_{E*} = N(4-8\delta_S^{-4})/2$. The infinite series in Eq.~\eqref{eq:fstar} converges exponentially to its limit, on account of the
exponential drop-off in the frequency at which higher-order ladders occur. Using the inflation, it is possible to bound the error incurred in the free energy from truncating the ladder series at some finite $M$ (the
proof is simple but not particularly insightful), 
\begin{align}
    \left|\delta f_*^M \right| < \frac{1}{24(1+4\delta_S)\delta_S^{2M+1}} \big(\Gamma_M+\zeta\big), \nonumber \\
    \Gamma_M = 3^M(1+\delta_S)(4\ln a_1^+ + 2\ln b_1^+), \nonumber \\
    \zeta = (\delta_S-1)(4\ln a_1^+ - 2\ln b_1^+) + 2\ln 2.
\end{align}
where we introduce the maximum and minimum eigenvalues of $\mathcal{A}_1$ as $a_1^+$ and $a_1^-$ respectively, and those of $\mathcal{B}_1$ as
$b_1^+$ and $b_1^-$. Taking for example the first 40 terms in the ladder
summation, we find $f_* = -0.06884471896847(17)$.

To calculate the partition functions in the presence of an aligning interaction (see Eq.~\eqref{eq:aligning_ints}), we define new transfer matrices
expressed in a basis of plaquette states. For the square plaquette, we take the three basis states as 

\begin{figure}[h!]
\includegraphics[width=0.7\linewidth]{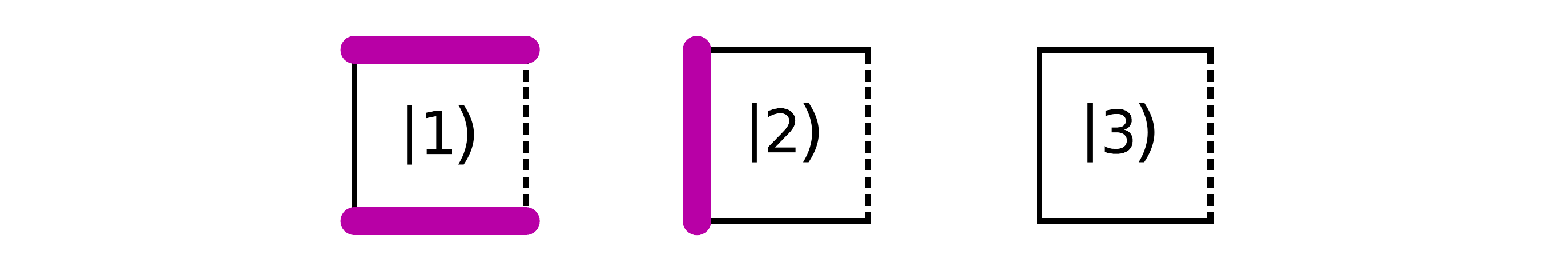}
\label{fig:plaq_basis}
\end{figure}

which leads to
\begin{equation}
    \mathcal{P}'_{\text{int}} = \begin{pmatrix} 0&0&\kappa\\1&\kappa&0\\1&1&0 \end{pmatrix}.
\end{equation}
The weights are given by $w = \kappa \equiv e^{-V/T}$ if $|i,j)$ is a flippable dimer configuration, $w=1$ if $|i,j)$ is a non-flippable configuration,
and $w=0$ if $|i,j)$ is not a valid (hard-core) dimer configuration. Only whole plaquettes i.e.\ those not including the dotted edge, can be counted as
flippable --- this avoids overcounting. The matrices for the rhombi sections are defined in the same basis, mapping from the square on the right of
the rhombi to the one on the left, summing over intermediate states: 

\begin{equation}
    \mathcal{Q}'_{\text{int}} = \begin{pmatrix} 0&0&\kappa \\ 1&\kappa&\kappa^2 \\ 1&\kappa&\kappa \end{pmatrix}, \hspace{.5cm} \mathcal{R}'_\text{int}
    = \begin{pmatrix} 0&0&\kappa \\ 1&\kappa&2\kappa^2 \\ 1&\kappa&\kappa+\kappa^2 \end{pmatrix}.
\end{equation}
Note we are again working in the $s=0$ block, and the staggered configurations give a
contribution of $+2$. Multiplication of these transfer matrices according to the ladder structures gives the analytic curves in
Fig.~\ref{fig:ladd_cv}.

Transfer matrices can also be employed to compute more complicated correlations. Here we outline how to compute monomer-monomer correlations, but
the extension to dimer-dimer and higher order correlators is not more difficult. Returning first to the periodic ladder, it can be observed that the
sector number $s$ changes by $\pm1$ either side of a monomer (below).

\begin{figure}[h!]
\includegraphics[width=0.7\linewidth]{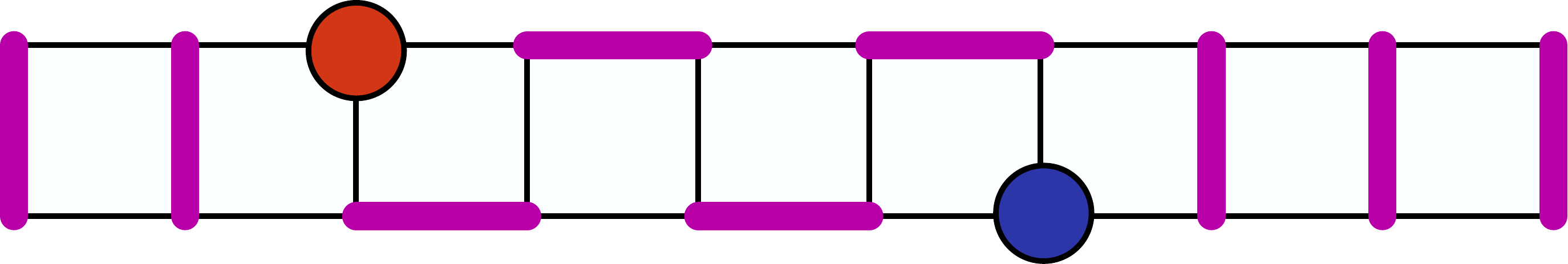}
\label{fig:monomerladder}
\end{figure}
 
The monomers therefore provide a matrix element between the staggered $s=\pm1$ sectors and the $s=0$ sector, which were previously disconnected in the
space of perfect matchings. As noted in the main text, the monomer-monomer correlation function is just the partition function with two monomers at
$x$ and $y$, $Z(x,y)$ (up to normalisation). A monomer $\mathcal{U}$ ($\mathcal{V}$) residing on the upper (lower) black vertex of the graphical square-square transfer matrix will act as 
\begin{equation}
    \mathcal{U} = \begin{pmatrix} 0&0&0&1\\0&0&0&0\\0&0&0&0\\1&0&0&0\end{pmatrix}, \hspace{.5cm} \mathcal{V} = \begin{pmatrix}
0&0&1&0\\0&0&0&0\\1&0&0&0\\0&0&0&0\end{pmatrix}.
\end{equation} 
The correlation function is then calculated by replacing $\mathcal{P}$ by $\mathcal{U}$ or $\mathcal{V}$ at the position of the
monomers, and taking the trace over the transfer matrix product as usual. For example, for the figure above (assuming periodic boundary conditions),
we have 
\begin{equation} 
    \tilde{Z}(0,4) = \text{Tr}\big(\mathcal{P}^4\mathcal{U}\mathcal{P}^3\mathcal{V}\big) = 5. 
\end{equation}

The calculations for the aperiodic ladders go through almost the same. The only difference is that the monomers can reside in several distinct
environments along the ladder (any of the black vertices in the graphical diagrams). For example, a monomer
residing on the `point' of the rhombus in the $A$-section graphical diagram has the matrix 
\begin{equation}
    \mathcal{U} = \begin{pmatrix} 0&0&1&0 \\ 0&0&1&0 \\ 1&1&0&0 \\ 0&0&0&0 \end{pmatrix}.
\end{equation}

We are primarily interested in the asymptotic behaviour of the monomer-monomer correlator at large separations. The main physical point to note is that the monomers once again mix the staggered and $s=0$ sectors, with a string of staggered states stretching
between two monomers. Since it is the $s=0$ sector that primarily contributes to the entropy of the ladders, the monomer-monomer correlators are
exponentially suppressed in the monomer separation (i.e.\ in the length of the staggered string). 

From the idea of effective matching problem, we
expect the correlators for all odd-$i$ $\mathcal{H}_i$ regions to be controlled by the correlator of the $L_1$ ladder, and those for all even-$i$
regions to behave like the star. The asymptotics of the $L_1$ ladder are accessed by considering the
large-$n$ limit of monomer
correlations on the extended ladder $L_1(n)$ formed by ajoining $n$ identical $A$-sections. While the overall normalisation of the correlations
depends on the environments of the two monomers, we expect from the above discussion that the leading order behaviour of the correlator depends only on the largest eigenvalue of
the $\mathcal{A}_1$ matrix, i.e.
\begin{equation}
    Z_1(x,y) \propto \bigg(\frac{1}{a_1^+}\bigg)^{n} + \ldots
\end{equation}
where $x$ lies in $\mathcal{S}_0$ and $y$ in $\mathcal{S}_n$. With $a_1^+ = 2+\sqrt{3}$, we obtain the form quoted in Eq.~\ref{eq:confined_leading}. The same problem for the star gives the trivial result
that whenever two monomers are present, only one compatible dimer configuration exits. Thus, the star's monomer-monomer correlator is independent of the
monomer separation,
\begin{equation}
    Z_0(x,y) = 1. 
\end{equation}

%
\section{Perfectly matched $\mathcal{H}$ regions}
\label{app:proof_punctured}
\begin{figure}[t]
\includegraphics[width=\columnwidth]{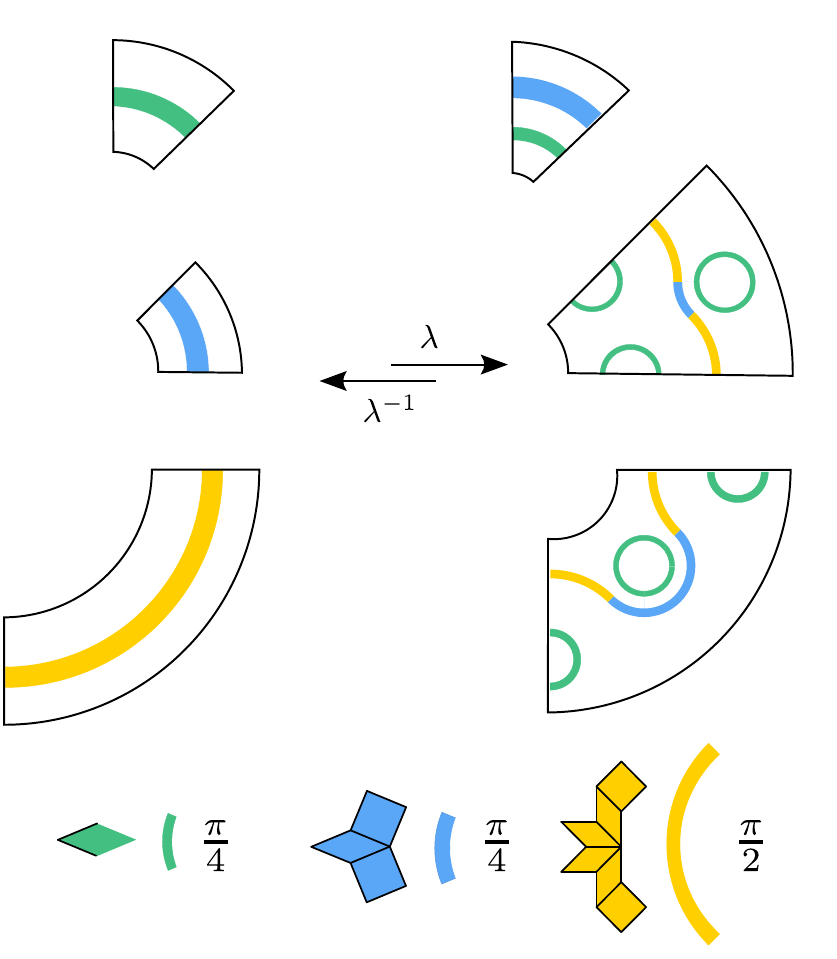}
\caption{Starting from $\mathcal{H}_1$, the order-1 ladder, an inflation rule $\lambda$ generates concentric, $D_8$-symmetric regions $\mathcal{H}_i=\lambda^i(\mathcal{H}_1)$
which can be perfectly matched, such that all sites in an $8_n$-empire belong to one of the $\mathcal{H}_i$ for some $i$. We have presented the inflation rules
in terms of ladder segments of Fig.~\ref{fig:abbtiles}. As before, the ladder segments are represented by coloured arcs of internal angles
$\pi/4$,$\pi/4$,$\pi/2$ for the segments $P$, $A$ and $B$ respectively. Each region $\mathcal{H}_i$ is composed of closed ladders as well as the links
connecting those ladders. For clarity, we have suppressed both the $8$-vertices located at the centre of the green circles (stars) and the
links between the ladder segments.}
  \label{fig:proofpunc1}
\end{figure}
\begin{figure*}
\setlength\fboxsep{0pt} \setlength\fboxrule{0.0pt} \fbox{\includegraphics[width=15cm]{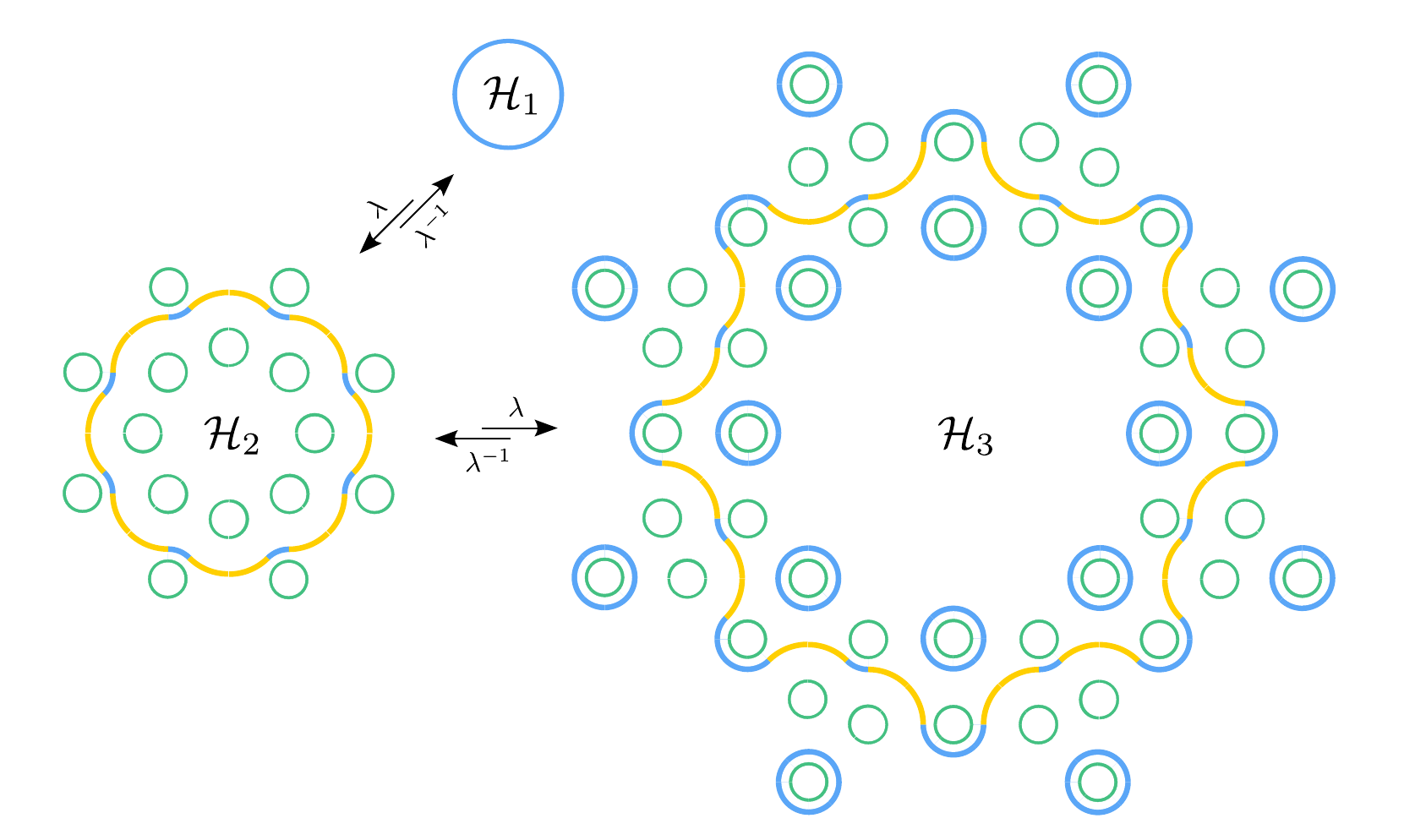}}
\caption{Using the inflation rules of Fig.~\ref{fig:proofpunc1}, we construct the first two inflated regions $\mathcal{H}_2$ and $\mathcal{H}_3$, starting from the
order-1 ladder $\mathcal{H}_1 \equiv L_1$. Ladder segments $P$, $A$, and $B$ are represented by coloured arcs as in
Fig.~\ref{fig:ladderinflate}. The $\mathcal{H}$ regions are composed of closed ladders as well as the links connecting those ladders. Each
$\mathcal{H}_i$ region admits a perfect matching. }
\label{fig:proofpunc2}
\end{figure*}

In Sec.~\ref{sec:memproof}, we showed that the \ABB~tiling  hosts a perfect matching, and membranes separate stars and ladders, which host dimers in
the perfect matching. Here, we consider punctured $8_n$-empires, obtained by removing the central $8_n$-vertex from the $8_n$-empire, and prove that
when such $8_n$-empires are terminated with certain $D_8$-symmetric boundary conditions  they host a perfect matching. Further, concentric membranes
around the absent central $8$-vertex separate perfectly matched components, which we label as $\mathcal{H}_i$. 
These statements can be proven as follows:
 
\begin{itemize}
  \item Starting with $\mathcal{H}_1$, which we take as the first closed ladder ($L_1$, in the language of Sec.~\ref{sec:memproof}) which surrounds the absent
      $8$-vertex and the star around it, there exists an inflation rule $\lambda$ such that $\mathcal{H}_{i+1}=\lambda^i(\mathcal{H}_1)$ are mutually exclusive
      $D_8$-symmetric regions, concentric with the central $8$-vertex, and all vertices of the punctured $8_n$-empire belong to one of
      the $\mathcal{H}_i$ for some $i$. In an $8_n$-empire, the central $8$-vertex is surrounded by a star (we define $\mathcal{H}_0$ to be the
      star, $\mathcal{H}_0 \equiv L_0$), which in turn is surrounded by $\mathcal{H}_1$, and a region $\mathcal{H}_i$ is surrounded by the region $\mathcal{H}_{i+1}$.
    The inflation rule $\lambda$ can be easily read off from the inflation of the ladder tiles in Fig.~\ref{fig:ladderinflate}. Vertices in a ladder are no
    longer entirely matched to vertices within the same ladder (dimers are placed on the membranes). Howerver, it is convenient to describe the inflation rules in terms of the
    ladder segments. A region $\mathcal{H}_i$ is composed of certain closed ladders that follow from the inflation, as well as the links connecting them,
    which can be read off from Fig.~\ref{fig:abbvertices} (previously membranes on the \ABB\ tiling). We present the $\lambda$ inflation rule in
    Fig.~\ref{fig:proofpunc1}. From $\lambda$, all regions $\mathcal{H}_i$ can be constructed starting from $\mathcal{H}_1$. The first two regions constructed using the inflation rules of Fig.~\ref{fig:proofpunc1} are displayed in Fig.~\ref{fig:proofpunc2}.

  \item Second, each component $\mathcal{H}_i$ generated by the inflation rule $\lambda$ can be perfectly matched. Such a perfect matching can be constructed
      following the arguments of Sec.~\ref{sec:perfect_matching_proof} used to construct a perfect matching of the \AB~tiling: $\mathcal{H}_i$ contains ladder
      segments which can be perfectly matched, using the dimer-decorated ladder tiles displayed in Fig.~\ref{fig:abbtiles}. The $8$-vertices (located
      at the centres of green circles in Figs.~\ref{fig:proofpunc1} and \ref{fig:proofpunc2}, not shown)  now lie at the \emph{vertices} of the
      component $\mathcal{H}_{i-2}=\lambda^{-2}(\mathcal{H}_i)$, with edge-lengths larger by a factor of $\delta_s^2$. Each edge
      of the larger $\mathcal{H}_{i-2}$ component implies an odd-length alternating path between the corresponding $8$-vertices in $\mathcal{H}_i$, and can be augmented
      to match the $8$-vertices. This follows directly from applying the augmenting paths for the dimer-inflation tiles,
      Fig.~\ref{fig:dimerinflrules}, to connect up the (now reintroduced) $8$-vertices in the ladder tiles (Fig.~\ref{fig:abbtiles}). Since both the
      star and order-1 ladder can be perfectly matched, all $\mathcal{H}_i$ can be perfectly matched.
      If boundary conditions are imposed on an $8_n$-empire such that all vertices outside the largest component $\mathcal{H}_n$ are excluded, then
      the truncated empire hosts a perfect matching. Now we argue the existence of membranes within the perfect matching.

  \item  Each component $\mathcal{H}_i$ has the property that if the central $8$-vertex is (say) a $\mathcal{U}$-vertex, all vertices on the inner
      boundary of $\mathcal{H}_i$ (towards the central $8$-verex) are $\mathcal{V}$-vertices while those on the outer boundary are
      $\mathcal{U}$-vertices. This can be seen by first noting that this is true for $\mathcal{H}_1$. If $\mathcal{H}_i$ has ladder segments with
      $\mathcal{U}$-vertices at a boundary, inflations of Fig.~\ref{fig:proofpunc1} result in segments with $\mathcal{U}$-vertices at the boundary.

      In a perfect matching, vertices in the smallest component $\mathcal{H}_0$ (the star surrounding the absent $8$-vertex) must be perfectly matched to vertices within $\mathcal{H}_0$. 
      $\mathcal{H}_0$ has an equal number (8) of  $\mathcal{U}$- and $\mathcal{V}$-vertices, with  only $\mathcal{U}$-vertices having edges connecting
      them to the rest of the graph. In a perfect matching, this constrains the $\mathcal{U}$-vertices to match to $\mathcal{V}$-vertices on the
      interior of $\mathcal{H}_0$, lest the $\mathcal{V}$-vertices remain unmatched. For the component $\mathcal{H}_1$,  the outer boundary has  only $\mathcal{U}$-vertices having edges connecting them to vertices in $\mathcal{H}_2$. At the inner boundary, only $\mathcal{V}$-vertices have edges connecting them to vertices in $\mathcal{H}_0$, but these edges cannot be matched as vertices in $\mathcal{H}_0$ are always matched to vertices within $\mathcal{H}_0$. This implies that vertices in $\mathcal{H}_1$ are matched to other vertices in $\mathcal{H}_1$ in all perfect matchings.
       
      Extending this argument implies that for all components $\mathcal{H}_i$, vertices of $\mathcal{H}_i$ are matched to vertices within $\mathcal{H}_i$. $\mathcal{H}_i$ are the perfectly matched components $G^{s}_i$ of the Dulmage-Mendelsohn decomposition reviewed in Appendix~\ref{app:memreview}.

  \end{itemize}
   We have shown that if we choose boundary conditions of the punctured $8_n$-empire  which exclude vertices lying outside the outermost component
   $\mathcal{H}_n$,  then the punctured $8_n$-empire hosts a perfect matching with $n$ concentric membranes.

\section{The directed loop algorithm}
\label{app:numericalmethods}
%

In this appendix we summarize the directed-loop algorithm for sampling dimer configurations/maximum matchings, as originally introduced in Refs.~\cite{Sandvik_Moessner} and \cite{Alet_etal}. Note that  this approach  works both for maximum as well as perfect matchings, so we develop the discussion without specializing to the perfect-matching case. Given a maximum matching, the algorithm generates new maximum matchings as follows:
\begin{enumerate}
\item \underline{Start} with a maximum matching. Randomly pick a matched vertex $s_{0}$. Let $s_0$ be matched to $s_1$.   Set $s_{i}=s_0$ and $s_{j}=s_{1}$. 
\item \underline{Pivot:} Randomly choose a neighbour of $s_j$, say $s_k$. (The probability governing this choice is described below.) Remove the dimer on the edge $(s_i,s_j)$ and place a dimer on the edge $(s_j,s_k)$. This is the elementary step of the update, where a dimer `pivots' over the vertex $s_j$ from the edge $(s_i,s_j)$ to the edge $(s_j,s_k)$. Note that in the first step, the dimer-pivoting move leaves an extra monomer  at $s_0$.
\item \underline{Stop} if $s_k$ hosted a monomer before the pivot.
\item \underline{Grow:} if $s_k$ hosts a dimer $(s_k,s_m)$ with $s_m\neq s_j$, the intermediate configuration has a monomer at the starting vertex $s_0$, and an \emph{antimonomer} (two dimers touching a site) at the vertex $s_k$.  Set $s_i=s_k, s_j=s_m$, and {go to Step 2 (Pivot)}.   
\end{enumerate}
Intuitively, each step of  the algorithm creates a monomer-antimonomer defect-pair and moves the antimonomer around until it annihilates with another monomer.  The value of  $s_k$ when the procedure terminates determines which of  two possible updates have been implemented: $s_k=s_0$ corresponds to flipping the dimer-occupancies in a closed alternating path of edges (a loop update), while $s_k \neq s_0$ corresponds to transporting a monomer from $s_k$ to $s_0$ (string update).

The utility of the algorithm lies in the fact that any two maximum matchings can be connected by a sequence of loop or string updates, and so it can sample the whole configuration space of maximum matchings.

To ensure {detailed balance} in the space of maximum matchings generated by the update,  we implement it in the enlarged configuration space that includes all maximum matchings as well as the intermediate configurations generated by the algorithm. The latter correspond to configurations which have an extra monomer-antimonomer pair relative to a maximum matching.
Imposing detailed balance  at each step of the update yields the transition probabilities $P_{ij;jk}$ for a dimer to pivot on a vertex $j$ from the edge $(s_i,s_j)$ to the edge $(s_j,s_k)$:
\begin{align}
  \label{eq:db1}
  w_{ij} P_{ij;jk} &= w_{jk} P_{jk;ij} \\
  \label{eq:db2}
  \sum_{(jk)}P_{ij;jk}&=1.
\end{align}
$w_{ij}$ is the weight contributed to the partition function by a dimer on the edge $(s_i,s_j)$.
In general, the system of equations \eqref{eq:db2} subject to the constraints \eqref{eq:db1} is underdetermined. It is necessary to look for solutions which minimize the probabilities $P_{ij;ij}$ of the loop retracing itself. Such solutions can be found using linear programming techniques~\cite{Wessel_Alet_Troyer}. If the dimers are non-interacting, as is the case in most of this paper, the backtracking probabilities can be set to zero. In this case, for an $n$-coordinated vertex $s_j$, $P_{ij;jk}=1/(n-1)$ for all edges $(s_j,s_k)\neq (s_i,s_j)$. 

Since the algorithm respects detailed balance in the extended configuration space with the monomer-antimonomer pair created while making the loop, it
affords access to the partition function $Z_{\rm ma}$.  This involves configurations with the same number of dimers as the maximum matching, but with
one additional monomer and one additional antimonomer. The loop update samples this partition function with the correct weights. This is very nearly
the quantity we are interested in when understanding questions of confinement, though there one usually considers a closely related partition function
$Z_{\rm mm}$, which involves two more monomers than the maximum matching.  In fact, the loop-construction procedure outlined above could equivalently
be described as creating two monomers in a maximum matching, and propagating one of the monomers until it annihilates with another monomer to give a
new maximum matching. To be precise, a step where the dimer pivots on the vertex $s_j$ from the edge $(s_i,s_j)$ to the edge $(s_j,s_k)$ can be
equivalently described in terms of a monomer hopping from the vertex $s_j$ to the vertex $s_m$ (which is matched to $s_k$), while another monomer is
fixed at the starting vertex $s_0$. The dimer on the edge $(s_k,s_m)$ is moved to  the edge $(s_j,s_k)$ during this hop. However, the detailed balance
equations are satisfied with respect to the partition function $Z_{\rm ma}$ instead of $Z_{\rm mm}$. To sample from $Z_{\rm mm}$ correctly we weight
each intermediate configuration generated in the loop update (with monomers at $s_j$ and $s_0$) with a factor of $(\sum_k w_{jk})^{-1}$. Measurement of $Z_{\rm mm}$ closely corresponds to the monomer correlations as discussed in  the main  text.

\begin{figure*}
\setlength\fboxsep{0pt} \setlength\fboxrule{0.0pt} \fbox{\includegraphics[width=12cm]{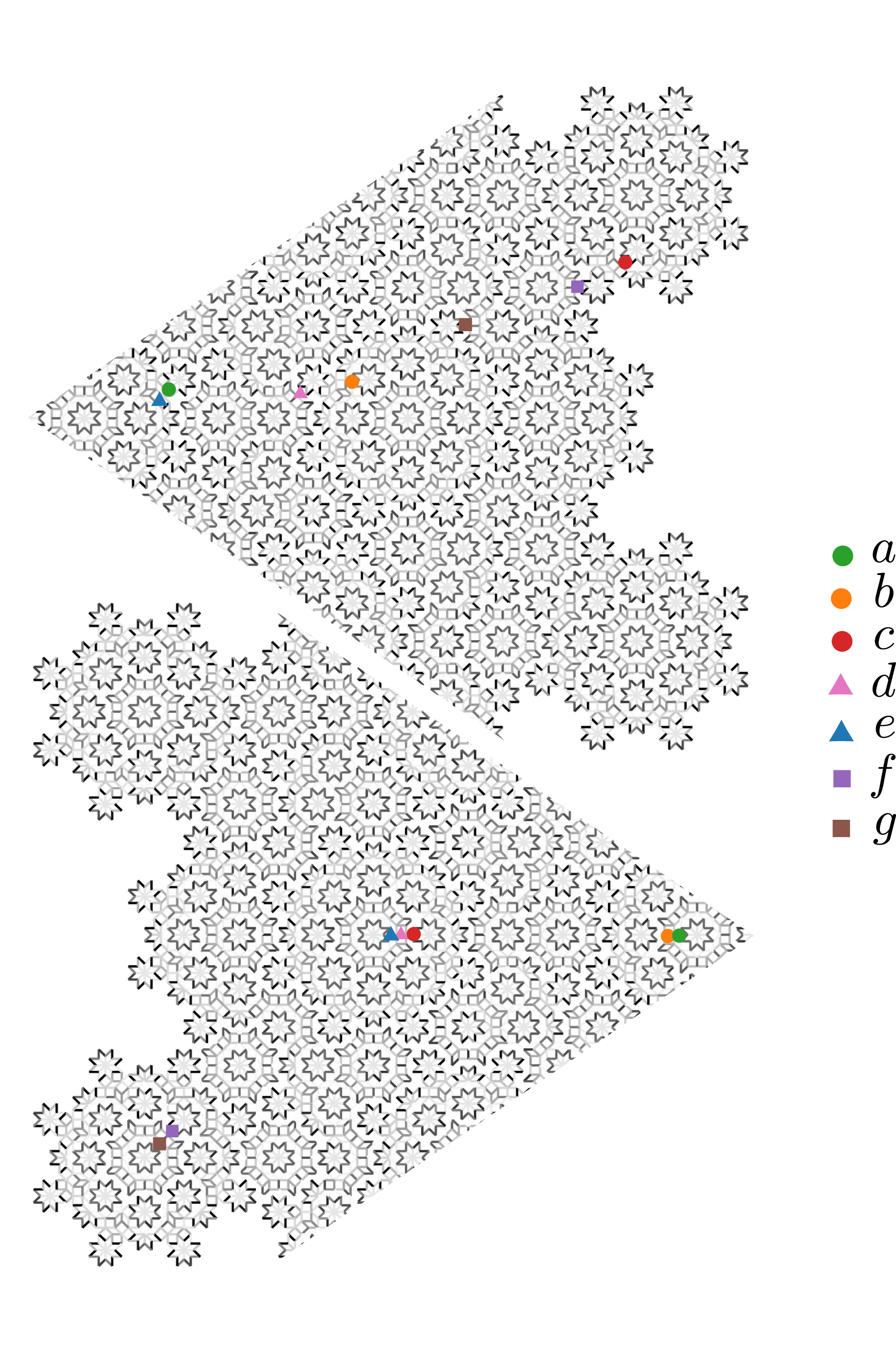}}
\caption{Top: The set of source edges $e_0$ considered in Fig.~\ref{fig:ddcorrsplaw}, which are representative of edges for which connected
correlations of dimers decay as power laws. Bottom: The set of source edges $e_0$ considered in Fig.~\ref{fig:ddcorrsbounded}, for which connected
correlations of dimers are bounded within pseudomembranes. The colour of the edges indicate dimer occupation density to reveal the structure of
pseudomembranes. We use the $D_8$-symmetry to choose (and display) source edges within a wedge--- eight of these wedges make up the whole sample. }
\label{fig:sourceedges}
\end{figure*}

%
\section{Source edges for dimer correlations}
\label{app:legend}
%

To investigate connected correlations of dimers, we investigated the quantity $C_{\mathrm Max}(e_0,x)$, the maximum absolute value of the dimer correlation function at a distance of $x$ edges from $e_0$, in Sec.~\ref{sec:dimercorrs}. Fig.~\ref{fig:ddcorrsplaw} shows slow decay of $C_{\mathrm Max}(e_0,x)$, consistent with power laws, for many different source edges $e_0$. Fig.~\ref{fig:ddcorrsbounded} shows that for some other choices of source edges, the connected correlations are bounded within pseudomembranes. For a $8_4$-vertex considered in Sec.~\ref{sec:dimercorrs}, we label the source edges considered in Fig.~\ref{fig:ddcorrsplaw} and Fig.~\ref{fig:ddcorrsbounded} in Fig.~\ref{fig:sourceedges}.

\end{appendix}
%

\bibliography{references}
\end{document}